\documentclass[online]{aa}  
\usepackage[pdftex]{color,xcolor}
\usepackage{graphicx}
\usepackage{amstext} 
\usepackage{amsmath} 
\usepackage{amssymb} 
\usepackage[figuresright]{rotating}
\usepackage{afterpage}
\usepackage{txfonts}

\begin{document} 
   \title{The ESPRI project: astrometric exoplanet search with PRIMA}
   \subtitle{I. Instrument description and performance of first light observations\thanks{\textit{Part of this work is based on technical observations collected at the European Southern Observatory at Paranal, Chile. Public data can be downloaded at \url{http://www.eso.org/sci/activitiess/toBeInserted}.}}} 
\author{J.~Sahlmann \inst{1}
		\and T.~Henning\inst{2} 
		\and D.~Queloz\inst{1}
		\and A.~Quirrenbach\inst{3} 
		\and N.M.~Elias II\inst{3} 
		\and R.~Launhardt\inst{2}
		\and F.~Pepe\inst{1}
		\and S.~Reffert\inst{3}
		\and D.~S\'egransan\inst{1}
		\and J.~Setiawan\inst{2}
		\and R.~Abuter\inst{4}
		\and L.~Andolfato\inst{4}
		\and P.~Bizenberger\inst{2}
		\and H.~Baumeister\inst{2}
		\and B.~Chazelas\inst{1}
		\and F.~Delplancke\inst{4}
		\and F.~D\'erie\inst{4}
		\and N.~Di Lieto\inst{4}
		\and T.P.~Duc\inst{4}
		\and M.~Fleury\inst{1}
		\and U.~Graser\inst{2}
		\and A.~Kaminski\inst{3}
		\and R.~K\"ohler\inst{2,3}
		\and S.~L\'ev\^eque\inst{4}
		\and C.~Maire\inst{1}
		\and D.~M\'egevand\inst{1}		
		\and A.~M\'erand\inst{5}
		\and Y.~Michellod\inst{6} 
		\and J.-M.~Moresmau\inst{4}
		\and M.~Mohler\inst{2}
		\and A.~M\"uller\inst{2}
		\and P.~M\"ullhaupt\inst{6}
		\and V.~Naranjo\inst{2}
		\and L.~Sache\inst{7}
		\and Y.~Salvade\inst{8}		
		\and C.~Schmid\inst{4}
		\and N.~Schuhler\inst{5}
		\and T.~Schulze-Hartung\inst{2}
		\and D.~Sosnowska\inst{1}
		\and B.~Tubbs\inst{2}
		\and G.T.~van Belle\inst{9}
		\and K.~Wagner\inst{2}	
		\and L.~Weber\inst{1}
		\and L.~Zago\inst{10}	
		\and N.~Zimmerman\inst{2}
		}
\institute{Observatoire de Gen\`eve, Universit\'e de Gen\`eve, 51 Chemin Des Maillettes, 1290 Versoix, Switzerland\\
		\email{johannes.sahlmann@unige.ch}	
	\and	
		Max-Planck-Institut f\"ur Astronomie, K\"onigstuhl 17, 69117 Heidelberg, Germany
	 \and 
		 Landessternwarte, Zentrum f\"ur Astronomie der Universit\"at Heidelberg, K\"onigstuhl 12, D-69117 Heidelberg, Germany
	 \and
		 European Southern Observatory, Karl-Schwarzschild-Str. 2, 85748 Garching bei M\"unchen, Germany
	\and
		 European Southern Observatory, Alonso de C\'ordova 3107, Vitacura-Santiago, Chile
	 \and	
	 	Automatic Control Laboratory, Ecole Polytechnique F\'ed\'erale de Lausanne, Switzerland
	 \and	
		 Laboratoire de Syst\`emes Robotiques, Ecole Polytechnique F\'ed\'erale de Lausanne, 1015 Lausanne, Switzerland	
	 \and 
		 Ecole d'ing\'enieur ARC, 2610 St. Imier, Switzerland
	 \and
		 Lowell Observatory, 1400 West Mars Hill Road, Flagstaff, Arizona, 86001, USA 
	\and
		Centre Suisse d'Electronique et Microtechnique, 2007 Neuch\^atel, Switzerland			 
		       }

\date{Received 15 October 2012 / Accepted 5 December 2012} 			

\abstract 
{The ESPRI project relies on the astrometric capabilities offered by the PRIMA facility of the Very Large Telescope Interferometer for discovering and studying planetary systems. Our survey consists of obtaining high-precision astrometry for a large sample of stars over several years to detect their barycentric motions due to orbiting planets. We present the operation's principle, the instrument's implementation, and the results of a first series of test observations.}
{We give a comprehensive overview of the instrument infrastructure and present the observation strategy for dual-field relative astrometry in the infrared $K$-band. We describe the differential delay lines, a key component of the PRIMA facility that was delivered by the ESPRI consortium, and discuss their performance within the facility. This paper serves as reference for future ESPRI publications and for the users of the PRIMA facility.}
{Observations of bright visual binaries were used to test the observation procedures and to establish the instrument's astrometric precision and accuracy. The data reduction strategy for the astrometry and the necessary corrections to the raw data are presented. Adaptive optics observations with NACO were used as an independent verification of PRIMA astrometric observations.}
{The PRIMA facility was used to carry out tests of astrometric observations. The astrometric performance in terms of precision is limited by the atmospheric turbulence at a level close to the theoretical expectations and a precision of 30 $\mu$as was achieved. In contrast, the astrometric accuracy is insufficient for the goals of the ESPRI project and is currently limited by systematic errors that originate in the part of the interferometer beamtrain that is not monitored by the internal metrology system.}
{Our observations led to defining corrective actions required to make the facility ready for carrying out the ESPRI search for extrasolar planets.}

\maketitle

\section{Introduction}
High-precision astrometry will become a key method for the detection and physical characterisation of close-in ($<$10 AU) extrasolar planets thanks to the onset of instruments that promise a long-term accuracy of 10--100 micro-arcseconds ($\mu$as) and the associated surveys of large stellar samples. So far, the study of exoplanet populations has been dominated by the radial velocity technique (e.g. \citealt{Mayor:2011fj}) and by transit photometry (e.g. \citealt{Borucki:2011qy}). The application of astrometry was mostly limited to the characterisation of particular objects with very massive companions \citep{Zucker:2001ve, Pravdo:2005fu, Benedict:2010ph, Lazorenko:2011lr, Reffert:2011lr, Sahlmann:2011lr, Anglada-Escude:2012vn}. The population of very massive planets and brown dwarf companions was studied with {\small HIPPARCOS} astrometry, which resulted in an observational upper mass limit of $\sim$35 Jupiter masses ($M_J$) for the formation of close-in planets around Sun-like stars and a robust determination of the frequency of close brown-dwarf companions of G/K dwarfs \citep{Sahlmann:2011fk}. To detect the barycentric motion of a nearby Sun-like star caused by a close Jupiter-mass planet, an astrometric accuracy of better than one milli-arcsec (mas) per measurement is needed \citep{Black:1982kx, Sozzetti:2005qy, Sahlmann2012PhD}. At present, only a few instruments are capable of satisfying this requirement. Among them are the {\small HST-FGS} \citep{Benedict:2001rz}, infrared adaptive optics observations \citep{Gillessen:2009lr}, optical seeing-limited imaging with large telescopes \citep{Lazorenko:2009ph}, and optical interferometry.\\ 
Construction of optical interferometers was in part motivated by their capability of performing precise (a few mas) global astrometry \citep{Shao:1990qq} and very precise ($\sim$10 $\mu$as) narrow-angle relative astrometry \citep{Shao1992}. The latter requires a dual feed configuration to observe two stars simultaneously, which was demonstrated at the Palomar Testbed Interferometer ({\small PTI}) \citep{Lane:2003zl}. Using the {\small PTI} infrastructure but observing sub-arcsecond-scale binary stars within the resolution limit of one single feed, precisions of tens of $\mu$as were obtained in this very-narrow angle mode \citep{Muterspaugh:2005lq}. The possibility of using infrared interferometry for astrometric detection of extrasolar planets was described by \cite{Shao1992}. Consequently, a demonstration experiment was set up at the {\small PTI} \citep{ColavitaPTI1999} and the interferometric facilities at the Keck and Very Large Telescope ({\small VLT}) observatories, which were being built at that time, included provisions for the dual-field astrometric mode. On the basis of the feasibility study by \cite{Quirrenbach:1998mi} for the {\small VLT} interferometer ({\small VLTI}), the development of this mode of operation was started and its implementation began with the hardware deployment at the observatory in 2008. The infrastructure for dual-feed operation and relative astrometry at the {\small VLTI} is named {\small PRIMA}, an acronym for phase-referenced imaging and micro-arcsecond astrometry. 
\subsection{The ESPRI project}\label{sec:tmp1}
The goals of the {\small ESPRI} project (extrasolar planet search with {\small PRIMA}, \citealt{Launhardt2008}) are to characterise known radial velocity planets by measuring the orbit inclination and to detect planets in long period orbits around nearby main-sequence stars and young stars, i.e. in a parameter space which is difficult to access with other planet detection techniques. The {\small ESPRI} consortium consists of three institutes: Max Planck Institut f\"ur Astronomie Heidelberg, Observatoire Astronomique de l'Universit\'e de Gen\`eve, and Landessternwarte Heidelberg. A detailed description of the science goals, organisation, and preparatory programme of {\small ESPRI} is given in a accompanying paper (Launhardt et al., in prep). Formally, the {\small ESPRI} consortium contributes to the {\small PRIMA} facility with the differential delay lines ({\small DDL}), the astrometric observation preparation software ({\small APES}), and the astrometric data reduction pipeline, all of which are eventually delivered to {\small ESO}, hence become publicly available. In return, {\small ESPRI} obtains guaranteed time observations ({\small GTO}) to carry out its scientific programme. In practice, {\small ESPRI} also contributes significantly to the commissioning of the {\small PRIMA}~astrometry mode, which includes making the facility functional, carrying out the observations, and reducing and analysing the data.

\noindent
The paper is organised as follows: In Sect.~\ref{sec:principles}, we discuss the measurement principle and the interferometric baseline definition. The {\small PRIMA} facility is described in Sect.~\ref{sec:primafacility} and the design and performance of the differential delay lines is presented in Sect.~\ref{sec:ddl}. The astrometric data reduction and modelling is introduced in Sects.~\ref{sec:red} and \ref{sec:axmodelling}, respectively. The results in terms of measurement precision and accuracy are summarised in Sect.~\ref{sec:prec}. We conclude in Sect.~\ref{sec:concl}. Auxiliary information is collected in the appendices.
\section{Principles of narrow-angle astrometry}\label{sec:principles}
\begin{figure}[h]\begin{center} 
\includegraphics[width = 0.49\linewidth,trim = 4cm 6.5cm 13cm 5cm, clip=true]{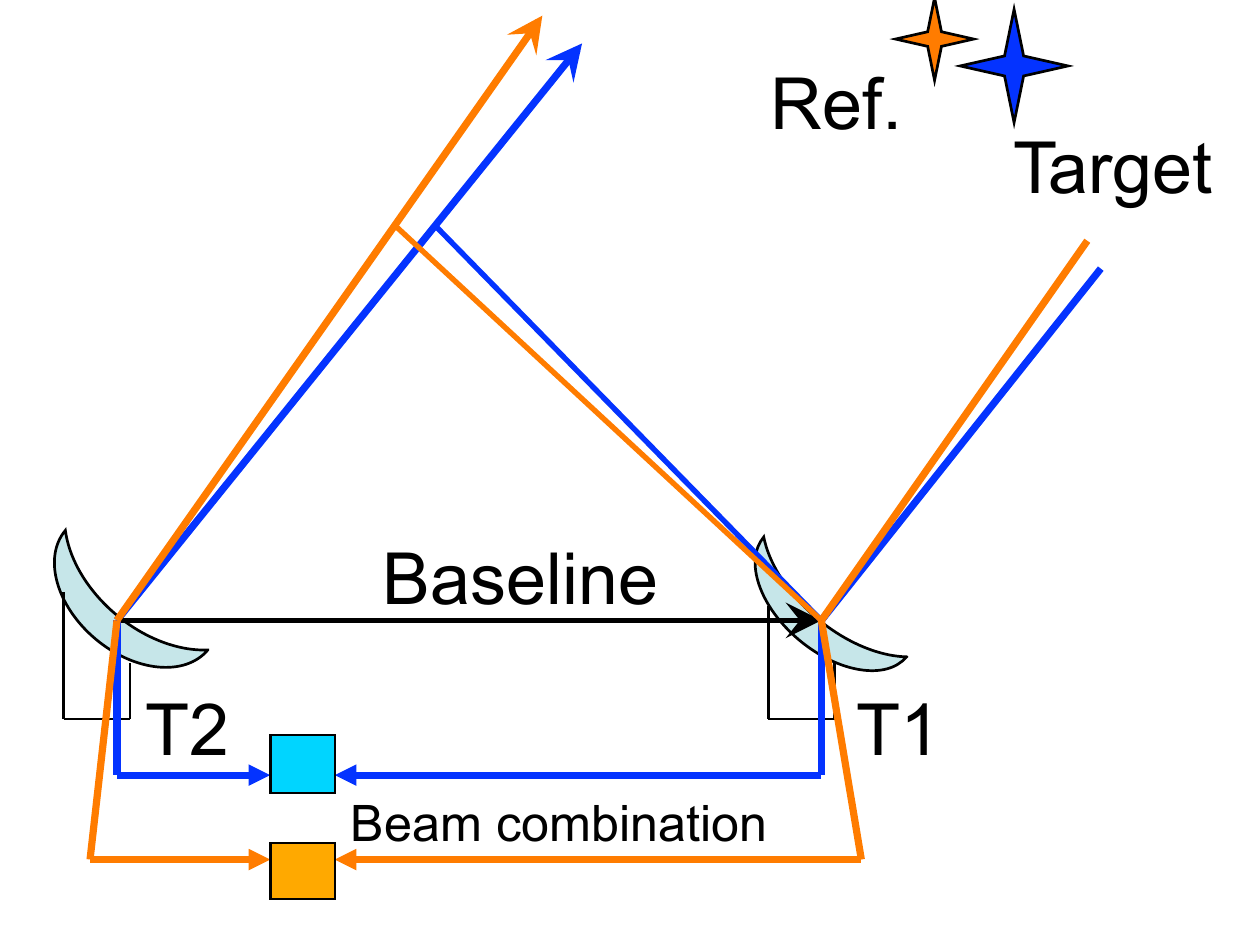}
\includegraphics[width = 0.49\linewidth,trim = 3cm 0cm 0.5cm 0cm, clip=true]{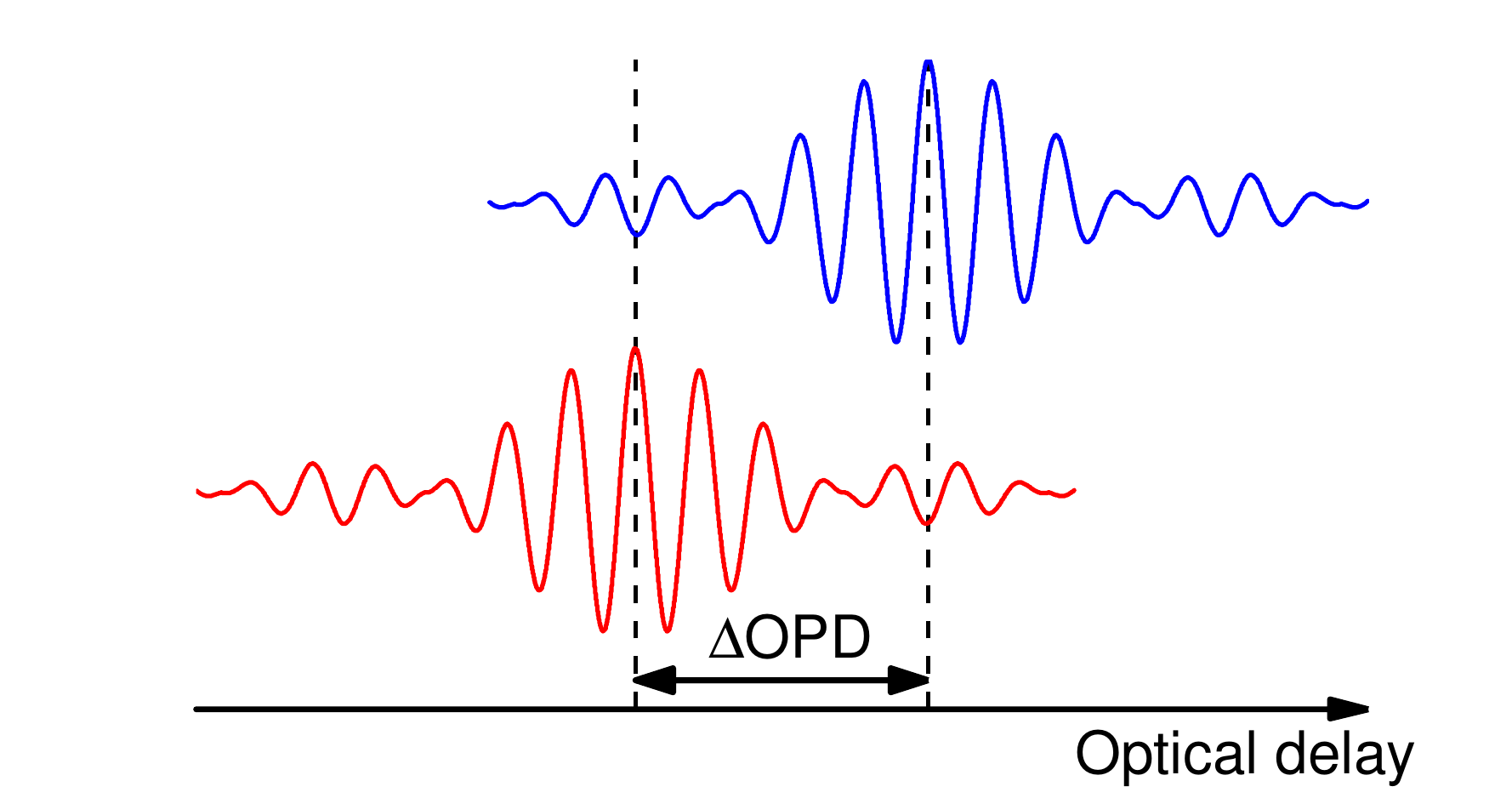}
\caption{Principle of {\small PRIMA} astrometry: reference and target star are observed simultaneously with two telescopes (\emph{left}). Light of both stars is made to interfere and the location of the two interference fringe patterns is measured in delay space (\emph{right}). The fringe separation $\Delta$OPD is proportional to the projected separation of the stars.}\label{fig:primasketch}\end{center}
\end{figure}

The measurement principle of narrow-angle relative astrometry with an interferometer is to observe two stars simultaneously and to measure the relative position of the two fringe patterns in delay space (Fig.~\ref{fig:primasketch}). 

\noindent
In stellar interferometry, the astrometric information is encoded in the fringe position measured in delay space. The internal delay that is necessary to observe interference therefore contains information on the star's position, hence a series of delay measurements can be used for the determination of accurate stellar positions (e.g. \citealt{Shao:1990qq, Hummel:1994rm}). The relation between the optical delay $w$ and the stellar position defined by the unit vector $\vec{s}$ in direction of the star is
 \begin{equation}\label{eq:OPDform2}
      w = \vec{B}  \cdot  \vec{s},
 \end{equation}
where $\vec B = \vec T_2 - \vec T_1$ is the baseline vector connecting two telescopes with coordinates $\vec T_i$. When observing two stars identified by unit vectors $\vec s_1$ and $\vec s_2$ simultaneously, the differential delay $ \Delta w$ can be written as difference of the respective $w$-terms
      \begin{equation}\label{eq:dopdmodel0}
      \Delta w   =  w_2 - w_1 =   \vec{B} \cdot \vec{s_2} -  \vec{B} \cdot  \vec{s_1}=   \vec{B} \cdot \Delta \vec{s}.
      \end{equation}
The potential for high-precision astrometry of this observation mode stems from the combination of a large effective aperture and the fact that the noise originating in atmospheric piston motion is correlated over small fields \citep{Shao1992}. In conventional imaging astrometry, the achievable astrometric precision depends on the aperture size $D$ of the telescope \citep{Lazorenko:2009ph}. \cite{Shao1992} showed that in the case of dual-field interferometry, where $D$ is effectively replaced by the projected baseline length $B_p$, and under the narrow angle condition
\begin{equation}\label{espri001}
\Theta h < B_p\;,
\end{equation}
where $\Theta$ is the field size and $h$ is the turbulence layer height, the astrometric error $\sigma_a$ due to atmospheric turbulence is 
\begin{equation} \label{eq:colav1}
\sigma_a  =  q_\mathrm{site} \, \frac{\Theta}{B_p^{2/3}\,T^{1/2}} \hspace{5mm} \mathrm{or} \hspace{5mm} 
\sigma_a^2  \sim   \frac{\Theta^2}{B_p^{4/3}\,T}, 
\end{equation}
with $B_p$ in metres, $\Theta$ in radians, and $T$ in hours. The parameter $q_\mathrm{site}$ depends on the atmospheric profile and in the case of a Mauna Kea model is 300\arcsec\, m$^{2/3}$ h$^{1/2}$ rad$^{-1}$. Based on Eq.~\ref{eq:colav1}, the expected atmospheric limit to the astrometric precision for a baseline length of 100~m, a separation between the stars of 10\arcsec, and an integration time of 1\,h is $\sim$10\,$\mu$as.

\noindent
In the horizontal coordinate system, the elevation $E$ and azimuth $A$ of a star at local sidereal time $t_s$ are related to the right ascension $\alpha$ and declination $\delta$ in the equatorial coordinate system by the hour angle $h_a = t_s - \alpha$ and the equations
\begin{equation} 
\begin{array}{rrlcl}
 \sin E        &=&   \sin \delta \sin L & +&  \cos \delta \cos h_a \cos L \\[3pt]
 \cos E\cos A &=&   \sin \delta \cos L & - & \cos \delta \cos h_a \sin L  \\[3pt]
 \cos E\sin A &=&                        & - & \cos \delta \sin h_a\\[3pt]
\Rightarrow  \tan A &=&  \multicolumn{3}{c}{\frac{\displaystyle -\cos \delta\sin h_a }{\displaystyle  \sin \delta \cos L - \cos \delta \cos h_a \sin L }},\\
\end{array}
\end{equation}
where $L$ is the observer's geographic latitude. The unit vector $\vec{s}$ in direction of the star is 
\begin{equation}
 \vec{s}= 
\left(
\begin{array}{l}
  \cos E\, \cos A\\
  \cos E\, \sin A\\
  \sin E
\end{array}
\right)
\end{equation}
and the baseline vector $\vec B$ in the horizontal coordinate system is \begin{equation} \label{eq:BNEE}
\vec{B} = 
\left(
\begin{array}{l}
  B_\mathrm{North}\\
  B_\mathrm{East}\\
  B_\mathrm{Elev}\\
\end{array}
\right),
\end{equation}
where the component $B_\mathrm{North}$ is the ground projection of $\vec B$ measured northward,  $B_\mathrm{East}$ is the ground projection of $\vec B$ measured eastward, and $B_\mathrm{Elev}$ is the elevation difference of $\vec B$ measured upward. The instantaneous optical path difference (OPD) $w$ is thus \citep{Fomalont:1974fj, Dyck:uq}
 \begin{equation}\label{eq:OPDform22}
 \begin{split}
      w &= \vec{B} \cdot \vec{s}\\         
          &= \cos E\,\cos A\,\, B_{North}  + \cos E\, \sin A \,\, B_{East} + \sin E\,\, B_{Elev}
          \end{split}
 \end{equation}
and the differential OPD $ \Delta w$ can be written as 
      \begin{eqnarray}\label{eq:dopdmodel}
\begin{split}
         \Delta w         & =  B_\mathrm{North} \, \left( \cos{E_{2}}\, \cos A_{2} - \cos E_{1}\, \cos A_{1}  \right) \\
                  &\, +  B_\mathrm{East}  \, \, \, \left( \cos E_{2}\, \sin A_{2} - \cos E_{1}\, \sin A_{1} \right)   \\
                  &\, +  B_\mathrm{Elev} \, \, \, \left( \sin E_{2}               - \sin E_{1}               \right). \\
                  \end{split}
      \end{eqnarray}
In the equatorial system the baseline coordinates are
\begin{equation}\label{bb2}
\left( \begin{array}{c}
             B_{x}\\
	     B_{y}\\
	     B_{z}
\end{array} \right)      
       =
      \left( \begin{array}{rrr}
          - \sin{L} &  0& \cos{L} \\
0& -1&  0\\
 \cos{L}&0& \sin{L}\\
          \end{array}
      \right)
       \cdot
       \left(\begin{array}{l}
  B_\mathrm{North}\\
  B_\mathrm{East}\\
  B_\mathrm{Elev}\\
\end{array}\right)      
\end{equation}
and the $u$-$v$-$w$-coordinate system describing the tangential plane in the sky are given by
      \begin{equation}\label{eq:uvw}
      \left(
          \begin{array}{c}
              u\\
              v\\   
              w
          \end{array}
      \right)
       =
      \left(
          \begin{array}{rrr}
           \sin H &  \cos H& 0 \\
 -\sin \delta \cos H& \sin \delta \sin H&  \cos \delta\\
  \cos \delta \cos H&-\cos \delta \sin H& \sin \delta\\

          \end{array}
      \right)
       \cdot
      \left(
          \begin{array}{c}
             B_{x}\\
	     B_{y}\\
	     B_{z},
          \end{array}
      \right)
      \end{equation}
yielding the optical path difference 
 \begin{equation}\label{eq:OPDform}
       w =  \cos \delta \cos H \; B_x -\cos \delta \sin H \; B_y+ \sin \delta \; B_z.
 \end{equation}
For completeness, we note that the projected baseline length is $B_p = \sqrt{u^{2}+v^{2}}$ and the projected baseline angle $\theta_{\bf B}$ is defined by $ \tan \theta_{\bf B} = v/u$ and that the {\small VLTI} baseline $\vec B_\mathrm{VLTI}$ is defined using a different convention from Eq.~\ref{eq:BNEE}:
\begin{equation}\label{bb1}
\vec B_\mathrm{VLTI} = 
\left(\begin{array}{l}
  B_\mathrm{South}\\
  B_\mathrm{West}\\
  B_\mathrm{Elev}\\
\end{array} \right) 
= 
\left(\begin{array}{r}
  -1\\
  -1\\
  1\\
\end{array}\right) \odot \vec{B},
\end{equation}
where the $\odot$-symbol indicates element-wise multiplication.
      
\subsection{Interferometric baselines}
In the theoretical description above, the interferometric baseline $\vec B$ is a well defined quantity. In practice, its definition is not straight-forward because the telescopes and other optical elements are moving during the observation and the simple definition as \emph{the vector connecting telescope $T_1$ and $T_2$} becomes ambiguous. Furthermore, {\small PRIMA} is used to observe two stars simultaneously and effectively realises two interferometers at the same time, thus making a clarification in the interpretation of the baseline necessary. The following conceptual description neglects any effect related to off-axis observations, optical aberrations, refraction, and dispersion and applies in the horizontal coordinate system. The device allowing us to determine the astrometric baseline is a laser metrology that monitors the optical path lengths of the optical train travelled by the stellar beams inside the interferometer. The two terminal points defining the monitored optical path of each beam are the metrology endpoints. Let $L_i$ be the optical path length measured with the metrology in beam $i$ and let $w = L_i-L_j$ be the instantaneous measured optical path difference between the two arms of an interferometer observing a star defined by the pointing vector $\vec s$. Note that the metrology yields the internal delay only in the case of ideal fringe tracking, i.e. when the fringe packet as seen by the fringe sensor is centred at all times. Thus for a real system, measurements by the fringe tracking system and their potential systematics have to be considered when using metrology readings.

\subsubsection{Wide-angle astrometric baseline}
The unique vector $\vec B_\mathrm{Wide}$ that relates the metrology measurement of optical path difference $w$ to the pointing vector $\vec s$ at all times via the equation
  \begin{equation}\label{eq:wab01}
w = \vec B_\mathrm{Wide} \cdot \vec s + c
\end{equation}
is called the wide-angle astrometric baseline, where $c$ is a constant term. The wide-angle astrometric baseline can be determined by measuring the metrology delays $w_{1...k}$ when observing a set of $1...k$ stars with coordinates $\vec s_{1...k}$, selected such that the resulting system of equations~\ref{eq:wab01} allows for the non-degenerate solution for the three components of $\vec B_\mathrm{Wide}$ (e.g. \citealt{Shao:1990qq, Buscher:2012zr}).\\
The pivot point of an ideal telescope is defined as the intersection between the altitude and azimuth axes and remains at a fixed position in the horizontal coordinate system at all times. It is the relative position of the two pivot points that determines the optical path lengths travelled by the stellar beams inside the interferometer. Thus, ideally we would like to make the metrology endpoints at one end coincide with the pivot points and at the other end be located at the location of beam combination. In this configuration, Eq.~\ref{eq:wab01} would be exact. In practice, several complications occur. First, due to the telescope design, misalignments, or flexures, the pivot point may not exist, be ill-defined, or does not remain fixed at all times. Second, the pivot point is hardly accessible and the metrology endpoints are located elsewhere in the beam train. Third, there may be a non-common path between the metrology and stellar beams, i.e.\ potential changes in stellar path length are not monitored by the metrology. Thus Eq.~\ref{eq:wab01} needs to be complemented with two terms
\begin{equation}\label{eq:wab02}
w + \epsilon= ( \vec B_\mathrm{Wide} - \vec \mu_1 + \vec \mu_2) \cdot \vec s + c,
\end{equation}
where $\epsilon$ is the OPD mismatch between the metrology and stellar beams caused by the non-common path, and $\vec \mu_i$ is the offset vector between the pivot point of telescope $i$ and the respective metrology endpoint. Both terms are time dependent and have to be modelled, but one term can be eliminated by placing the metrology endpoints either in the entrance pupils, then $\epsilon=0$, or at the pivot points, then  $\vec \mu_i = \vec \mu_j=0$.\\ 
The {\small PRIMA} facility realises a dual-feed interferometer and at first it can be modelled as two independent interferometers, denoted feed A and feed B, observing the star $\vec s_A$ and $\vec s_B$, respectively. Thus there are four stellar and four metrology beams and each interferometer has a wide-angle baseline defined by 
\begin{equation}\label{eq:wab03}
\begin{split}
 w_A  &= L_3 - L_1 = \vec B_\mathrm{A, Wide} \cdot \vec s_A +c_A\\
 w_B  &= L_4 - L_2 = \vec B_\mathrm{B, Wide} \cdot \vec s_B +c_B\\
\end{split}
\end{equation}\vspace{-5mm}
\subsubsection{Narrow-angle astrometric baseline} 
The goal of the experiment is to measure the angular separation vector $\Delta \vec s = \vec s_A-\vec s_B$ between the two targets in the sky. Although in principle we could use Eq.~\ref{eq:wab03} to achieve this task, the required accuracy on the wide-angle baseline knowledge becomes unachievable for the level of accuracy anticipated for $\Delta \vec s$. Instead, both interferometers can be tied together by a common metrology system measuring the difference of optical path differences $\Delta w = w_A - w_B$. The unique vector $\vec B_\mathrm{AX}$ that relates $\Delta w$ to the separation vector $\Delta \vec s$ at all times via the equation
  \begin{equation}\label{eq:nab01}
\Delta w = \vec B_\mathrm{AX} \cdot \Delta \vec s + c_\Delta
\end{equation}
is called the narrow-angle astrometric baseline. We can rewrite Eq.~\ref{eq:nab01} as
  \begin{equation}\label{eq:nab02}
\Delta w =  L_3 - L_1 - L_4 + L_2= \vec B_\mathrm{A, Wide} \cdot \vec s_A - \vec B_\mathrm{B, Wide} \cdot \vec s_B + c_A - c_B
\end{equation}
which shows that in order to satisfy Eq.~\ref{eq:nab01}, the identity $\vec B_\mathrm{A, Wide} = \vec B_\mathrm{B, Wide} = \vec B_\mathrm{AX}$ is a necessary condition, which requires that the metrology endpoints corresponding to the beams $L_1$/$L_2$ and $L_3$/$L_4$ coincide in telescope 1 and 2, respectively. We are still not done, because in practice the unknown terms introduced in Eq.~\ref{eq:wab02} have to be considered. We obtain
\begin{equation}\label{eq:nab03}
\begin{split}
\Delta w(t) + \Delta \epsilon(t) = &\hspace{3mm}\left[ \vec B_\mathrm{A,Wide} - \vec \mu_{A,1}(t) + \vec \mu_{A,2}(t)\right] \cdot \vec s_A(t)+ c_A\\ 
&- \left[ \vec B_\mathrm{B,Wide} - \vec \mu_{B,1}(t) + \vec \mu_{B,2}(t)\right] \cdot \vec s_B(t)+c_B
\end{split}
\end{equation}
where time dependencies are explicitly noted. Because of the requirement of coinciding metrology endpoints in each telescope, we have $\vec B_\mathrm{A, Wide} = \vec B_\mathrm{B, Wide}$ and $\vec \mu_{A,i} =  \vec \mu_{B,i} = \vec \mu_{i}$ and it follows
\begin{equation}\label{eq:nab04}
\Delta w(t) + \Delta \epsilon(t) = \left[ \vec B_\mathrm{AX} - \vec \mu_{1}(t) + \vec \mu_{2}(t)\right] \cdot \vec \Delta s(t) + c_\Delta,
\end{equation}
where $\Delta \epsilon(t)$ is the differential OPD caused by non-common path between stellar and metrology beams and $c_\Delta = c_A-c_B$ is a constant.\\
Because of the differential measurement, the requirements on the measurement accuracy of $ \vec B_\mathrm{AX}$ are relaxed, but the terms $\Delta \epsilon(t)$ and $\vec \mu_{i}(t)$ have to be minimised by the optical design and consequently modelled (Sect.~\ref{sec:sster}). The vectors $\vec \mu_{i}(t)$ can for instance be determined by accurate modelling combined with external measurement devices, which monitor the telescope motion \citep{Hrynevych:2004qy}. Alternatively, the narrow-angle baseline can be determined by measuring several star pairs with different and known separation vectors and solving the system of equations Eq.~\ref{eq:nab04} for $\vec B_\mathrm{AX}$.

\subsubsection{Imaging baseline}
The imaging baseline determines the orientation and value of the spatial frequency sampled with the interferometer, i.e.\ the $u$-$v$-coordinates (Eq.~\ref{eq:uvw}), and it is related to the sky-projected configuration of the interfering partial wavefronts. To distinguish between imaging and wide-angle baseline the following example can be useful: If half of one telescope aperture is masked during the observation of an on-axis source, this will not change the wide-angle baseline (the fringe position in delay space remains constant) but it will alter the imaging baseline, because the interfering wavefront portions are changing.  

\subsubsection{Model applied for initial PRIMA tests}
The theoretical description of the narrow-angle baseline given above does not strictly apply to {\small PRIMA} observations and several second-order terms have to be considered (e.g. \citealt{Colavita:2009qy}). Additionally, the wide-angle baseline of {\small PRIMA} $\vec B'_\mathrm{Wide}$ is determined using the delay line metrology that measures $w_{DL}$ (Sect.~\ref{sec:opdmodeldet}), whereas the differential delay is measured using the dedicated {\small PRIMA} metrology having different endpoints yielding the quantity $\Delta L$. For the analysis of the initial astrometric measurements relying on Eq.~\ref{eq:nab04}, we will assume that the unknown quantities vanish, i.e.\ $\Delta \epsilon(t) = 0$ and  $\vec \mu_{i}(t) = 0$, that the endpoints of the differential metrology coincide, and that the measured quantities are given by $\Delta w(t) = \Delta L(t)$ and $\vec B_\mathrm{AX} = \vec B'_\mathrm{Wide}$. Those assumptions may not necessarily be fulfilled and we discuss potential effects in the text.
\begin{figure*}\begin{center} 
\sidecaption  
\includegraphics[width = 12cm,trim = 0mm 0mm 0mm 0mm, clip]{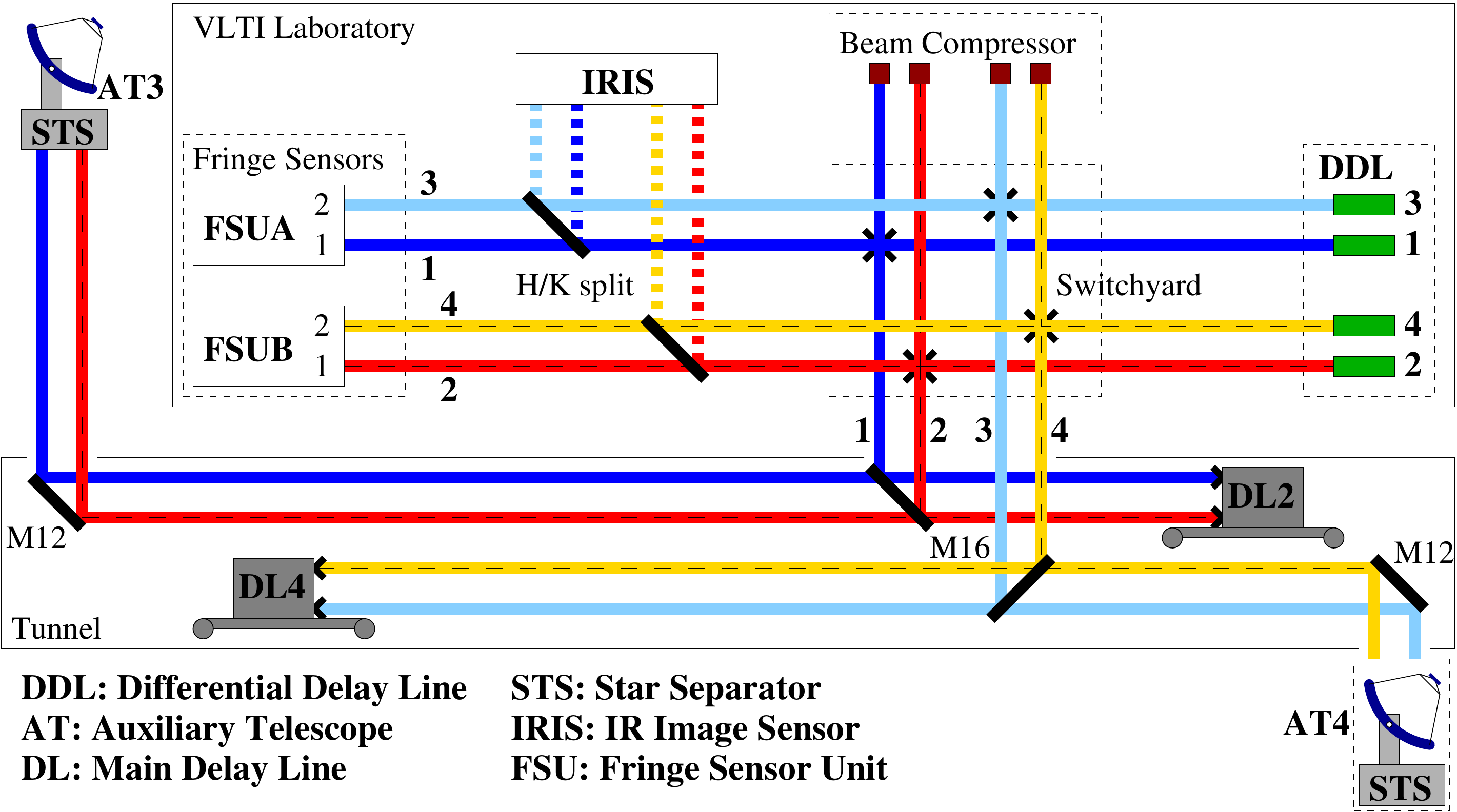}
\caption[{\small PRIMA} beam paths in the VLTI]{Schematic of the stellar beam paths for {\small PRIMA} astrometric observations showing the main subsystems and their relative locations (not to scale). The beams propagate through underground light ducts from the telescopes to the delay line tunnel from where they are relayed into the beam combination laboratory. The input channel number of each beam is shown in bold. The PRIMA metrology monitors the path lengths of the four beams between the respective endpoints in the FSU and STS. The two metrology beams that monitor the FSUB feed are shown as dashed lines.}\label{fig:labsketch}\end{center}
\end{figure*}
\section{PRIMA at the Very\,Large\,Telescope\,Interferometer}\label{sec:primafacility} 
The {\small PRIMA} facility consists of a considerable amount of subsystems, which are distributed both physically on the observatory platform and systematically over the VLTI control system. Its installation necessitated the enhancement of every VLTI building block, i.e.\ of the telescopes, the delay lines, the laboratory infrastructure, and the control system. {\small PRIMA} is a multi-purpose facility with several observing modes, which was added to the existing VLTI framework under the requirement not to impact the already operating instruments. Therefore {\small PRIMA} cannot be considered an astronomical instrument in the classic sense. 
General descriptions of {\small PRIMA} are given by \cite{Delplancke2006} and \cite{Belle2008}. A detailed description of the {\small PRIMA} fringe sensors and an assessment of their performance  for fringe-tracking is given by \cite{Sahlmann:2009kx}. The astrometric instrument that uses the {\small PRIMA} facility is named {\small PACMAN} \citep{Abuter:2010lr}. The status of the VLTI and its subsystems is given by \cite{Haguenauer2008,Haguenauer:2010uq}.

\noindent
The various subsystems of {\small PRIMA} were tested individually at the Paranal observatory since August 2008. Testing of the {\small PRIMA} facility began in July 2010, when dual-feed fringes were recorded for the first time. Astrometric observations became possible in January 2011 and {\small PACMAN} 'first light' was achieved on January 26, 2011. For {\small PRIMA} astrometry observations, the interactions between the {\small PRIMA} and {\small VLTI} subsystems are numerous and flawless interplay is required for the basic functionality. Furthermore, the accuracy goal for {\small PRIMA} astrometry sets stringent requirements on the performance of every subsystem, and measurement biases can originate virtually anywhere along the beam train. Thus, a detailed description of the {\small VLTI-PRIMA} system is required and given below.

\subsection{Design goal for the astrometric accuracy}\label{sec:designgoal}
The design goal for {\small PRIMA} was that measurement errors introduced by instrument terms shall be smaller than 10\,$\mu$as for observations of two objects in a field of $\sim$1\arcmin~and with a baseline of $\sim$100~m, i.e.\ inferior to the atmospheric limit for 1 h of integration. We can use Eq.~\ref{eq:nab01} to estimate how this requirement translates into the required accuracy of measured quantities. For simplicity we set the astrometric accuracy to $\sigma_\Theta = 10\,\mu$as for the separation between two targets located in a $\Theta = 10\arcsec$~field. The relative accuracy is $\sigma_\Theta/\Theta = 10^{-6}$ and thus the astrometric baseline $\vec B_\mathrm{AX}$ has to be known at this accuracy level, corresponding to $100\,\mu$m for a baseline length of 100~m. This also illustrates why we cannot get away with measuring the two wide-angle baselines and using Eq.~\ref{eq:wab03}, because the accuracy requirement on $\vec B_\mathrm{Wide} $ would be orders of magnitude more stringent, thus not reachable. The expected differential delay is of the order of 5~mm, thus the required accuracy in the differential delay measurement is 5~nm for optical path lengths reaching several hundred metres.

\subsection{PRIMA subsystems and beam train}
A schematic overview of the {\small PRIMA-VLTI} system is shown in Fig.~\ref{fig:labsketch}. Below, a short description of the involved subsystem is given in order of incidence, which can be used together with the schematic to follow the path of an incoming stellar beam. The main subsystems are briefly described in the next sections.
\begin{itemize}
  \item Telescope: Two {\small VLTI} Auxiliary Telescopes are used.
  \item Star separator: Splits the image plane between the two targets and generates separated output beams for each of them. 
  \item M12: A passive folding mirror located in the delay line tunnel that sends the stellar beams coming from an auxiliary telescope light duct towards a main delay line.
 \item Main delay line: The main delay line is used for fringe tracking on the primary star. The carriage moves on rails and is capable of introducing optical delay in both feeds with nanometre accuracy over $\sim$120\,m and with a high bandwidth. 
 \item M16: These are configurable mirrors that fold the beams into the laboratory feeding the desired input channel.
 \item Beam compressor: Each beam is downsized from 80~mm to 18~mm by passing through a parallel telescope.
 \item Switchyard: A set of configurable mirrors, which for {\small PRIMA} astrometry fold the beams towards the differential delay lines.
 \item Differential delay line (DDL): One of the four DDLs is used to compensate for the differential delay between the two star feeds and for fringe tracking on the secondary star.
 \item $H/K$-dichroic mirror: Folds the $H$-band light towards the infrared image sensor (IRIS) and transmits the $K$-band light towards the fringe sensor unit.
 \item Fringe Sensor Unit (FSU): Combines two $K$-band input beams from one star and detects the fringe signals. The twin sensors FSUB and FSUA are driving the primary and secondary fringe tracking loops, respectively. 
 \item IRIS: Images the $H$-band beams and measures the point-spread-function (PSF) position and motion for feedback to the control system \citep{Gitton:2004wd}.
 \end{itemize}
In total, the stellar beams are reflected on 38 optical surfaces before being injected into the single-mode fibres of the FSU, see Table~\ref{tab:reflection}. These are 13 reflections more than in the single-feed case, for which \cite{Sahlmann:2009kx} estimated a total effective transmission (including effects of injection fluctuations) in $K$-band of $4\pm1$~\%. If we assume an average reflectivity of 0.98 or 0.95 per mirror, the additional decrease in transmission for the nominal dual-feed case is 23~\% or 49~\%, respectively. A detailed analysis and description of the measured transmission in dual-feed is outside the scope of this work.

\subsection{Auxiliary telescopes, derotator, and star separator}
\begin{figure}\begin{center} 
\includegraphics[height = 8cm,trim = 0mm 0mm 0mm 0mm, clip]{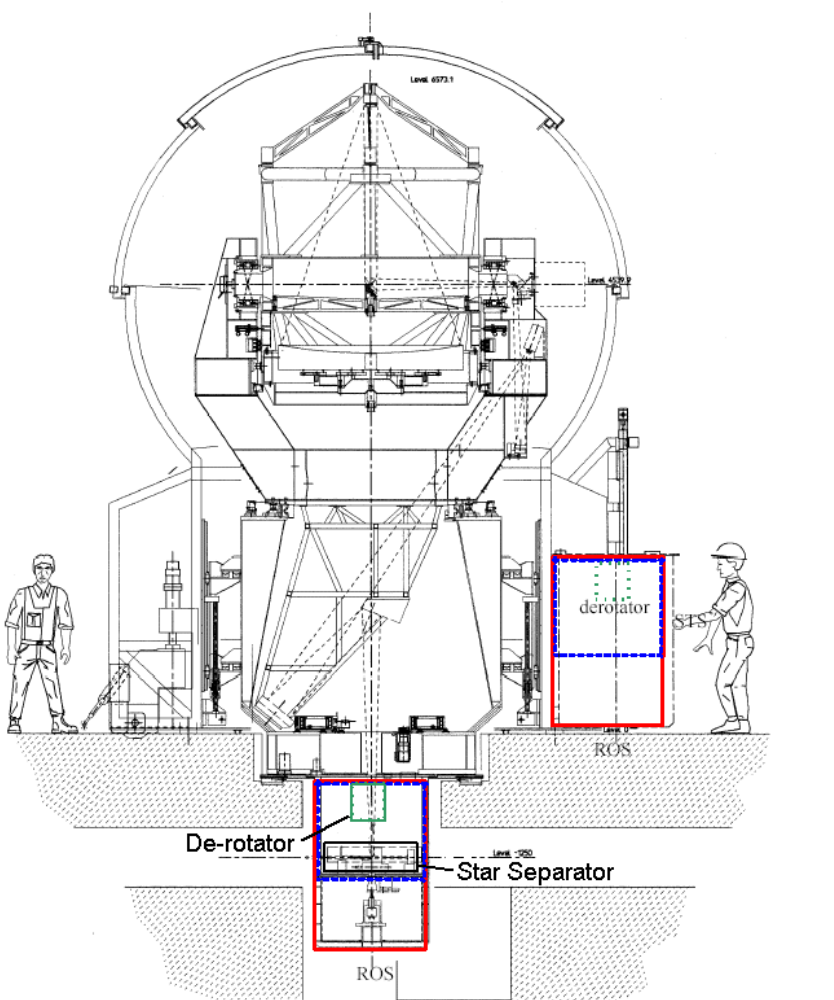}
\caption{The cross-section of an auxiliary telescope indicating the location of the derotator and the star separator below the main telescope structure. From \cite{Nijenhuis:2004lr}.}\label{fig:ATsketch}\end{center}
\end{figure}

\begin{figure}\begin{center} 
\includegraphics[height = 8.5cm,trim = 10mm 0mm 5mm 0mm, clip]{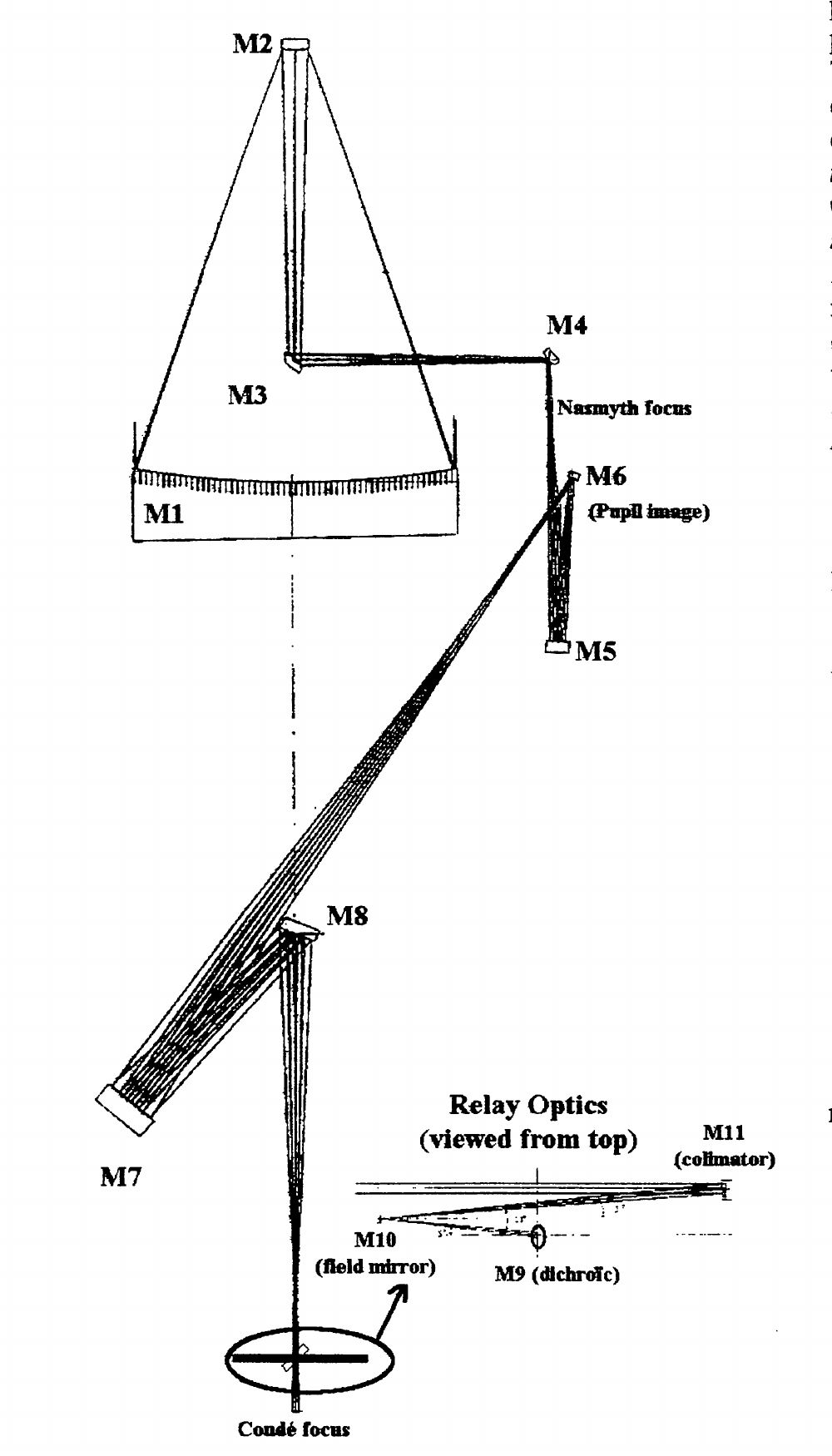}
\caption{Beam tracing inside the auxiliary telescope indicating the mirror designations. From \cite{Delrez:2001fk}. The derotator is located between M8 in the AT and M9 in the STS, see also Fig.~\ref{fig:STSsketch}.}\label{fig:ATbeamsketch}\end{center}
\end{figure}

\begin{figure}[h!]
  \begin{center}  
\includegraphics[width = \linewidth, trim = 0mm 7mm 0mm 5mm, clip=true]{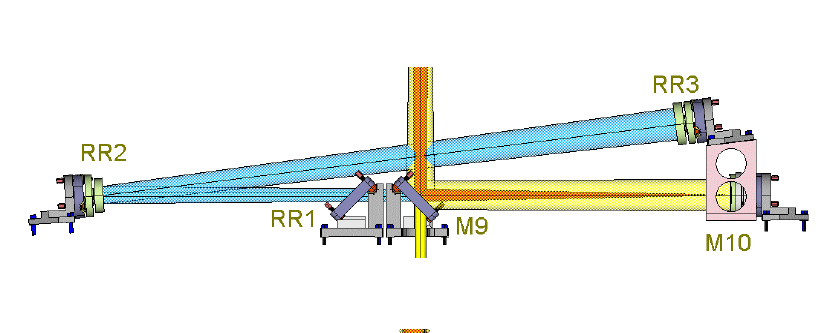}
\caption{Optical layout of the metrology endpoint. This side view shows the stellar beam coming from the telescope (above) being reflected towards M10 and the metrology beams coming from the laboratory being reflected on themselves by the assembly of the spherical mirrors RR2 and RR3 and the compensation plate RR1. In this view, the metrology beams enter horizontally from the right, after reflection off M10. The light that is seen to traverse M9 downwards is the visible portion of the stellar beam propagating towards the fast quadcell image sensor.}
\label{fig:endpointsketch}\end{center}
\end{figure}

The Auxiliary Telescopes (AT, \citealt{Koehler:2002lr, Koehler:2006rr}) of the VLTI have a 1.8~m diameter primary mirror in altitude-azimuth mount and a coud\'e beam train as shown in Figs.~\ref{fig:ATsketch} and \ref{fig:ATbeamsketch}. Several actuators are used to manipulate the stellar beam: the secondary mirror M2 can be controlled in lateral position (X,Y), in longitudinal position for focus adjustment (Z), and in tip and tilt. A first-order image stabilisation system is implemented to attenuate the fast atmospheric image motion: a quadcell sensor based on avalanche photo diodes is located below the star separator and receives the visible light transmitted by the M9 dichroic mirror (Fig.~\ref{fig:endpointsketch}). It measures the image centroid and sends offsets to the fast tip-tilt mirror M6 located upstream, hence realising a closed-loop control system for image motion, capable of very low-order wavefront control. The derotator is located below the telescope and above the star separator. Its role is to generate a fixed field image for the star separator, i.e.\ to remove the field motion caused by the Earth rotation. The derotator is implemented as a reflective K-prism assembly mounted on a motorised rotation stage, such that a 180\degr~derotator motion results in 360\degr~field rotation. 
\begin{figure}
\begin{center}
\includegraphics[width = 0.7\linewidth,trim = 10mm 0mm 10mm 0mm, clip=true]{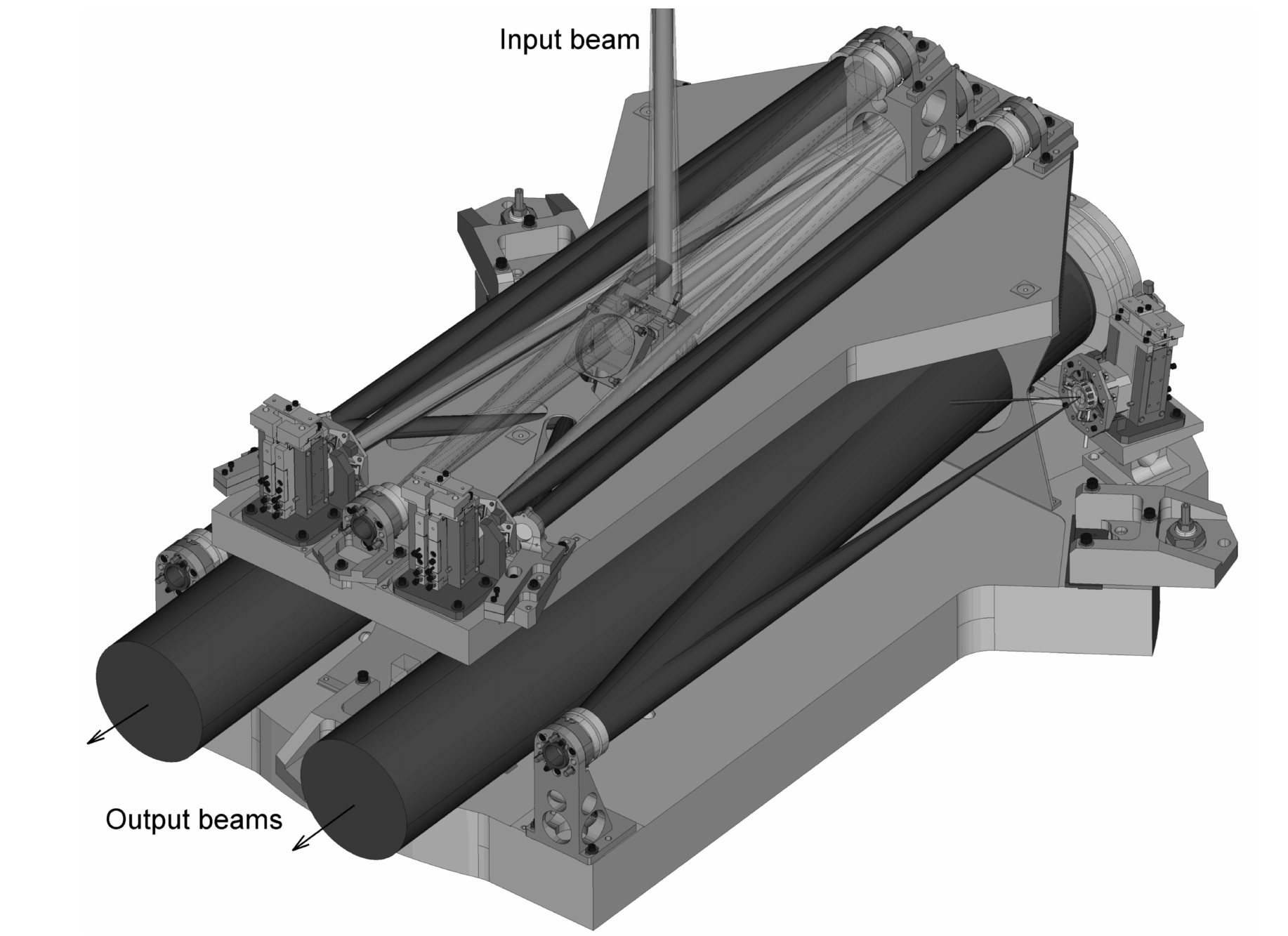} 
\includegraphics[width = \linewidth,trim = 5mm 0mm 0mm 10mm, clip=true]{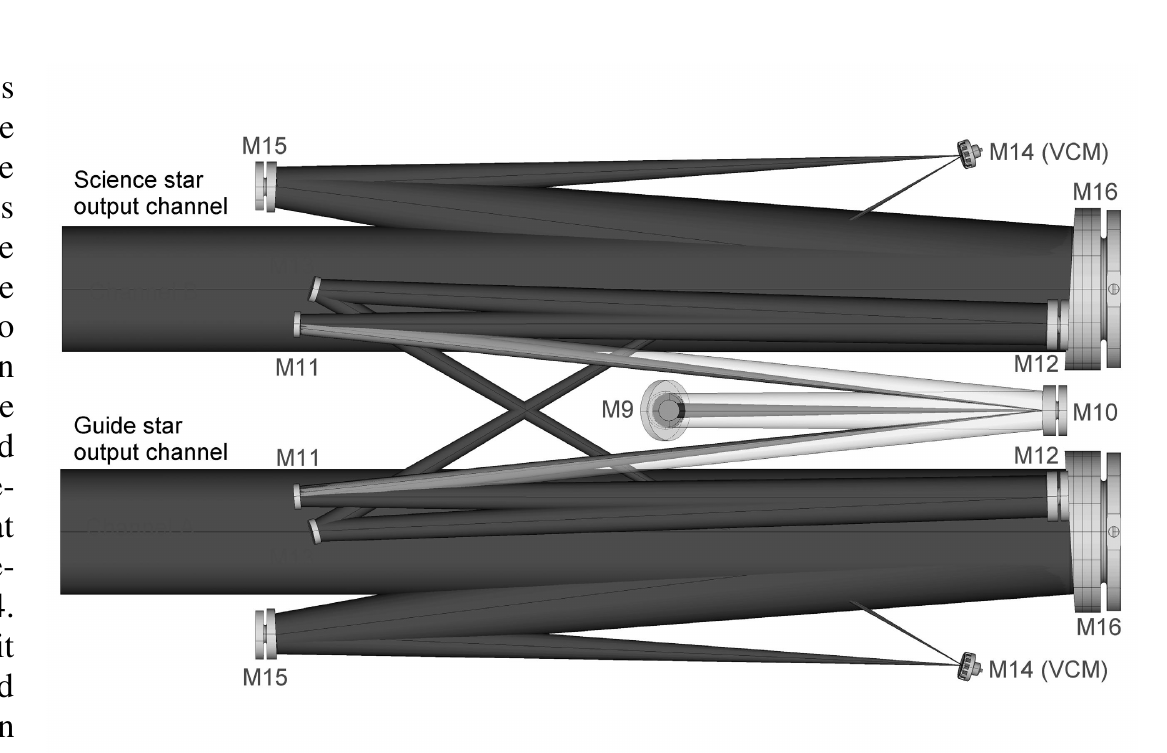}
\caption{\emph{Top}: Optical layout of one star separator. \emph{Bottom}: Top view of the STS optical layout identifying the mirror names. M11 (FSM) is the field selector mirror acting on the image and M14 (STS-VCM) is the variable curvature mirror acting on the pupil. From \cite{Nijenhuis:2004lr}.}\label{fig:STSsketch}\end{center}
\end{figure}

\noindent
The star separator \citep{Nijenhuis:2008cy} has three main functionalities: (a) it splits the field between the two observed objects and generates two output beams, each containing the light from one sub-field; (b) it supplies the end-point for the {\small PRIMA} metrology; (c) it controls the image position measured downstream in the laboratory and the lateral pupil position. 
\begin{description}
  \item[Field splitting:] After reflection on M9, the infrared stellar light is focused on M10, which is a roof mirror that separates the field in two and generates two beams (Fig.~\ref{fig:STSsketch}). The telescope and the derotator are adjusted such that the middle position between the two objects is located on the M10 edge and the objects are imaged on either side of the edge (Fig.~\ref{fig:normalswapped}). The telescope guiding and the derotator motion ensure that this situation is maintained during the observation. 
  \item[Metrology end point:] The {\small PRIMA} metrology beams originate downstream in the laboratory and overlap at the location of the dichroic M9, which they traverse towards an assembly of two spherical mirrors and a compensation plate (Fig.~\ref{fig:endpointsketch}), that serve to reflect the beams on themselves. The metrology beams are folded back into the stellar beams, traverse M9 again, and return towards the laboratory. The endpoint is realised by RR3 in Fig.~\ref{fig:endpointsketch}, which is where by design both metrology beams have total overlap (RR3 is in a pupil plane).
  \item[Image and pupil positioning:] M11 is the first mirror downstream of the field separator and is located in a pupil plane. It is mounted on a piezo tip-tilt stage which is used to position the image location in the laboratory, i.e.\ for fine-pointing the object. M11 is called the field selector mirror (FSM). M14 is located in an image plane and mounted on a piezo tip-tilt stage, which is used to adjust the output pupil lateral position. This mirror is also used to set the pupil's longitudinal position, defined by the mirror's radius of curvature. It is planned to achieve this dynamically with a variable curvature mirror (VCM), but it was not implemented at the time of writing and a fixed curvature mirror is used instead. Still, M14 is named STS-VCM.
\end{description}
The location of the metrology end points in the star separator shows that there is a substantial non-common path between the stellar and metrology beams inside the telescope. The stellar light path from the primary mirror M1 to the dichroic M9 and in particular inside the derotator is not monitored by the {\small PRIMA} metrology. Conversely, the retro-reflector assembly of three optical elements is monitored by the metrology, but is not part of the stellar beam path.

\subsection{Main delay lines}
\begin{figure}[h]\begin{center} 
\includegraphics[width = \linewidth,trim = 7mm 0mm 0mm 0mm, clip]{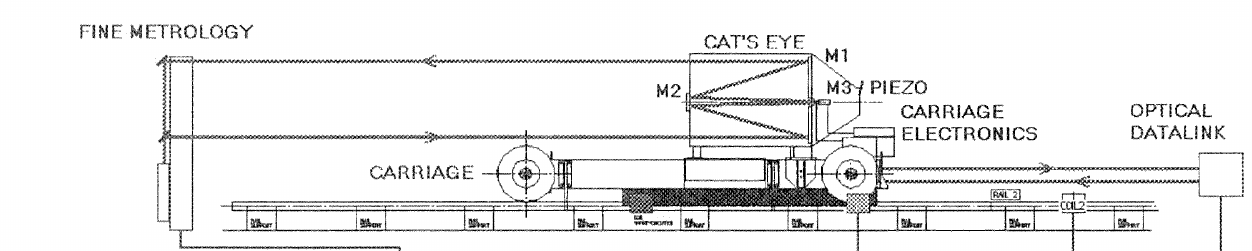}
\caption{Drawing of a main delay line showing the rail, the carriage holding the cat's eye assembly, and the delay line metrology beams. The wheelbase is 2.25~m. The tertiary mirror M3 is the variable curvature mirror DL-VCM, which is mounted on a fast piezo stage. From \cite{Hogenhuis:2003qy} .}\label{fig:DLsketch}\end{center}
\end{figure}

The {\small VLTI} main delay lines \citep{Hogenhuis:2003qy} are precision carts carrying a cat's eye-type reflector and used to introduce variable optical delay inside the interferometer. The rail length of $\sim$60~m limits the maximum optical delay to $\sim$120~m per delay line. Real-time delay control at nanometre level over the full range is achieved with a two-stage system composed of a coarse mechanism moving the full cart and a fine piezo actuator, combined with an internal metrology system measuring the position of the cart along the rail (Fig.~\ref{fig:DLsketch}). The main delay line has dual-feed capability, i.e.\ it accepts two beams which will experience the same optical delay. A variable curvature mirror (DL-VCM, \citealt{Ferrari:2003qe}) is part of the cat's eye assembly and it is dynamically adjusted along the delay line trajectory to keep the longitudinal pupil position constant in the laboratory and to preserve the field-of-view of the interferometer. It is realised by a mechanism consisting of a steel mirror connected to an over-pressure chamber of tunable pressure.

\subsection{Differential delay lines}
There is one differential delay line (DDL) for each of the four {\small PRIMA} beams, i.e.\ the optical delay can be adjusted for each beam individually. The DDL concept is very similar to the main delay lines with the difference that they are kept under vacuum. A motor for coarse actuation is combined with a fine piezo actuator and an internal laser metrology to achieve optical path length control at nanometre level and high bandwidth. A detailed description of the DDL is given in Sect.~\ref{sec:ddl}.

\subsubsection{PRIMA metrology}
The {\small PRIMA} metrology \citep{Leveque:2003fk, Schuhler2007, Sahlmann:2009kx} is a 4-beam, frequency-stabilised infrared laser metrology with the purpose of measuring the internal differential optical path difference (DOPD) between the two {\small PRIMA} feeds. The metrology beams have a diameter of 1~mm at injection and propagate within the central obscuration of the stellar beams, which originates from the telescope secondary mirror. Metrology endpoints are given by the fringe sensors' beam combiners and the star separator modules (Fig.~\ref{fig:endpointsketch}). The {\small PRIMA} metrology system gives access to two delays: the differential delay $\Delta L$ between both feeds, which is the main observable for astrometry and is named {\small PRIMET}, and the optical path difference of one feed $\Delta L_{B}$ corresponding to FSUB (Sect.~\ref{sec:fsu}), hence is named {\small PRIMETB}. Both are based on incremental measurements after a fringe counter reset, which is triggered by the instrument, meaning that the metrology has no pre-defined zero-point. The {\small PRIMA} metrology beams are also used to stabilise the lateral stellar pupil position. A quadcell sensor located on the fringe sensors' optical bench measures the lateral position of the metrology return beams and stabilises their positions with the STS-VCM mirror actuators in closed loop. Under the condition that the metrology and stellar beams are co-aligned and superimposed, the lateral pupil positions of the four stellar beams are stabilised. Details on the metrology measurement architecture are given in Table~\ref{tab:PRIMETdef}. 
\begin{table}
\caption{Details of the {\small PRIMA} metrology system.}
\label{tab:PRIMETdef}  \centering  
\small
\begin{tabular}{l c c c c c} 	
\hline\hline %
Feed && \multicolumn{2}{c}{FSUA} & \multicolumn{2}{c}{FSUB} \\
\hline
Input channel\tablefootmark{a} & & IP3 & IP1 & IP4 & IP2\\
Polarisation && p &s &p&s\\
Monitored path &(m) & $L_3$& $L_1$& $L_4$& $L_2$\\
$\delta \nu$\tablefootmark{b} &(MHz)& $+38.65$& $+38.00$& $-39.55$& $-40.00$\\
OPD & (m)& \multicolumn{2}{c}{$\Delta L_A = L_3-L_1$} &\multicolumn{2}{c}{$\Delta L_B = L_4-L_2$}\\
Beat frequency & (kHz)& \multicolumn{2}{c}{650} &\multicolumn{2}{c}{450}\\
Diff. delay &(m)&\multicolumn{4}{c}{$\Delta L =  \Delta L_A -\Delta L_B$}\\
\hline
Constants:\\
$\Delta \nu_P$\tablefootmark{c} &(MHz)& \multicolumn{4}{r}{78.1}\\
$\nu_P$\tablefootmark{d} &    (MHz)          & \multicolumn{4}{r}{227\,257\,330.6}\\
$\lambda_P$\tablefootmark{e} & (nm)      & \multicolumn{4}{r}{1\,319.176183} \\
\hline 
\end{tabular} 
\tablefoot{\tablefoottext{a} {The input channel (IP) refers to the physical location of the beam at the switchyard level. There are eight input channels at VLTI.} \tablefoottext{b} {Frequency shift relative to the laser frequency $\nu_P$.} \tablefoottext{c} {Frequency shift between both feeds introduced to avoid crosstalk.}  \tablefoottext{d} {Calibrated laser frequency.} \tablefoottext{e} {Laser vacuum wavelength.}} 
\end{table}
\subsection{Fringe sensor unit}\label{sec:fsu}
The stellar beams are combined by the fringe sensor unit (FSU, \citealt{Sahlmann:2009kx}). There are two combiners, named FSUA and FSUB, each receiving two beams of one stellar object. The FSU contains the interface for the injection and extraction of the {\small PRIMA} metrology beams. The output delay measurements and fluxes are used for real-time control and recorded by the astrometric instrument, constituting the scientific data. Because of its central role within the {\small PRIMA} facility, the detailed spectroscopic and photometric characterisation of the FSU is required. The calibration of the sensors is necessary to optimise their performance in the control system and to minimise systematic effects on the astrometric measurement. An exhaustive description of the FSU is given by \cite{Sahlmann:2009kx}.

\subsection{Critical aspects of the PRIMA-VLTI control system}
Several control loops acting on the stellar beams via the opto-mechanical components listed above are required to make {\small PRIMA} observations possible and they are listed in Table~\ref{tab:controlloops}. The fast communication between distributed subsystems relies on a fibre network realising a kHz-bandwidth shared memory \citep{Abuter2008, Sahlmann:2009kx}. When observing, the number of active loops totals at 19, if we consider equivalent instances for different beams, which illustrates the complexity required to orchestrate the facility. All control loops are discussed in this work except for the fringe tracking loops that are of critical importance for the efficiency of the observations. The fringe tracking loops are implemented in the OPD controller (OPDC) and the differential OPD controller (DOPDC).  So far, both controllers were operating independently and with identical algorithms and parameters, which use both group and phase delay measurements of the FSU to track on zero group delay and rely on the FSU delivered S/N to switch between three states: \emph{search}, \emph{idle}, and \emph{track} \citep{Sahlmann:2009kx, Sahlmann:2010fk}. When closed, the inner phase controller tracks a time varying target $\Phi^\mathrm{target}$ designed to maintain the group delay at zero. The discretised expressions prescribing the command $RT^\mathrm{offset}$ (\emph{real time offset}) sent to the delay line in addition to the predicted sidereal motion are given in Eqs.~\ref{eq:ftk22} and \ref{eq:ftk222}, where $\phi$ and $GD$ is the unwrapped phase (in rad) and group delay (in m) delivered by the fringe sensor, respectively, and $\kappa = 0.2586267825 \cdot10^{-6}$ is a constant. 
\begin{equation}
\label{eq:ftk22}
\Phi^\mathrm{target}_k = \Phi^\mathrm{target}_{k-1}  + p_0\, GD_k 
\end{equation}
\begin{equation}
\label{eq:ftk222}
RT^\mathrm{offset}_k = RT^\mathrm{offset}_{k-1} + p_1\, \kappa \left(\Phi^\mathrm{target}_k - \phi_k \right)
+ p_2\, \kappa \left(\Phi^\mathrm{target}_{k-1} - \phi_{k-1} \right)
\end{equation}
The subscript $k$ indicates the value at each time step and increases by one every 500 $\mu$s, i.e.\ the controller has a sampling rate of 2 kHz. The control gains $p_0=0.001$, $p_1 = 0.007$ and $p_2 = 0.063$ were determined for robust operation under varying atmospheric conditions and were kept unchanged during the complete test campaign. The outer group delay loop is thus implemented as integral controller and the inner phase loop realises a proportional-integral controller.
The fringe tracking strategy for {\small PRIMA} astrometry observations can be optimised by coupling both control loops and by adapting the gains of the differential controller, which is an ongoing activity at the time of writing. 

\subsection{Observing with the astrometric instrument}\label{sec:PRIMAobservations}
\begin{figure}
\begin{center}
\includegraphics[width = 0.48\linewidth,trim = 0cm 0.5cm 0cm 0cm, clip=true]{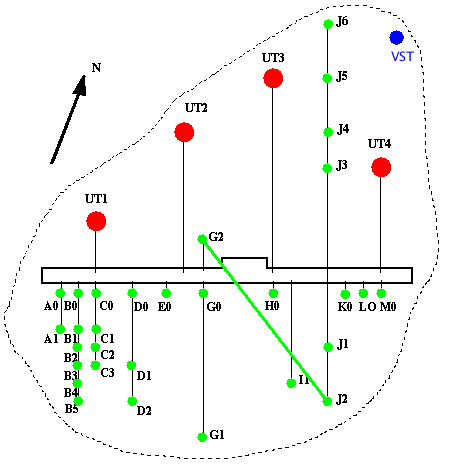}
\includegraphics[width=0.44\linewidth,trim = 16.6cm 20.2cm 2.5cm 3.5cm, clip=true]{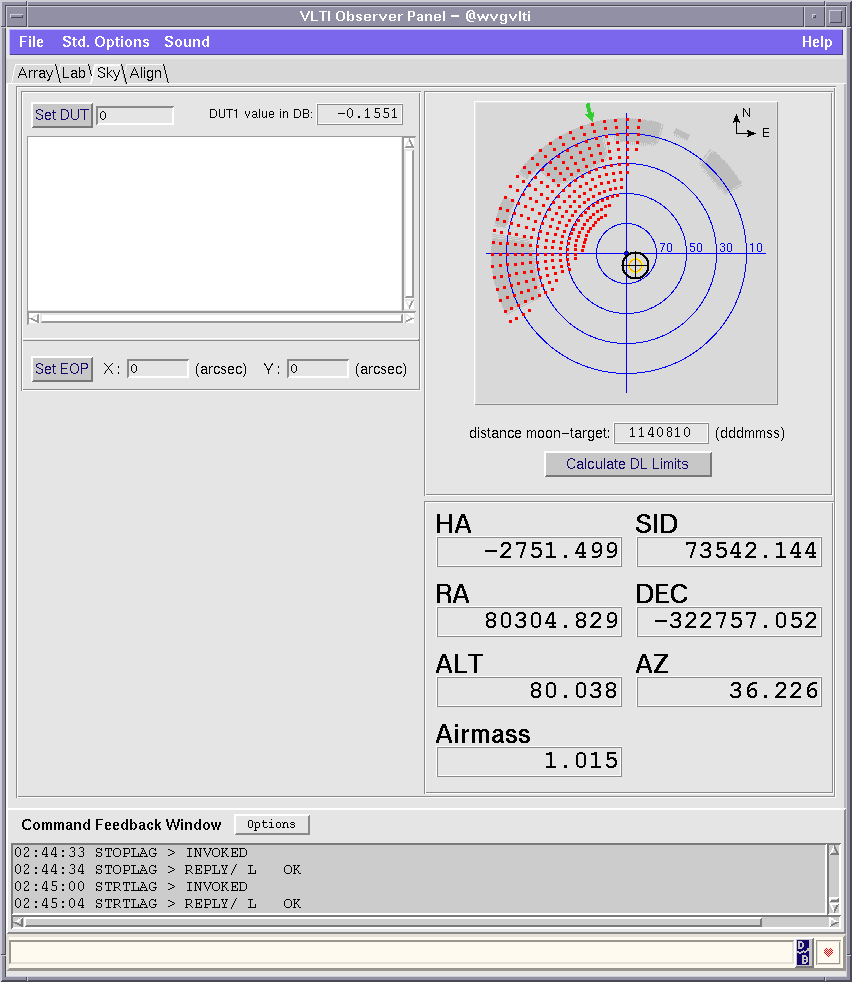}
\caption{\emph{Left}: Layout of the Paranal observatory platform indicating the baseline used during the commissioning as shown by the APES software. \emph{Right}: Not accessible regions in the sky due to delay line limits (dotted regions) and UT shadowing (grey patches) for the baseline G2-J2 as shown by the VLTI observer panel. Labels mark the elevation angle.}\label{fig:vlti}\end{center}
\end{figure}

The astrometric instrument of {\small PRIMA} is named {\small PACMAN} \citep{Abuter:2010lr} and is materialised by a computer, connected to the interferometer control system and the rapid fibre link between the real-time computers. The instrument executes the observation blocks by commanding the interferometer control system to preset the system to the specified configuration, acquire the target with telescopes and delay lines, and eventually by triggering the data recording. So far, {\small PRIMA} astrometric commissioning observations were made with the auxiliary telescope AT3 in station G2 and the telescope AT4 in station J2, resulting in a baseline length of $\sim\!91.2$~m (Fig.~\ref{fig:vlti}). The main delay lines DL2 and DL4 were used and despite of their long stroke, observations in the north-western sky were prohibited because the available internal optical delay is not sufficient. 

\subsubsection{Normal-swapped sequence}
\begin{figure}
\begin{center}
\includegraphics[width = \linewidth,trim = 0cm 0cm 0cm 0cm, clip=true]{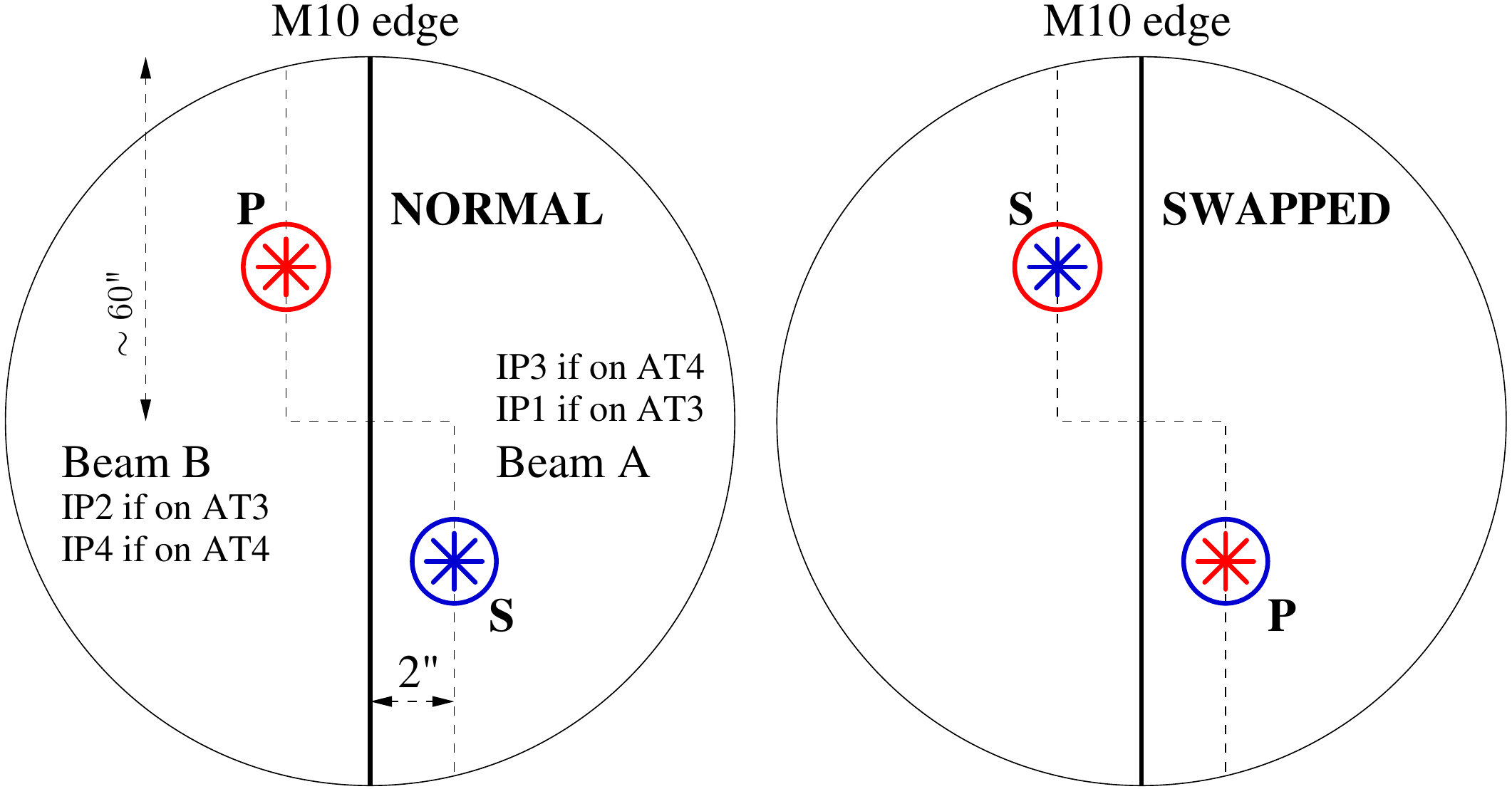}
\caption{To illustrate the implementation of the swap procedure, the M10 roof mirror is shown as seen by the technical CCD of the auxiliary telescope. \emph{Left}: In the normal mode configuration, the light of the primary star P falls onto the left side of M10 and feeds IP2 or IP4, thus FSUB, whereas the light of the secondary star feeds FSUA. The derotator ensures that the objects are kept on the dashed line at a distance equivalent to 2\arcsec~ from the mirror edge. \emph{Right:} The situation in swapped mode, where the derotator has rotated the field by 180\degr. The primary star is fed to FSUA and the secondary star to FSUB.}
\label{fig:normalswapped}\end{center}
\end{figure}

\begin{figure}
\begin{center}
\includegraphics[width = 0.8\linewidth,trim = 0cm 0cm 0cm 0.2cm, clip=true]{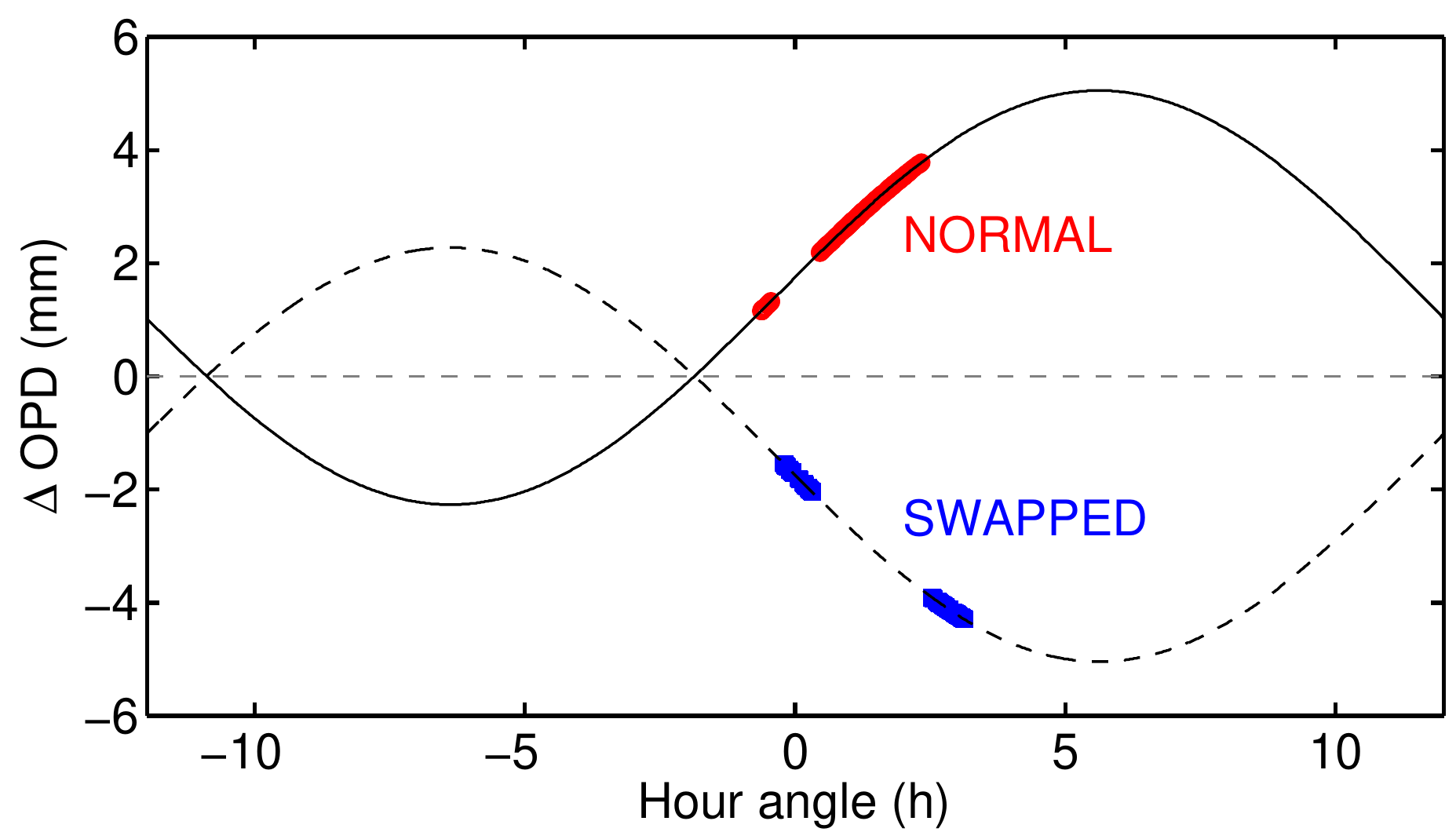}
\caption{The expected differential delay for HD\,10360 in normal (solid curve) and swapped mode (dashed curve) over 24 hours. Red and blue symbols show the data of the demonstration run discussed below. The metrology zero point has been subtracted and is indicated by the horizontal dashed line. The delay amplitude varies between 0-5~mm.} \label{fig:normalswapped24h}\end{center}
\end{figure}

The observation strategy to calibrate the metrology zero-point and to minimise differential errors between the two feeds is to obtain a sequence of normal and swapped observations. The physical swap procedure is executed by the derotator located between telescope and star separator. It rotates mechanically by 90\degr~resulting in 180\degr~field rotation so that the primary and secondary object fall on either side of the STS-M10 roof mirror and switch position after the operation (Fig.~\ref{fig:normalswapped}). Consequently, the external differential delay between the two objects is inverted, which has to be compensated internally by the DDL. The typical amplitude of differential delay is tens of mm with a slow dependence on hour angle, see Fig.~\ref{fig:normalswapped24h}.
\begin{figure}[h!]\begin{center}
\includegraphics[width=0.24\linewidth,trim = 7.1cm 25.2cm 2.5cm 4.2cm, clip=true, angle=0]{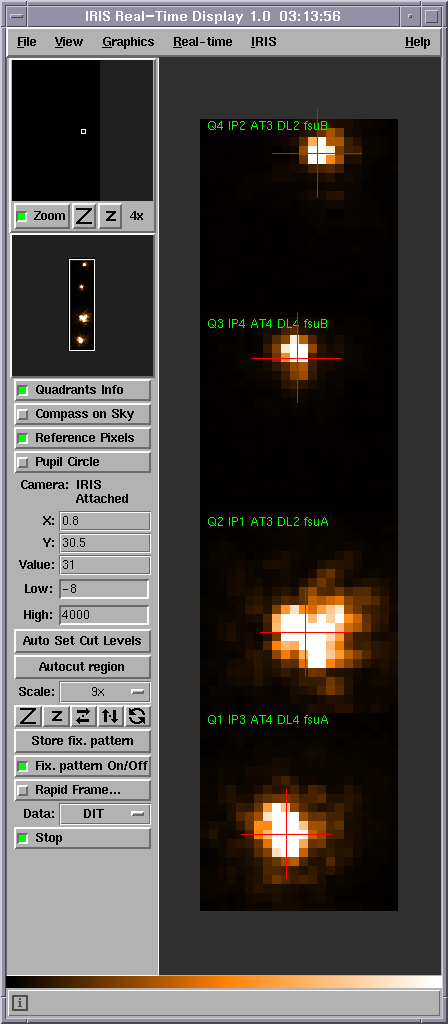}
\includegraphics[width=0.24\linewidth,trim = 7.1cm 18.2cm 2.5cm 11.2cm, clip=true, angle=0]{20569f14}
\includegraphics[width=0.24\linewidth,trim = 7.1cm 11.2cm 2.5cm 18.2cm, clip=true, angle=0]{20569f14}
\includegraphics[width=0.24\linewidth,trim = 7.1cm 4.2cm 2.5cm 25.2cm, clip=true, angle=0]{20569f14}
\caption{IRIS $H$-band images of HD\,66598\,A (\emph{two right panels}) and HD\,66598\,B (\emph{two left panels}). The IRIS quadrant, VLTI input channel number, telescope, delay line, and fringe sensor are indicated at the top of each panel. The magnitude difference between the stars is $\Delta m_H \sim 1.8$ and the exposure time was 0.2~s for all IRIS quadrants. The red cross in each panel indicates the guiding pixel, i.e.\ the target for the image stabilisation control loop, defined as the position of optimal alignment with the FSU. Because of fast atmospheric turbulence, the IP4 beam was not well aligned at the time of exposure.}\label{fig:irisexample}\end{center}
\end{figure} 

\subsubsection{Planning and definition}
The definition of astrometric observations with {\small PRIMA} follows the standard ESO scheme. Observation blocks (OB) are prepared with the P2PP\footnote{\tiny \url{http://www.eso.org/sci/observing/phase2/P2PPTool.html}} tool and transferred to the broker for observing blocks (BOB) on the instrument workstation, which executes the sequence of observing templates defined by the parameter settings in the OB. The {\small PRIMA} astrometric observation preparation software APES\footnote{\tiny \url{http://obswww.unige.ch/~segransa/apes/tutorial.html}} allows the user to plan the observation blocks of target stars defined in a user-provided catalogue and to export them to text files, which can be loaded in P2PP. 

\subsubsection{Instrument calibration and alignment}
The only component of the {\small PACMAN}  instrument that necessitates regular calibration are the fringe detectors (FSU). During the test runs, FSUA and FSUB are calibrated daily on a thermal source inside the laboratory \citep{Sahlmann:2009kx} with the goal of determining the photometric and spectral response of the system. The opto-mechanical components of the FSU are stable and require only minor re-alignments in monthly intervals due to seasonal temperature variations inside the laboratory\footnote{These concern the cold camera image actuators and the fibre positioners at injection level.}. To reach the required astrometric accuracy level of $10^{-6}$, the spectral response of the {\small PRIMA-VLTI} system has to be known during the observations and an on-sky calibration procedure has been devised \citep{Sahlmann:2009kx}, but has not been commissioned at the time of writing.\\
The alignment of the fringe sensor as self-contained system is stable, but the co-alignment with the optical axes of the interferometer can be disturbed by moving optical components during the day, thus it has to be verified. Before the night, both the laboratory guiding camera IRIS and the fringe detectors FSUA and FSUB are aligned on the beams generated by the laboratory light source. Because during observation the guiding camera will stabilise the image on the so-defined positions (the guiding pixels shown in Fig.~\ref{fig:irisexample}), it is guaranteed that the stellar beams are also stabilised on the FSU. Thermal variations could disturb the co-alignment, but this alignment strategy has proven efficient during many observing nights and usually does not necessitate intra-night corrections.

\subsubsection{Target acquisition}
The first observing template handles the target acquisition. The telescopes are pointed and the delay lines slew to the predicted position of zero total delay. The movable mirrors of the VLTI are configured to feed the four beams into the fringe sensor units. The integration times of the FSU and IRIS detectors are set to the value appropriate for the object magnitudes. When the telescope control loops for image stabilisation and guiding are closed, the stellar beams are acquired with the IRIS infrared camera and the laboratory guiding is enabled, using the STS-FSM mirror actuator to stabilise the PSF image for the FSU. In a similar fashion, the pupil stabilisation loop using the {\small PRIMA} metrology beams and the STS-VCM mirror actuator is enabled. At this stage the telescopes are guiding, the delay line and the differential delay line follow the predicted trajectories\footnote{At VLTI, the delay compensation is done with only one delay line (or DDL) moving at the time. The other delay line (or DDLs) is kept fixed during an observation.}, and the facility is ready to begin observing.

\subsubsection{Observation sequence}
The start of an astrometric observation is defined by the reset of the {\small PRIMA} metrology fringe counters. The astrometric measurement relies on the uninterrupted validity of the internal differential delay measured with this metrology, thus another reset or loss of it, e.g. a metrology glitch, marks the end of the usable observation data. Since the metrology zero-point is unknown, it has to be determined by a calibration step consisting of exchanging the roles of the primary and the secondary object, i.e.\ the swap procedure. The alternating observation in swapped and normal states reduces adverse effects on the astrometry, e.g. caused by dispersion, because errors common to both states are removed and only the differential terms remain (the metrology zero point can be seen as such a common mode error). The {\small PRIMA} astrometric data acquisition sequence can be broken down into conceptually equal blocks, which are executed either in normal or in swapped mode in the following order:
\begin{enumerate}
  \item Photometric calibration: sky-background, sky-flat, and combined flat exposures are taken to calibrate the FSU \citep{Sahlmann:2009kx}. These steps use the STS-FSM mirrors to apply an offset from the star in order to measure e.g. the sky background level. The IRIS and FSU camera backgrounds are taken simultaneously.
  \item Fringe detection scan: The actual fringe position in the primary and secondary feed differ by typically less than 1~mm from the model prediction. To facilitate the start of fringe tracking, a scan in OPD of $\sim$5\,mm is performed with the main delay line while recording FSU data. The processing of the resulting file yields the fringe positions in delay space of both primary and secondary feed and an estimate of the fringe S/N in the respective feed. The OPD control loop is then closed and fringes are tracked with the main delay line.
  \item Scanning observations: While fringe tracking on the primary star with the main delay line, a series of fast scans across the fringes of the secondary star is performed with one DDL and recorded (typically 400 scans). The data can be used both for astrometry and to measure the spectral response of the {\small PRIMA-VLTI} system. The scanning observation is optional and not always executed.
  \item Tracking observations: The secondary fringe tracking loop is closed with one DDL and data is recorded in dual-fringe tracking (typically 5 min of data). This represents the standard astrometry data.
\end{enumerate} 
After these 4 steps, the swap or unswap procedure is executed, which consists of opening the fringe tracking and beam guiding loops and of turning the field by 180\degr, thus sending the light of stellar objects into the respective other feed. After closing the telescope and laboratory guiding loops again, the sequence of steps 1.-4. is repeated. An astrometric measurement becomes possible after two sequences, i.e.\ when at least one observation in normal mode and one in swapped mode have been made, since the zero-point of the metrology can be determined. The basic characteristics and differences of the normal and swapped operation modes are listed below:
\begin{description}
  \item[\textbf{Normal mode:}] FSUB is used to track the primary star fringes with the main delay line via OPDC. FSUA is used to track the secondary star fringes with DDL1 via DOPDC. The internal differential OPD is controlled with DDL1, which can be used for fast scanning or fringe tracking.  
  \item[\textbf{Swapped mode:}] FSUA is used to track the primary star fringes with the main delay line via OPDC. FSUB is used to track the secondary star fringes with DDL2 via DOPDC. The internal differential OPD is controlled with DDL2, which can be used for fast scanning or fringe tracking.  
\end{description}

\subsubsection{Hardware inadequacies affecting the instrument performance}
All astrometric observations reported herein were obtained between November 2010 and January 2012. During this time, the {\small PRIMA} -VLTI subsystems exhibited the following hardware problems, which did not impede astrometric observations, but significantly reduced the instrument performance in terms of accessible stellar magnitude range and the data quality. 
\begin{description}
  \item[Optical aberrations of AT3-STS:] The two stellar beams coming from AT3-STS suffered from optical aberrations, which an observer can interpret as different focus positions. At the time of writing, the optical aberrations had not been corrected, yet. In practice, the best focus position of an auxiliary telescopes is found by the operator using the remote adjustment of the secondary mirror and the image quality as seen on IRIS. In the {\small PRIMA} case and because there is one common focus actuator for both beams (the telescope secondary mirror), an intermediate focus position had to be determined by minimising the optical aberrations of both beams apparent on IRIS. This is problematic because of the fast injection degradation with de-focus and the fast temporal change of focus position. During {\small PRIMA} observations the AT foci are adjusted approximately every 30 minutes.
  \item[Pupil vignetting of stellar beams:] The shape of the {\small PRIMA} pupils can be measured with a pupil camera located between the switchyard and the fringe sensors. Figure~\ref{fig:pupilexample} shows an example taken in November 2011. 
  \begin{figure}\begin{center}
\includegraphics[width=0.49\linewidth,trim = 0cm 8cm 0cm 0cm, clip=true, angle=0]{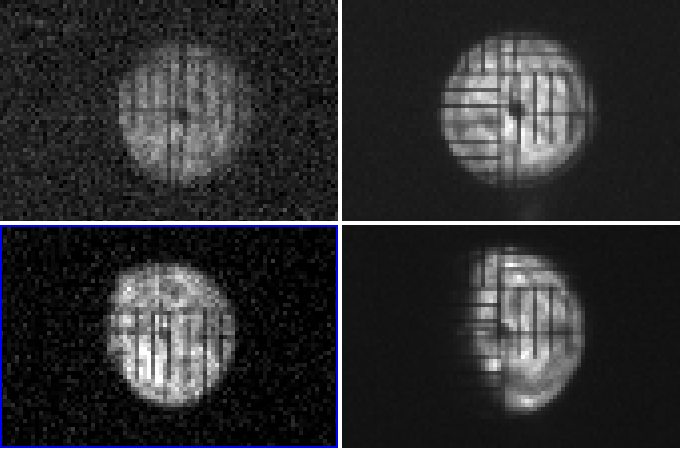}
\includegraphics[width=0.49\linewidth,trim = 0cm 0cm 0cm 8cm, clip=true, angle=0]{20569f15}
\caption{{\small PRIMA} pupils measured after the acquisition of HD\,10360 in swapped mode on November 19, 2011. The input channels are IP1 - IP4 from left to right. While all pupils show signs of vignetting, the two rightmost coming from AT4 are strongly obscured.}\label{fig:pupilexample}\end{center}
\end{figure} 
  \item[APD assembly of AT4-STS for image correction:] The tip-tilt correction system of the {\small VLTI} auxiliary telescopes has been upgraded by implementing a new assembly of the lenses in front of the avalanche photo diodes, resulting in improved image correction quality especially in good seeing conditions \citep{Haguenauer:2010uq}. The AT3-STS has undergone the upgrade, whereas AT4-STS was operating with the old system, thus operating in non-optimal conditions.
  \item[FSUA fibre transmission and cold camera alignment:] During the integration of the FSU at the VLTI, a degraded transmission of FSUA compared to FSUB was noticed \citep{Sahlmann:2009kx}. Additionally, the cold camera alignment of FSUA was not optimised resulting in a distorted spectral response function. Eventually, the fluoride glass fibre assembly of FSUA was exchanged in March 2011 and the cold camera was aligned, which improved the camera throughput\footnote{The cold camera flux loss in March 2009 was 13 \% and 5~\%, compared to 10~\% and 4~\% in November 2011 for FSUA and FSUB, respectively. The uncertainties are 1~\%.} and spectral response.
\end{description}
\section{Differential delay lines for PRIMA}\label{sec:ddl}
When observing two stars with a dual-field interferometer, the differential optical delay $\Delta w$ between the stellar beams has to be compensated dynamically to make the simultaneous observation of both fringe packets possible. For PRIMA, this is realised with the differential delay lines (DDL), which were delivered by the ESPRI consortium. To make the system symmetric and minimise differential errors, there are four DDL, i.e.\ one per telescope and per star. In preparation of a potential extension of {\small PRIMA} to operation with four telescopes, the DDL system can accommodate up to eight DDL. A detailed description of the DDL before installation at the observatory was given by \cite{Pepe2008}. Here, we present a concise overview of their design and implementation and report on their performance at the VLTI observatory.

\subsection{Technical requirements}
The DDL have been designed to comply with the technical requirements set by their operation within the {\small PRIMA} facility and specified by ESO. 
The most relevant requirements are:
\begin{itemize}
  \item Stroke: To compensate for the differential delay between two fields separated by 2\arcmin~observed with a baseline of 200 m, a single DDL must be able to introduce an optical delay of $\pm$66 mm.
  \item Transfer function: A high actuation bandwith ($>$200 Hz) is required to act as fast actuator for optical path length control, e.g. to compensate for atmospheric or structural turbulence. 
\item Vacuum operation: The DDL must be operated in vacuum to minimise the effects of differential dispersion on the astrometric measurement (cf. Sect.~\ref{sec:MDLfit}). In this way, both stellar beams in one interferometer arm travel the same optical path length in air.
  \item Operation modes: ($i$) Blind tracking mode: The DDL must be able to follow a given trajectory at a rate of up to 200 $\mu$m/s, e.g. corresponding to the differential delay change due to Earth rotation. ($ii$) Active tracking mode: The DDL must be able to introduce fast delay corrections, e.g. to compensate for atmospheric piston in closed loop with a piston sensor when fringe tracking, optionally in addition to following a given trajectory. 
($iii$) Scanning mode: The DDL can be operated in a scanning mode executing a periodic triangular delay modulation, for instance to search for fringes or to obtain data covering the complete fringe envelope.
\end{itemize}

\subsection{Design}
The DDL was designed by the {\small ESPRI} consortium in close collaboration with ESO. The concept is based on Cassegrain-type retro-reflector telescopes (cat's eyes) with $\sim$20 cm diameter that are mounted on linear translation stages. A stepper motor at the translation stage provides the long stroke of up to 69 mm, whereas a piezo actuator at the M3 mirror in the cat's eye realises the fine adjustment over $\sim$10\,$\mu$m with an accuracy of 1 nm. Both actuators are driven with a combined control loop, such that the optical path length can be adjusted over the full range of 120 mm (twice the stroke length) with an accuracy of 2 nm. The DDL and their internal metrology system are mounted on a custom optical bench in non-cryogenic vacuum vessels. The internal metrology beams are launched and collected in front of the cat's eye (Fig.~\ref{fig:DDLsketch}). Vacuum windows are part of the optical system and are integrated in the vacuum vessel. The actuators are controlled with front-end electronics located close to the optical bench in the interferometric laboratory of the {\small VLTI}. The interface with the interferometer control system is implemented with Local Control Units (LCU) running standard {\small VLT} control software and located outside the laboratory. 

\subsubsection{Optical table and vacuum system}
\begin{figure}
\begin{center} 
\includegraphics[width = \linewidth,trim = 0mm 0mm 0mm 0mm, clip]{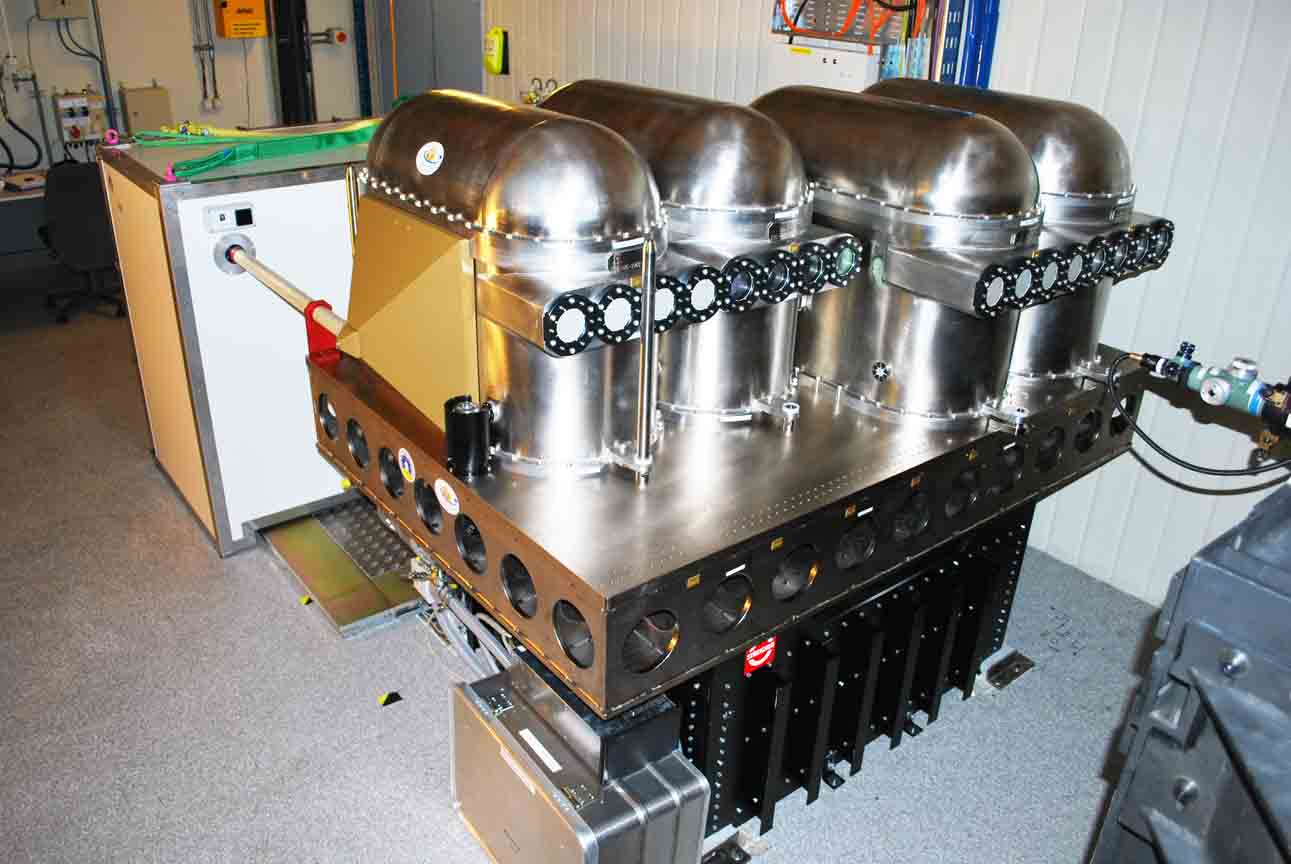}
\caption{PRIMA DDL installed in the VLTI laboratory. The optical table supporting four vacuum vessels is seen in the foreground. Each vessel has four circular optical windows of 66~mm diameter. The cabinet visible in the background contains the front-end electronics.}\label{fig:DDLtank}\end{center}
\end{figure}

The DDL are supported by a stiff optical table on which four vacuum vessels are installed. It is installed next to the switchyard table in the {\small VLTI} laboratory (cf. Figs.~\ref{fig:labsketch} and \ref{fig:DDLtank}). Each vessel can host two DDL, i.e.\ the vacuum system is prepared for up to eight units. As of February 2012, four DDL are installed and two vessels are therefore empty\footnote{At the time of writing, the construction of two additional DDL to complement the VLTI facility is underway.}. A pumping system is installed for vacuum maintenance. The front-end electronics are installed in an actively cooled cabinet located next to the optical table. The cabinet and the table have independent fixations to the laboratory floor, to avoid that vibrations from the liquid cooling system and cooling fans inside the cabinet are transmitted to the DDL table. 

\subsubsection{Cat's eye optics}
\begin{figure}
\begin{center}   
\includegraphics[width = \linewidth,trim = 0mm 15mm 0mm 1mm, clip]{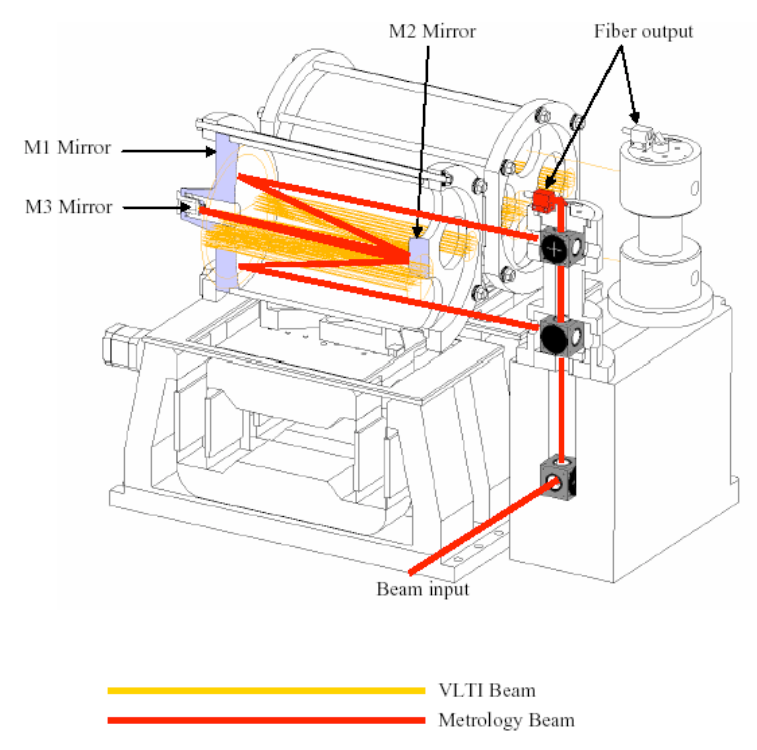}
\caption{Schematic and longitudinal section of two DDL. The translation stage supporting the optics is visible below the telescope tubes. The M3 mirror is mounted on a fast piston and tip-tilt piezo actuator. Stellar beams are shown in yellow and internal metrology beams are red.}\label{fig:DDLsketch}\end{center}
\end{figure}

The cat's eye retro-reflector is realised with three mirrors and five reflections, which result in a horizontal shift between input and output beam of 120 mm. Because the stellar pupil's longitudinal position in the beam combination laboratory has to be identical with and without DDL in the beamtrain and due to the pupil configuration present at the VLTI, the magnifications of individual DDL are not identical but were adapted by adjusting the curvature of M3. All other optical parameters are identical for the four DDL. The parabolic primary mirror, the hyperbolic secondary mirror, and the spherical tertiary mirror are mounted to the telescope tube. M3 is attached to a three-piezo actuator capable of piston and tip-tilt adjustment (Fig.~\ref{fig:DDLsketch}).

\subsubsection{Translation assembly}
The mechanical translation assembly consists of a motorised translation stage providing the 70 mm stroke and a piezoelectric short-stroke actuator for M3. The main translation stage is a guided mechanism composed of two arms where each arm is a compensated parallelogram constituted of four prismatic blades. This custom system realises a rigid, though highly accurate one-dimensional displacement mechanism and is driven by a stepper motor used as DC motor (\emph{Ultramotion Digit}). A piezo-electric platform (customised \emph{Physik Instrumente S-325}) was chosen to support M3. This actuator has three degrees of freedom (tip, tilt, and piston) and makes it possible to both compensate for slowly varying lateral pupil shifts caused by imperfections of the main translation stage and to introduce fast piston changes. The mechanical piston stroke is 30 $\mu$m, which corresponds to a differential optical stroke of $\pm30\mu$m, and the mechanical tip/tilt range is $\pm4$ milli-rad. 

\subsubsection{Internal metrology}\label{sec:met}
The purpose of the internal metrology is to measure the instantaneous optical position of the DDL, thus the optical delay introduced in the stellar beam. The system is based on commercially available technology for displacement measurement (\emph{Agilent}), which is also used in the main delay lines. Each single DDL has its own metrology receiver such that its position can be measured independently. A folding mirror directs the metrology laser beam into the vacuum vessel where it is split in two beams, one for each of the two DDL enclosed in the vessel. Each beam feeds a Mach-Zehnder type interferometer, whose interference signals are detected with optical receivers. The front-end electronics convert optical to electric signals and the back-end electronics (VME boards) compute the interferometric phase with a resolution of $2\pi/256$, corresponding to an OPD resolution of 2.47 nm.\\
After several months of operation at the observatory, opto-mechanical drifts were detected in the metrology system which made regular alignment necessary. Consequently, a slightly modified design leading to a more robust system has been devised and will replace the current system in 2013. 

\subsubsection{Instrument and translation control}
The instrument control hardware is composed of VLT standard components and the software complies with the VLT common software package for instrumentation. The real-time control algorithms are coded in the ESO software architecture TAC (tools for advanced control). To achieve optimal control of the two-stage system composed of the piezo actuator and the motor, their respective controllers are interfaced by an 'observer'. In this way, a non-degenerate closed loop control system is realised, where the feedback is given by the laser metrology measurement. The first resonance of the motorised translation occurs at $\sim$100 Hz. Because this stage does not need to be very fast, we avoid excitation of this mode by limiting the bandwidth of the motor stage to $\lesssim$10 Hz and by low-pass filtering the reference fed to the motor controller. The motorised stage therefore off-loads the piezo at low frequency. Because the mirror attached to the piezo-electric stage is very light ($\sim$3.5 g), the amplifier (\emph{PI} E-509) and the input capacitance of the piezo itself are determinant for the system bandwidth. 

\subsection{Laboratory performance}
\begin{figure}[h!]\begin{center}  
\includegraphics[width = \linewidth,trim = 0mm 0mm 0mm 0mm, clip]{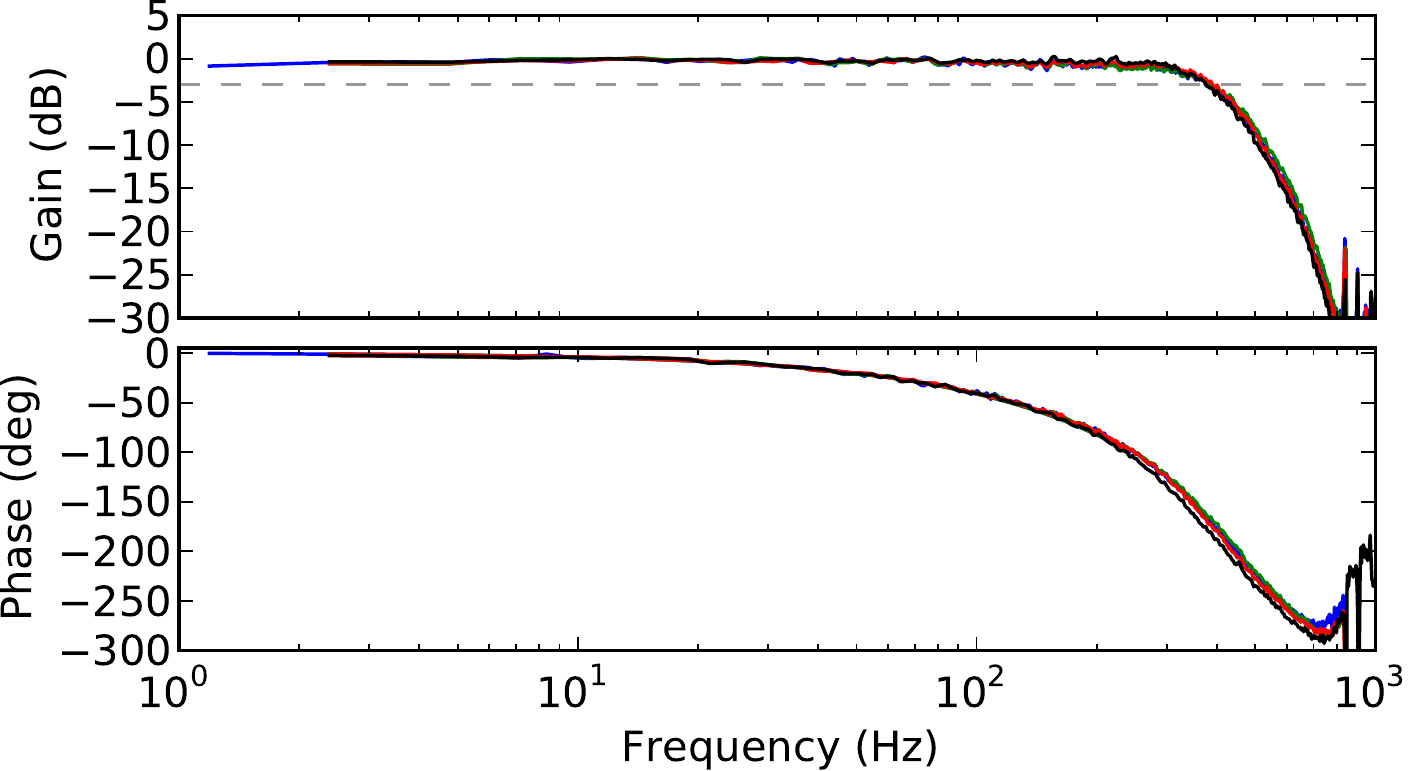}
\caption{DDL transfer function gain (\emph{top}) and phase (\emph{bottom}) as a function of frequency. Different colours identify the four DDL units. The dashed line in the top panel indicates the $-3$\,dB threshold.}\label{fig:DDLTF}\end{center}
\end{figure}

The DDL system was thoroughly tested before delivery to the observatory to confirm that it complies with the technical requirements and the detailed optical performance is reported in \cite{Pepe2008}. We therefore briefly summarise the most important values: The throughput of the DDL's optical system composed of cat's eyes and windows is shown in Table~\ref{tab:DDLtrans}.
\begin{table}[b]
\caption{Total optical transmittance of the DDL system}
\label{tab:DDLtrans}  \centering  
\small
\begin{tabular}{c r r r r} 	
\hline\hline %
Bandpass  & \multicolumn{4}{c}{Transmittance\tablefootmark{a}} \\
($\mu$m) & Requirement & Windows (2$\times$) & Cat's eye & Total\\
 \hline
$0.6 - 1.0$&	$0.60- 0.80$&	0.86&	0.80&	0.69\\
$1.0- 2.0 $&	$> 0.80$&	0.93&	0.90&	0.84\\
$2.0 - 2.5 $&	$> 0.85$&	0.93&	0.93&	0.86\\
$2.5 - 28 $& 	$> 0.90$&(no windows)&	0.91&	0.91\\
$1.319    $&$> 0.80$&	0.99&	0.90	&0.89\\
\hline 
\end{tabular} 
\tablefoot{\tablefoottext{a} {Theoretical throughputs are calculated from the transmittance curves of individual components provided by the manufacturers.}}
\end{table}
The transfer function of the DDL is shown in Fig.~\ref{fig:DDLTF}. The responses of the single DDL units were made to match by tuning their control parameters and the actuation bandwidth is $380\pm10$ Hz with a quality factor of $\sim$0.7. This setup was chosen for increased robustness, but can further be tuned if the operation conditions at the observatory make it necessary.
The maximum velocity of the DDL is $5.38\pm0.01$ mm/s. A summary of the achieved performance is given in Table~\ref{tab:DDLsummary}. 
\begin{table}[h]
\caption{Technical requirements and achieved performance}
\label{tab:DDLsummary}  \centering  
\begin{tabular}{l r r r } 	
\hline\hline %
	&&Specification	&Measured\\
	\hline
Field of view in pupil &(\arcmin)&	$> 10$&$	> 10$\\
Delay range&(mm)&	$0- 120$&$	0- 137$\\
RMS wavefront error &(nm)&$	< 25$&$16\pm2$\\
Tilt &(\arcsec)&	$< 1.5$ & $0.3\pm0.1$\\
Differential tilt &(\arcsec)&	$< 0.75$&$	< 0.7$\\
Transmission &&	\multicolumn{2}{c}{See Table~\ref{tab:DDLtrans}}\\
Resolution &(nm)&	$< 2.5$	&$2.5$\\
Bandwidth & (Hz) &	$ > 200$ & $380\pm10$ \\
Lateral pupil stability &($\mu$m)&	$< 50$ PTV&	$60$ PTV\\
\hline 
\end{tabular} 
\end{table}

\subsection{Performance at the observatory}
After integration at the {\small VLTI}, it was verified that the DDL in stand-alone operation comply with the performance established in the laboratory in Europe. Thereafter they were introduced and commissioned as a part of the interferometer opto-mechanical and control system. To become compatible with the operation of the {\small PRIMA} metrology, rate and acceleration limiters had to be introduced in software to limit the displacement speed and acceleration at an acceptable level, especially when large, step-like motion commands are sent by the control system. During the {\small PRIMA} dual-feed commissioning, the DDL were used routinely in all operation modes, i.e.\ to follow a predicted sidereal motion, to act as actuator for fringe tracking on the secondary star, and to perform fast triangular motion for fringe scanning. Their operation proved to be robust and reliable. In a single test observation, the DDL were also successfully used for fringe tracking on the primary star. Taking advantage of their excellent actuation bandwidth, the DDL could be used to correct high-frequency piston disturbances, e.g. structural vibrations, in the future. The advantage of having two fast piston actuators in sequence also offers the opportunity to test optimised control strategies for fringe tracking on primary and secondary star.\\
It was noticed that fast motion with large amplitudes beyond the DDL's specified working range can induce mechanical oscillations of one DDL system, which then propagate to the other DDL by mechanical coupling since all DDL are mounted on the same optical table. Those oscillations are caused by the stepper motor which is controlled without additional damping and they may impact operations because the internal metrology system can fail in these conditions.\\
In response to the requirements of the second-generation {\small VLTI} instrumentation, ESO ordered two additional DDL which are under construction at the time of writing. The necessary upgrade to the DDL metrology opto-mechanical system mentioned in Sect.~\ref{sec:met} will be made when those new DDL will be installed.
\section{PRIMA astrometry data reduction}\label{sec:red}
An astrometric observation is characterised by the sequence of normal and swapped exposures and requires at least one exposure in each mode. Raw data are therefore grouped into sets of files that correspond to one observation or \emph{astrometric run} and do not have reported {\small PRIMA} metrology glitches. Between January 2011 and March 2012, {\small PRIMA} was used to obtain $\sim$60 astrometric sequences of eleven different targets. During the first light mission, many short runs were recorded, whereas the later commissionings concentrated on acquiring long duration runs suitable for accurate model testing. For the purpose of commissioning, we developed a dedicated and highly flexible data reduction and analysis software package, which is briefly described below. Further details can be found in \cite{Sahlmann2012PhD}. The {\small PRIMA} data are sampled at kHz-rate and stored in binary FITS tables. Before fitting an astrometric run containing continuous data over tens of minutes, these data have to be reduced to a manageable size. To ease computations, all data tables are linearly inter- or extrapolated to a common time grid. Basic quality checks and verifications are made at this stage. These include verification of the FSU and {\small PRIMA} metrology sampling rates, verification of the {\small PRIMA} metrology status and glitch counters requiring data correction where applicable, and the detection of potential fringe-runaway events\footnote{As a consequence of inapt control thresholds for fringe tracking.}. The data reduction is accomplished by averaging over a timespan of the order of $\sim$1~s. Combined observables of several raw data tables, e.g. the astrometric observable, are computed before averaging to account for possible correlations. To minimise biases caused by time-averaging quantities which drift considerably during the averaging window, e.g. the differential delay and the delay line position, a model is subtracted from the measurement before averaging and added back thereafter. The model motions are based on the target input catalogue. The output of the data reduction step is one intermediate data file per run, which contains all necessary information for the detailed astrometry  analysis and model fitting. 

\subsection{PRIMA astrometry raw data format}
The {\small PACMAN}  instrument records {\small PRIMA} astrometry data in FITS files, containing a primary header and eleven binary tables. The most relevant entries are discussed herafter. The FSU table contains four data columns corresponding to the detector quadrants A-D and every column contains six subcolumns for the detected intensities in the six spectral pixel. It also contains the phase delay (PD) and group delay (GD), used as feedback signals for the fringe tracking loop, and the FSU S/N, used by the OPD controllers for mode switching \citep{Sahlmann:2009kx}. The {\small METROLOGY\_DATA} table contains the differential delay $\Delta L=\Delta L_A- \Delta L_B$ and the  {\small METROLOGY\_DATA\_FSUB} table contains  $- \Delta L_B$, i.e.\ the delay in the FSUB feed multiplied by $-1$. The corrections applied to the PRIMA metrology measurements are discussed in Appendix~\ref{sec:PRIMETcorr}. The OPD controller tables contain information about the controller state, the internal control signals, the setpoints sent to the delay line, and the metrology readings of all active differential and main delay lines. The time-stamps in all tables of one file are given relative to one common time defined by the {\small PCR\_ACQ\_START} header keyword. The sampling rate of the data in the D/OPDC, {\small PRIMA} metrology, and FSU table is 2~kHz, 4~kHz, and 1~kHz, respectively, where the latter can be smaller depending on the star magnitude. See \cite{Sahlmann2012PhD} for further details.

\subsection{Intermediate reduced data}
The relevant reduced data, i.e.\ the 1 second averages for all files of one astrometric run, are stored in an intermediate file. This decouples the data reduction from the data analysis step and facilitates the exchange and comparison of different reduction strategies. For every 1~s averaged sample, the file contains a timestamp, a mode identifier, which is an integer number that indicates the mode in which the data sample was taken (normal or swapped and dual-tracking or fast-scanning mode), and the observables with associated uncertainties, e.g. the astrometric observable, the delay line and differential delay line positions, the {\small PRIMA} metrology readings, and the file name corresponding to the data point.

\subsection{Dual fringe tracking data}
Tracking files are named {\small \texttt{PACMAN\_OBJ\_ASTRO[...]}} and contain dual-feed fringe tracking data. For the reduction, only samples in dual fringe tracking are kept and this criterion is based on the controller states of OPDC and DOPDC, both required to be '7'. The astrometric observable $AX_{obs} $ is a linear combination of the differential delay $\Delta L$ measured with the {\small PRIMA} metrology, the group delay $GD_A$ of FSUA, and the group delay $GD_B$ of FSUB, directly taken from the respective binary tables:
\begin{equation}\label{eq:axobs}
AX_{obs} = \Delta L - GD_A - GD_B,
\end{equation}
where the signs have been determined empirically. Figure~\ref{fig:summary} of the appendix shows an example of this first analysis step. 
Equation~\ref{eq:axobs} is sufficient for a first performance evaluation of the astrometric instrument. In the future, it could be improved for instance by considering the phase delay, which has lower noise compared to the group delay but has to be corrected for dispersion, and/or by implementing an optimised group delay algorithm. For the science-grade data reduction, it is envisaged to account for the near real-time spectral response of the beamtrain and to obtain optimal and unbiased estimators for the astrometric observable on the basis of the raw pixel counts. 

\subsection{Scanning data}
Scanning files are named {\small \texttt{PACMAN\_OBJ\_SCAN[...]}}\footnote{The filenames \texttt{PACMAN\_OBJ\_SPECTRUM[...]} and \texttt{PACMAN\_SKY\_VLTIRESPONSE[...]} contain equivalent data but were taken with a different template to obtain the star's spectrum and the interferometer's spectral response, respectively.} and contain data taken while fringe tracking on the primary and performing fast fringe scans on the secondary feed using a DDL. The basic reduction is slightly more complex than for the tracking data and it involves the following steps: (1) Scan identification and data association. (2) Interpolation on a regular OPD grid and flux normalisation. (3) Computation of the wavelet transform \citep{Torrence:1998fk}\footnote{A modified version of the code available at \url{http://paos.colorado.edu/research/wavelets/} was used.} for fringe localisation. (4) Elimination of low-quality scans based on S/N and primary fringe tracking quality criteria. The astrometric observable is now given by
\begin{equation}\label{eq:axobs2}
AX_{obs} = \Delta L_w - GD_{track},
\end{equation}
where $\Delta L_w$ is the {\small PRIMA} metrology value at the secondary's fringe position determined with the wavelet method and $GD_{track}$ is the average group delay of the primary measured by the tracking FSU during the scan.  
\begin{figure}[h!]\begin{center}
\includegraphics[width = \linewidth]{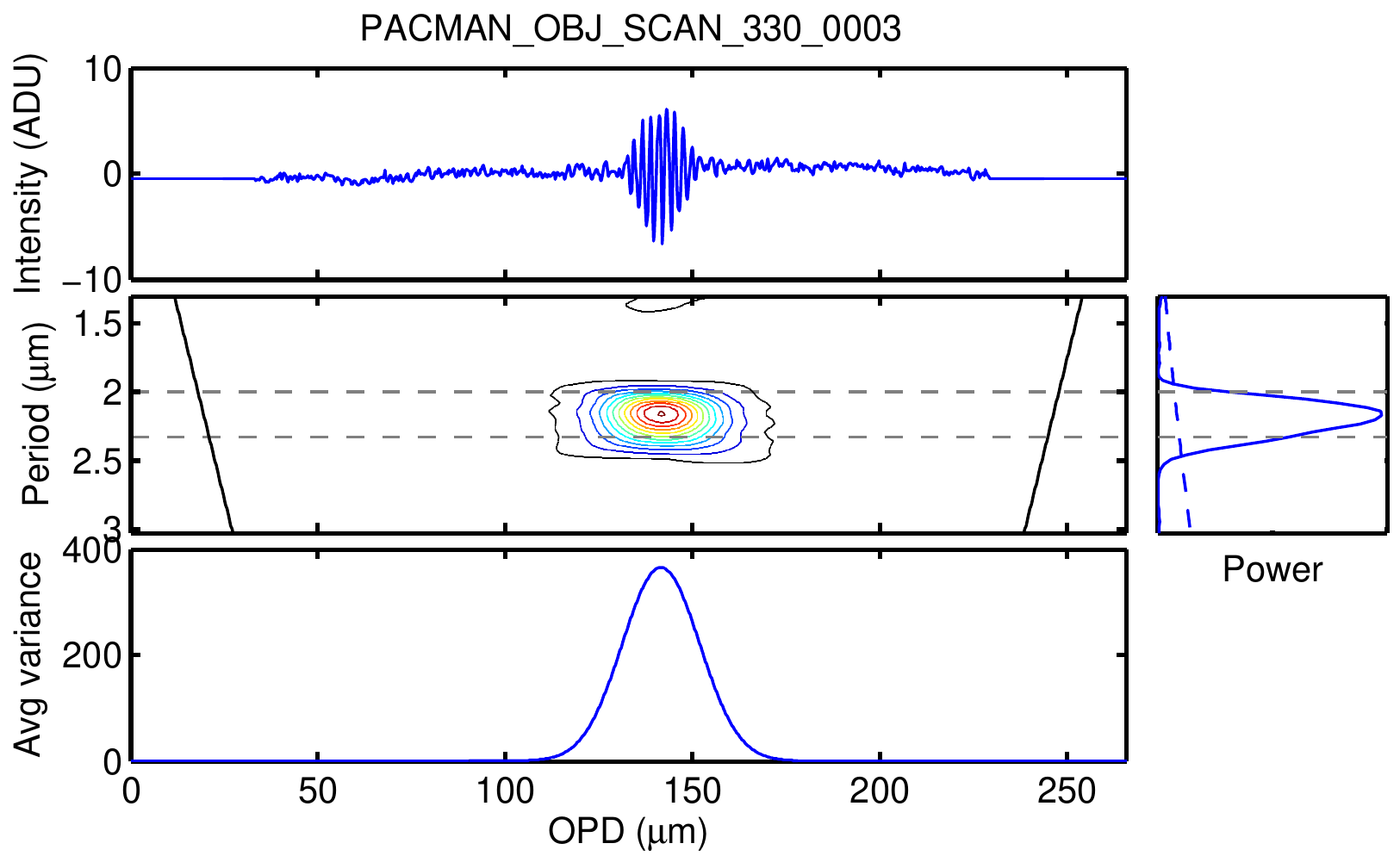}
\caption[Illustration of the wavelet analysis]{Illustration of the wavelet analysis. \emph{Top}: The time series showing the fringe signal (here A-C in white light). \emph{Middle}: The contour plot shows the fringe localisation by the wavelet transform in both delay and frequency space. Only data between the two horizontal dashed lines are considered. \emph{Bottom}: Projection of the wavelet power in the time domain. \emph{Right}: Projection of the wavelet power in the frequency domain.}\label{fig:wavelet}\end{center}
\end{figure}
\begin{figure}
\begin{center}
\includegraphics[width = 0.8\linewidth]{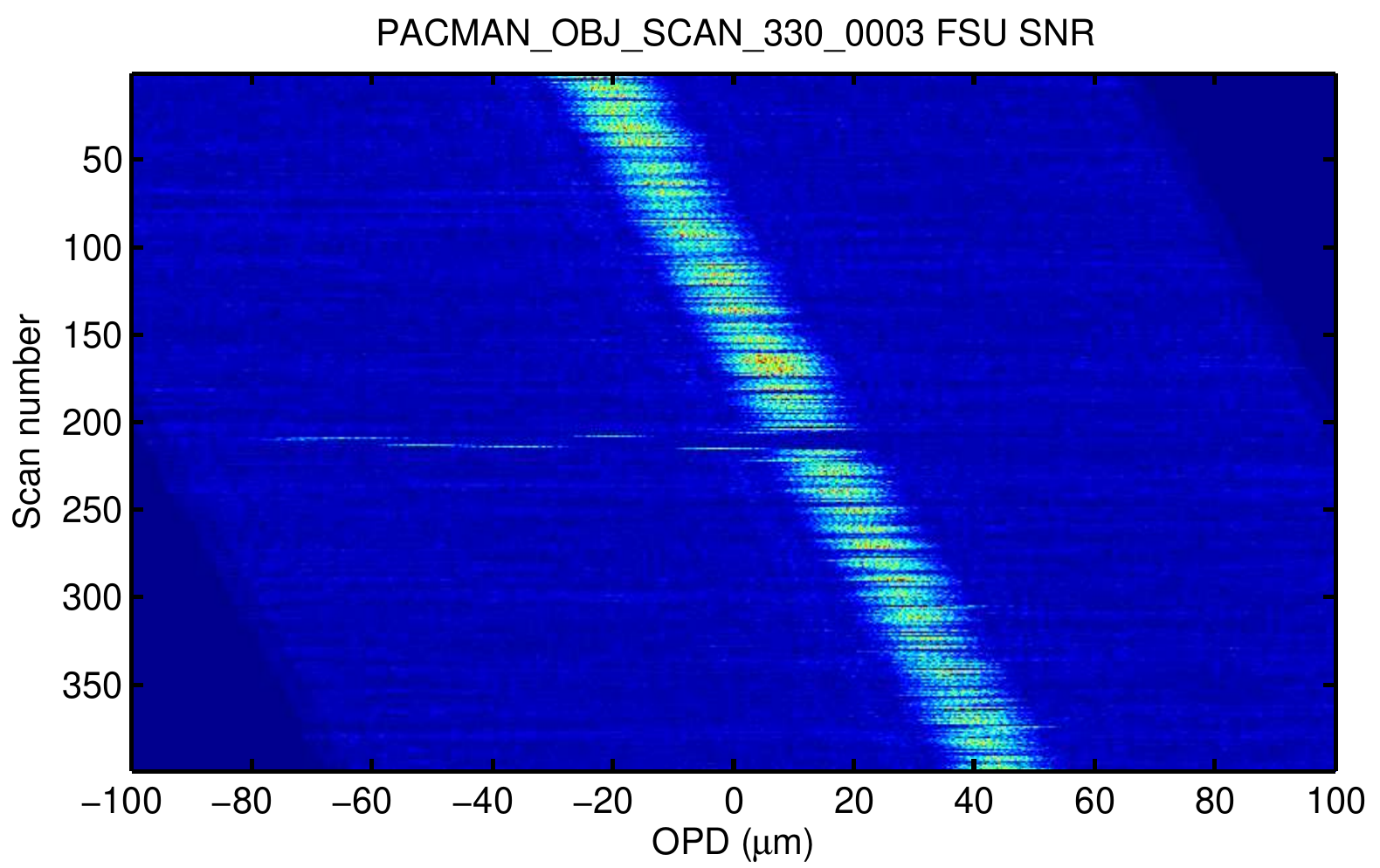}
\caption{Raw S/N delivered by the FSU in the scanning channel as a function of differential delay (measured with the {\small PRIMA} metrology) and scan number shown on the vertical axis. A loss of lock on the primary fringe tracking occurs at the scan 220.}\label{fig:scanning}\end{center}
\end{figure}\\ 
In the simplest implementation, the wavelet method is applied to the two white-light flux differences A--C and B--D, both pairs are close to phase-opposition, and the final fringe position is a combination of both results. Figures~\ref{fig:wavelet} and \ref{fig:scanning} illustrate the reduction procedure. The above algorithm is sufficient for a first performance evaluation and proved that astrometric data obtained in dual tracking or fringe scanning mode are equivalent at the $\mu$m-level, see e.g. Fig.~\ref{fig:axres1}, with the scanning data having a slightly higher noise. The scanning data are rich in information about the photometric fluctuations during the observations, the instantaneous phase-shifts and effective wavelengths of the FSU quadrants or pixels, and the dispersion effects on the fringe packet, which we have started to characterise.

\section{PRIMA astrometry data modelling}\label{sec:axmodelling}
We describe the data analysis principles that are applied to characterise the observations and eventually lead to the astrometric measurement of the target pair's separation. We first discuss the modelling of the main delay, before proceeding to the differential delay fitting. With the help of one observation sequence, we illustrate the individual steps of the data analysis up to the fit of a separation vector. Error bars are usually derived from Monte Carlo simulations. Three binary stars play an important role in the initial system characterisation and are therefore presented in Table~\ref{tab:targets} and briefly discussed below.
\begin{itemize}
  \item \object{HD 202730} is separated by 7\arcsec~from the secondary \object{GJ 9733 B}. The separation change measured over 150 years is well approximated with a linear motion and this binary is included in the \href{http://www.usno.navy.mil/USNO/astrometry/optical-IR-prod/wds/lin1}{Catalog of Rectilinear Elements}.
  \item \object{HD 10360} is separated by 11\arcsec~from the secondary \object{HD 10361}. Both stars are very bright in $K$-band and have nearly equal masses \citep{Takeda:2007ys}. This visual binary shows considerable orbital motion over the available separation measurements of 200 years and an orbital solution was given by \cite{van-Albada:1957lr}. This is the pair observed during the run used for illustration below. 
  \item \object{HD 66598} is separated by 36\arcsec~from the secondary \object{HD 66598 B}. So far, it is the widest pair observed with PRIMA. As indicated by the available measurements in the literature, this binary does not exhibit a significant separation change over the last 100 years.
\end{itemize}

\begin{figure}
\begin{center}
\includegraphics[width = 0.6\linewidth]{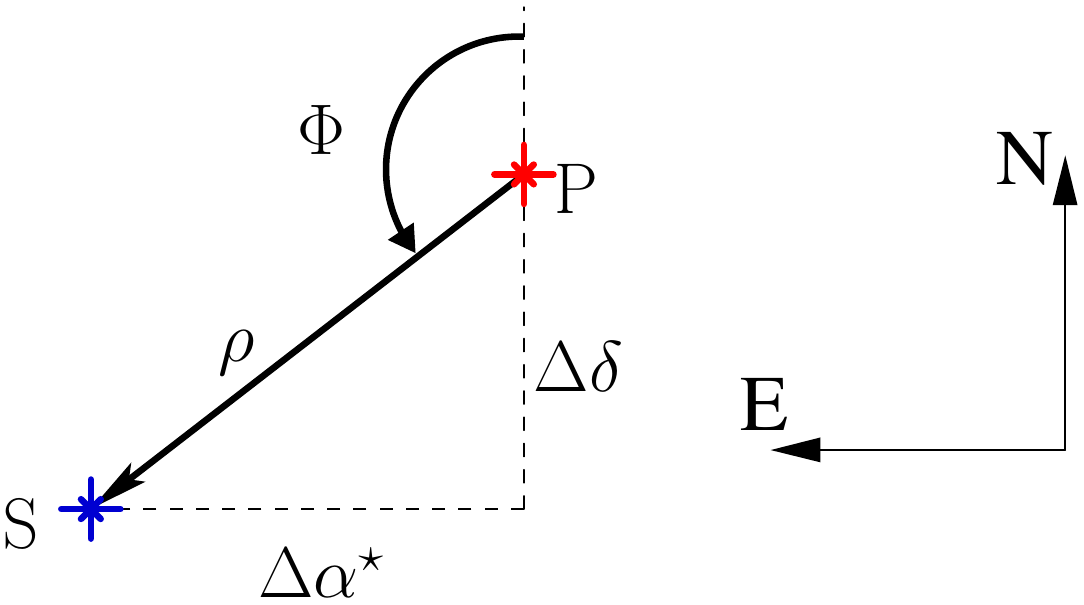}
\caption{The target pair separation $\rho$ and position angle $\Phi$ define the relative position of the secondary star S and primary star P. The separation vector points towards S and has the rectangular components $\Delta \alpha^\star$ and $\Delta \delta$. North is up and East is left.}
\label{fig:myPair}\end{center}\end{figure}

\subsection{Correction of atmospheric refraction and main delay fitting}\label{sec:MDLfit}
The VLTI-internal main delay is introduced by the main delay lines and the effect of the Earth rotation makes it a dynamic quantity. We have access to two measurements of its value, one is the internal metrology of the main delay line (Helium-Neon laser, $\lambda_{DL} \sim 633$~nm) and the other is the FSUB arm of the {\small PRIMA} metrology named {\small PRIMETB} (Nd:YAG laser, $\lambda_{P} \sim 1319$~nm), which in addition to the tunnel measures the optical path within the laboratory and the light ducts up to the retro-reflectors inside the star separators of the auxiliary telescopes (Fig.\,\ref{fig:labsketch}). Both have to be corrected for atmospheric refraction. For astrometric interferometry, the following terms related to chromatic and achromatric refraction have to be considered:
\begin{enumerate}
  \item Correction of laser metrology measurements: The purpose of a metrology system is to measure the optical path length between its endpoints. Usually, the vacuum laser wavelength is known and used to convert phases to delays. If air-filled beam trains are used, the effective laser wavelength is altered and the delays have to be corrected for the refractive index representative of the air inside the interferometer. 
  \item Effect of wavelength difference between stellar and metrology beams: If the stellar and metrology bandpasses are not identical, the chromaticity of the refractive index creates a mismatch between the optical path difference experienced by the stellar beam and the one determined with the metrology.
  \item Atmospheric refraction: For conventional imaging instruments, atmospheric refraction results in an offset between the true source vector $\vec s_0$ and the apparent, refracted source vector $\vec s$ (e.g. \citealt{Gubler:1998fk}). In the case of an interferometer, the first order term of this effect is removed if either vacuum delay lines are used or if the effects 1. and 2. are accurately corrected. This is because the geometric path difference $GPD$ is equal to the vacuum optical path difference $\vec B \cdot \vec s_0$ occurring outside the atmosphere, where $n=1$, and it is equal to the external optical path difference $n_{1,\mathrm{FSU}} \,\vec B \cdot \vec s$ experienced within the atmosphere of refractive index $n_{1,\mathrm{FSU}}$, where the second subscript indicates that the index is computed for the bandpass accepted by the instrument (e.g. \citealt{Daigne:2003uq}). To observe fringes, it is compensated by an optical path difference $n_{2,\mathrm{FSU}}\,D$ internal to the interferometer, where $D$ is the internal vacuum delay. The laser metrology system measures the internal optical path difference $L = n_{2,\mathrm{MET}}\,D$, where $n_{2,\mathrm{MET}}$ is taken at the metrology wavelength and is determined by the atmosphere at the delay lines, cf. Fig.~\ref{fig:GPD}. It follows that
  \begin{equation}\label{eq:disp1}
GPD = \vec B \cdot \vec s_0 = {n_{1,\mathrm{FSU}}}\,\vec B \cdot \vec s = n_{2,\mathrm{FSU}}\,D = \frac{n_{2,\mathrm{FSU}}}{n_{2,\mathrm{MET}}}\,L,
\end{equation} 
where $n_1$ and $n_2$ are determined for the detected stellar bandpass and the metrology wavelength, respectively, with the local atmospheric parameters. Note that the angle of $\vec s$ depends on $n_1$ through Snell's law. If the delay lines are evacuated, $n_2 = 1$ and $\vec B \cdot \vec s_0  = L$. Second order terms appear for instance because the zenith direction is different for two separated telescopes \citep{Mozurkewich:1988lr} and due to the elongation of the stellar images by chromatic refraction across the bandpass.
\begin{figure}[h!]\begin{center}
\includegraphics[width = \linewidth]{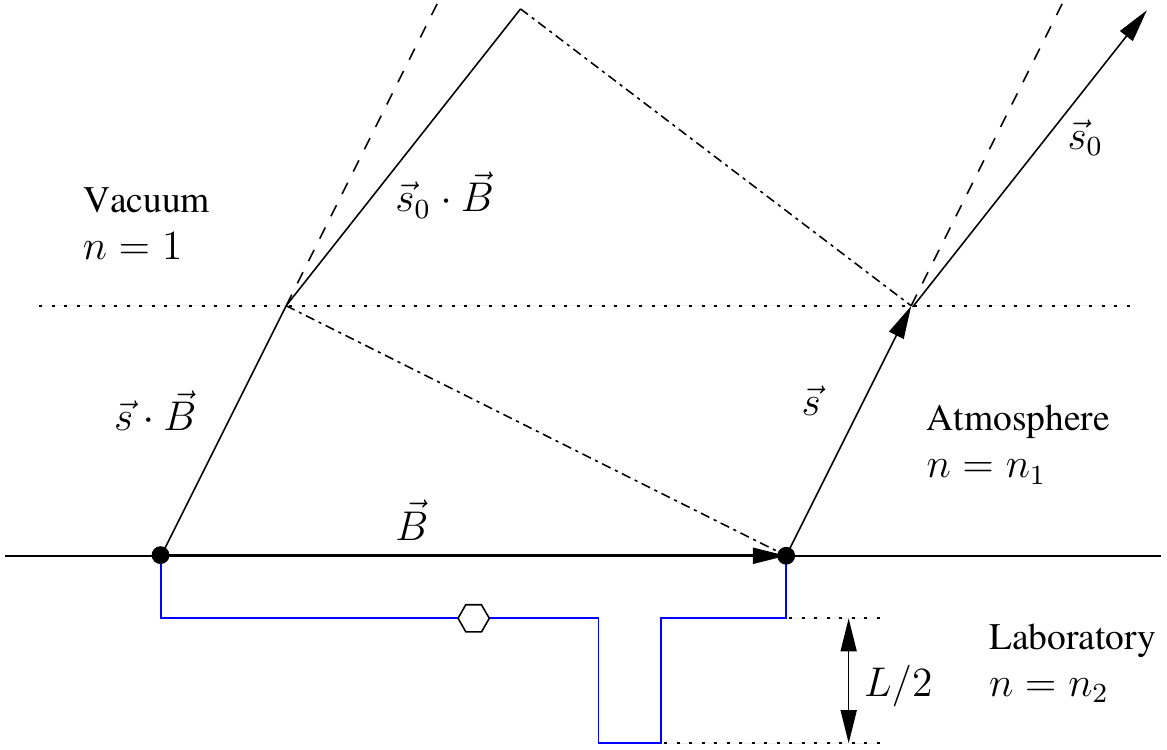}
\caption{Schematic of an interferometer with baseline $\vec B$ observing a source at true position $\vec s_0$ through a plan-parallel atmosphere of refractive index $n_1$.}
\label{fig:GPD}\end{center}\end{figure}

\item Dispersion effects due to chromatic refraction: When the external vacuum delay is compensated in air, the chromatic dependence of the index of refraction distorts the fringe packet and reduces fringe visibility \citep{Tango:1990rm}. Depending on how the astrometric measurement is performed, biases can occur.
\end{enumerate}
The first two effects are usually dominant and are discussed here. A couple of remarks are appropriate to underline the particularities of {\small PRIMA-VLTI}:
\begin{itemize}
  \item The vacuum wavelengths of all delay line metrology systems (Model \emph{Agilent} 5519-A) are identical $\lambda_{DL,vac} = 632.991354$ nm and are supposed to be stable within $2\cdot 10^{-8}$. However, the {\small VLTI} control system was implemented such that a modified wavelength $\lambda_{DL,mod} = 632.863000$ nm is applied instead, presumably to account for an 'average' refraction.
  \item The {\small PRIMA} metrology $\Delta L_B$ in the B feed ({\small PRIMETB}) has to be corrected for refraction in the infrared. It is measuring at $\sim$1.3$\,\mu$m and the accepted stellar bandpass is centred at $\sim$2.25$\,\mu$m.
  \item Conversely, the differential value $\Delta L$ does not need to be corrected, because the differential delay is introduced by the DDL that are under vacuum. The same applies to the delay value of the DDL internal metrology. This assumes that any effect on $\Delta L$ caused by gradients of temperature, pressure, and humidity between the two interferometer feeds outside of the DDL vacuum vessels can be neglected.
\end{itemize}
The refraction terms are calculated with the formulae of \cite{Birch:1993fk} and \cite{Mathar:2007zl} for visible and infrared wavelengths, respectively. We assumed dry air, i.e.\ neglected any dependence on humidity. Pressure and temperature readings were extracted from the file headers and we assumed that the pressure is global and given by the observatory monitoring system. There are considerable temperature gradients within the facility, which typically amount to $\sim$1 K along the tunnel and up to $\sim$4 K between ambient and inside the laboratory, depending on the season. Temperature variations during an observation are usually much smaller than the gradient. Refraction terms were computed for each file separately, but only the global average over the tunnel having four temperature sensors was used to estimate $n_{DL}$. For the terms applied to the {\small PRIMA} metrology and for the instrument bandpass, we used the global average obtained with nine temperature sensors between the outside and the laboratory, and we made the approximation $n_{\mathrm{FSU}}= n_{1,\mathrm{FSU}} \approx n_{2,\mathrm{FSU}}$. Table~\ref{tab:DLDDLwavelength} lists the calculated refractive index at the prominent wavelengths, e.g.\ $n_{DL}=1+2.025\cdot 10^{-4}$. Because the delay line system does not apply the vacuum wavelength, three corrections are necessary to convert the delivered metrology value $\Omega_\mathrm{DL}$ into the corrected value $\Omega'_\mathrm{DL}$
\begin{equation}\label{eq:disp3}
\Omega'_\mathrm{DL} =  \frac{\rho_{mod}}{ n_{DL}} \, \Omega_\mathrm{DL} \, n_{\mathrm{FSU}}= \frac{ \Omega_\mathrm{DL}}{n'}\,n_{\mathrm{FSU}},
\end{equation}
where $\rho_{mod} = \lambda_{DL,vac}/\lambda_{DL,mod} = 1+2.028\cdot 10^{-4}$. Similarly we get
\begin{equation}\label{eq:disp4}
\Delta L'_B =  \frac{\Delta L_B}{n_{P} }\,n_{\mathrm{FSU}} .
\end{equation} 
\begin{figure}
\begin{center}
\includegraphics[width = 0.8\linewidth,  trim = 0cm 0cm 0cm 0cm, clip=true]{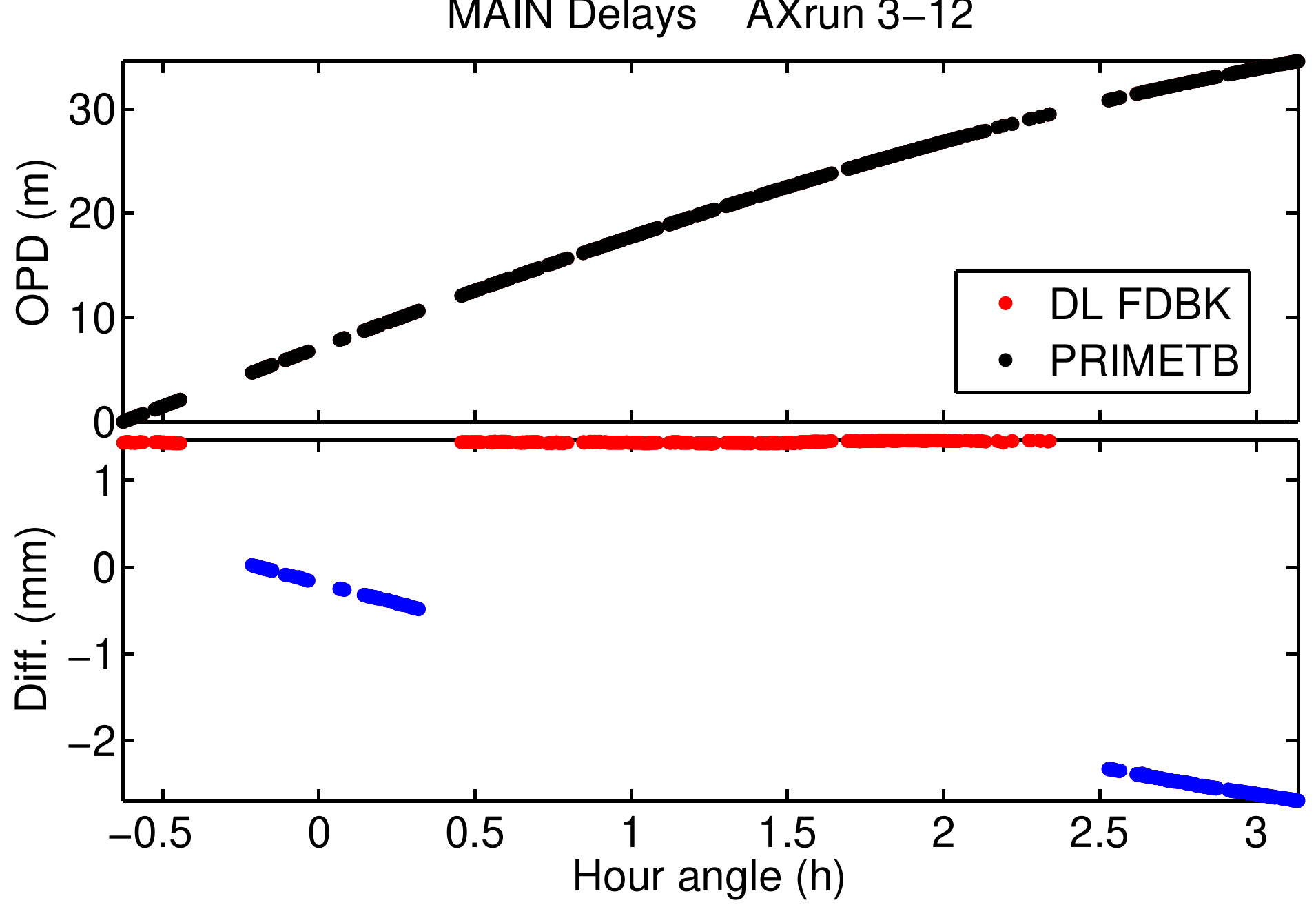}
\caption{Internal delay corrected for refraction measured with delay line and {\small PRIMA} metrologies (top) and their difference (bottom), which is shown in red and blue for normal and swapped mode data, respectively.}
\label{fig:DLrefrac}\end{center}
\end{figure}
\begin{figure}
\begin{center}
\includegraphics[width = 0.8\linewidth,  trim = 0cm 0cm 0cm 0cm, clip=true]{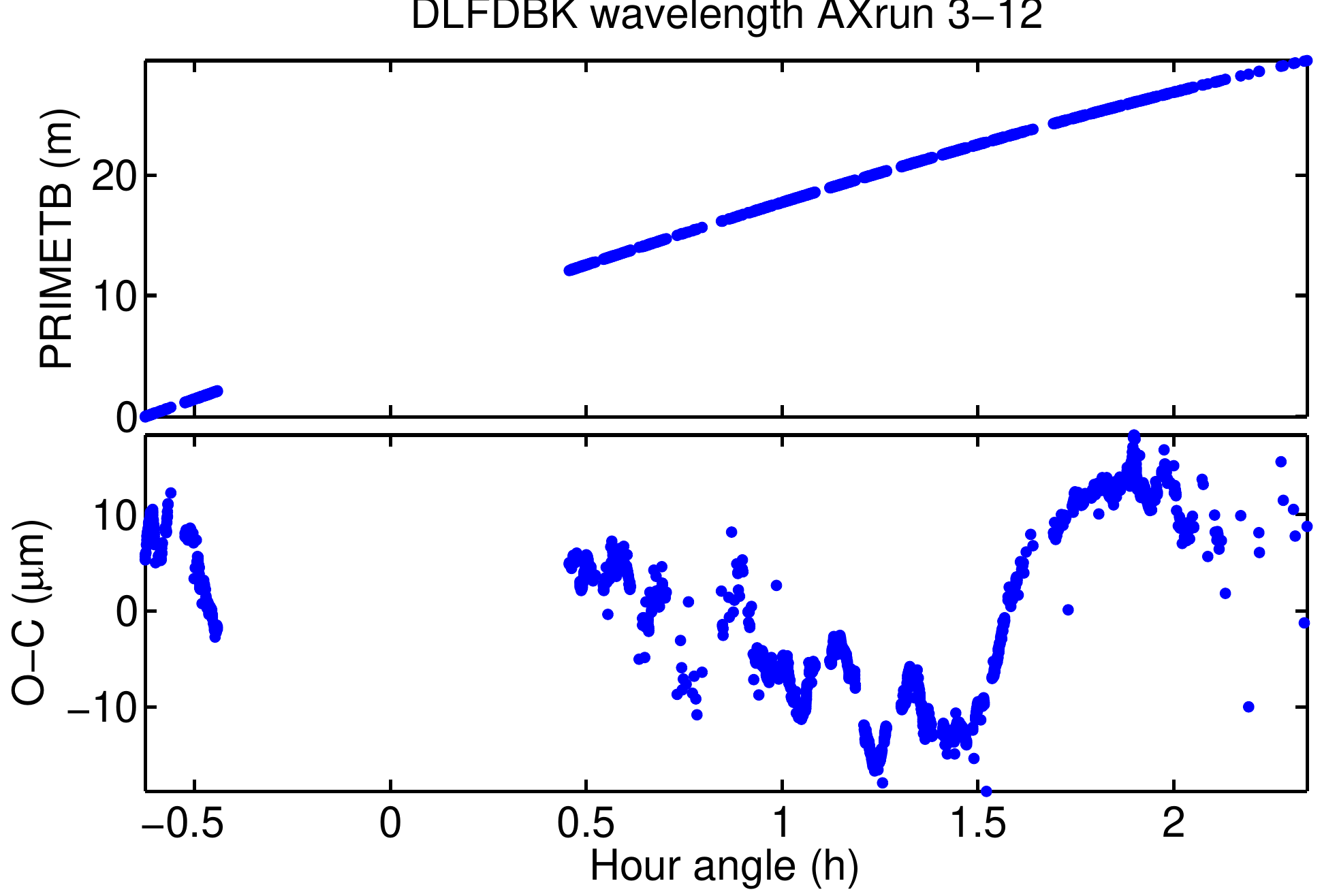}
\caption{{\small PRIMETB} internal delay (top) and the residuals after fitting the delay line delays to it (bottom) using the model Eq.~\ref{eq:DLrefraction}.}
\label{fig:DLrefrac2}\end{center}
\end{figure}
\noindent
Figures~\ref{fig:DLrefrac} and \ref{fig:DLrefrac2} show both corrected measurements and their difference, which is very small in normal mode. The variable offset in swapped mode appears because {\small PRIMETB} sees the differential delay introduced by the DDL in swapped mode. To quantify the agreement between both measurements in normal mode, we determined an average mismatch factor $n_\mathrm{corr}$ by adjusting the linear model
\begin{equation}\label{eq:DLrefraction}
{\Delta L'_B}= \gamma_\mathrm{corr}+  n_\mathrm{corr} \,\Omega'_\mathrm{DL},
\end{equation}
where $\gamma_{corr}$ is an offset. The value of $n_\mathrm{corr}-1$ is smaller than $1\cdot10^{-6}$ and comparable to the uncertainties in the applied refraction indices of Table~\ref{tab:DLDDLwavelength}, thus validating our approach. 

\subsection{Determination of the wide-angle baseline}\label{sec:opdmodeldet}
The relationship Eq.~\ref{eq:wab01} shows that the interferometric delay depends on both the target's position in the sky and on the wide-angle baseline vector $\vec B$. An astrometric measurement on the basis of a measured delay is thus possible if the baseline is known. In most cases, a simultaneous adjustment of both target coordinates and baseline is prohibited by the degeneracy of the model function. Consequently, the strategy consists of determining the baseline from a dedicated set of observations and to assume that the baseline calibration remains valid for the astrometric observations obtained thereafter. The principle of the baseline calibration is to observe a set of stars with accurately known coordinates and distributed over a preferably large sky-area. For each star, the internal delay corresponding to the fringe position is recorded. Thus, the corresponding system of equations~\ref{eq:wab01} can be solved for four unknowns: a global zero-point and the three components of the baseline vector. Currently, the only VLTI system capable of precisely measuring the internal delay is the delay line metrology, because the {\small PRIMETB} zero-point is not tied to a physical reference\footnote{The zero-point of the main delay line metrology is defined for each night by a mechanical reference in the tunnel. A similar procedure is imaginable for the PRIMA metrology in the future.}. 
\begin{table}
\caption{Targets used for baseline calibration}
\label{tab:ALL}  \centering  
\begin{tabular}{r r r r } 	
\hline\hline %
Nr. & Target & RA & DEC\\ 
 &  & (h:m:s) & (d:m:s)\\ 
\hline 
1 & \object{HIP 14146} & 03 02 23.50 & -23 37 28.10\\ 
2 & \object{HIP 20384} & 04 21 53.33 & -63 23 11.01\\ 
3 & \object{HIP 22449} & 04 49 50.41 & +06 57 40.60\\ 
4 & \object{HIP 24659} & 05 17 29.09 & -34 53 42.74\\ 
5 & \object{HIP 34088} & 07 04 06.53 & +20 34 13.08\\ 
6 & \object{HIP 36795} & 07 34 03.18 & -22 17 45.84\\ 
7 & \object{HIP 50799} & 10 22 19.58 & -41 38 59.86\\ 
8 & \object{HIP 50954} & 10 24 23.71 & -74 01 53.80\\ 
9 & \object{HIP 56647} & 11 36 56.93 & -00 49 25.49\\ 
\hline 
\end{tabular}
\end{table}

\begin{figure}	
\begin{center}
\includegraphics[width=0.5\linewidth]{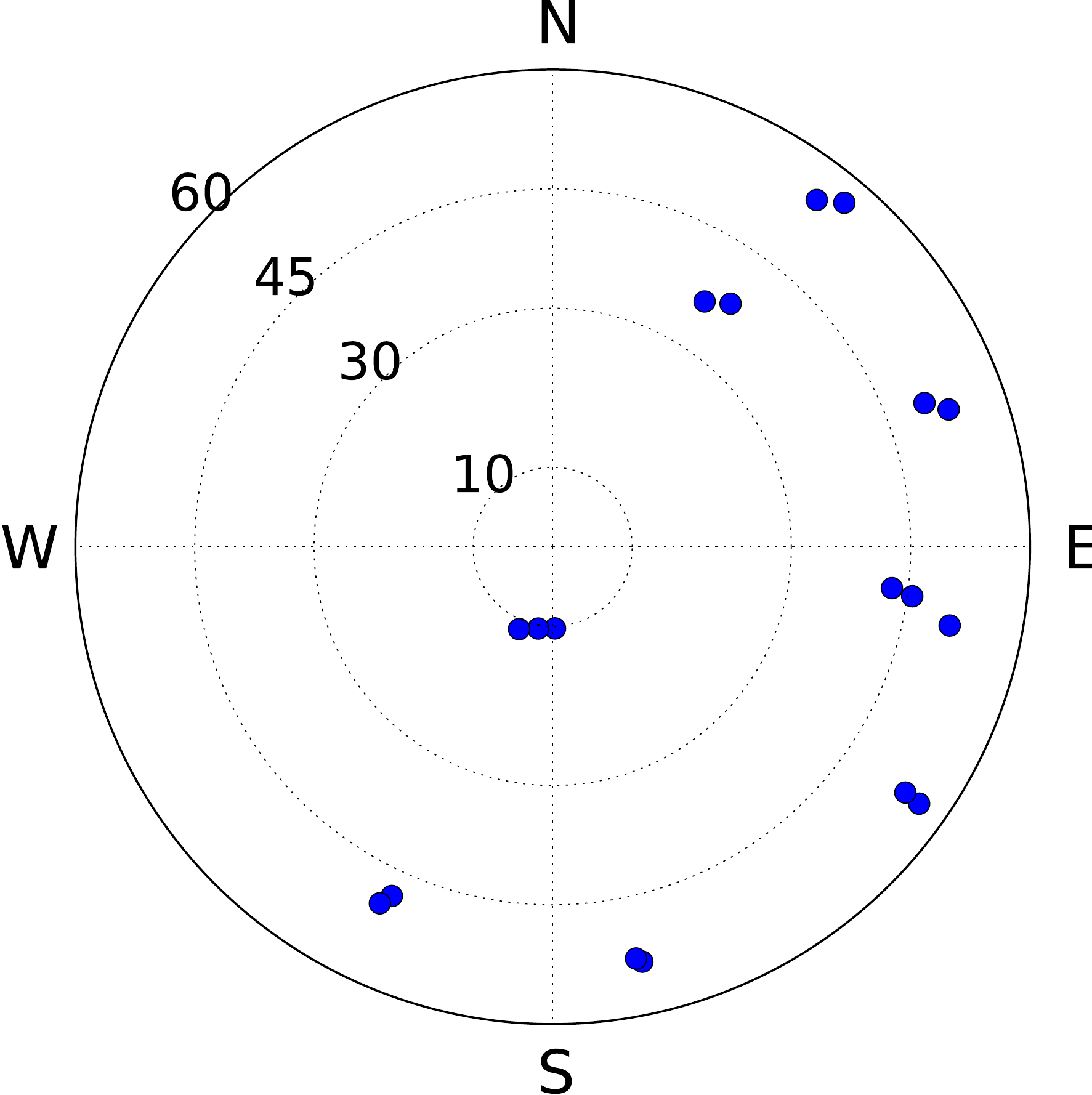}
\caption{Orientation and zenith angle of the nine observed stars.}\label{fig:BCAL}
\end{center}
\end{figure}
\begin{figure}	
\begin{center}
\includegraphics[width=0.9\linewidth,  trim = 0cm 0cm 0cm 0cm, clip=true]{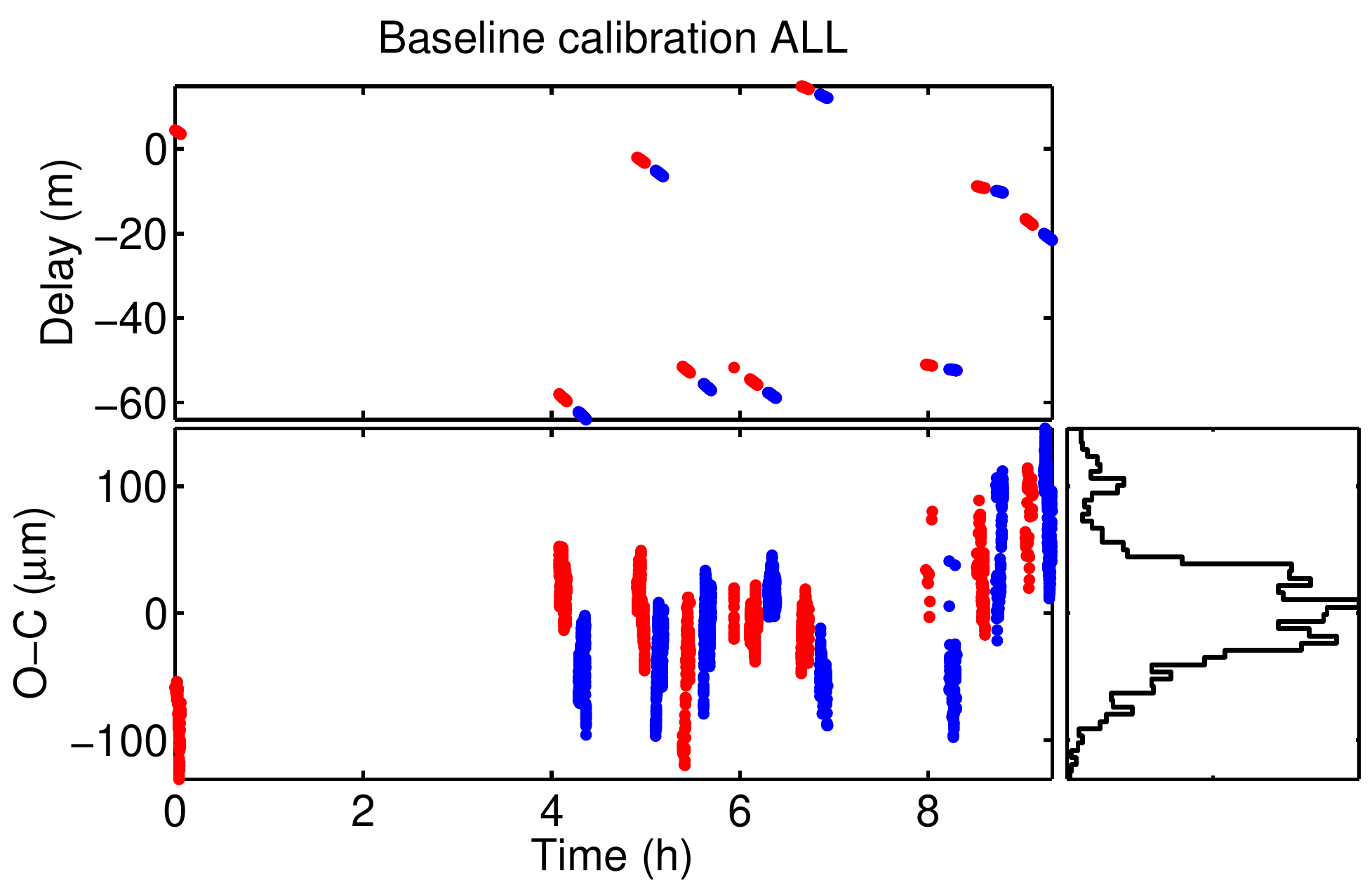}
\caption{\emph{Top}: The measured delays, where normal mode data is shown in red, swapped mode data is shown in blue. \emph{Bottom}: Residuals of the baseline fit as a function of time and their distribution. The RMS is 46 $\mu$m.}\label{fig:BCALresiduals}
\end{center}
\end{figure}

\begin{figure}	
\begin{center}
\includegraphics[width=0.7\linewidth]{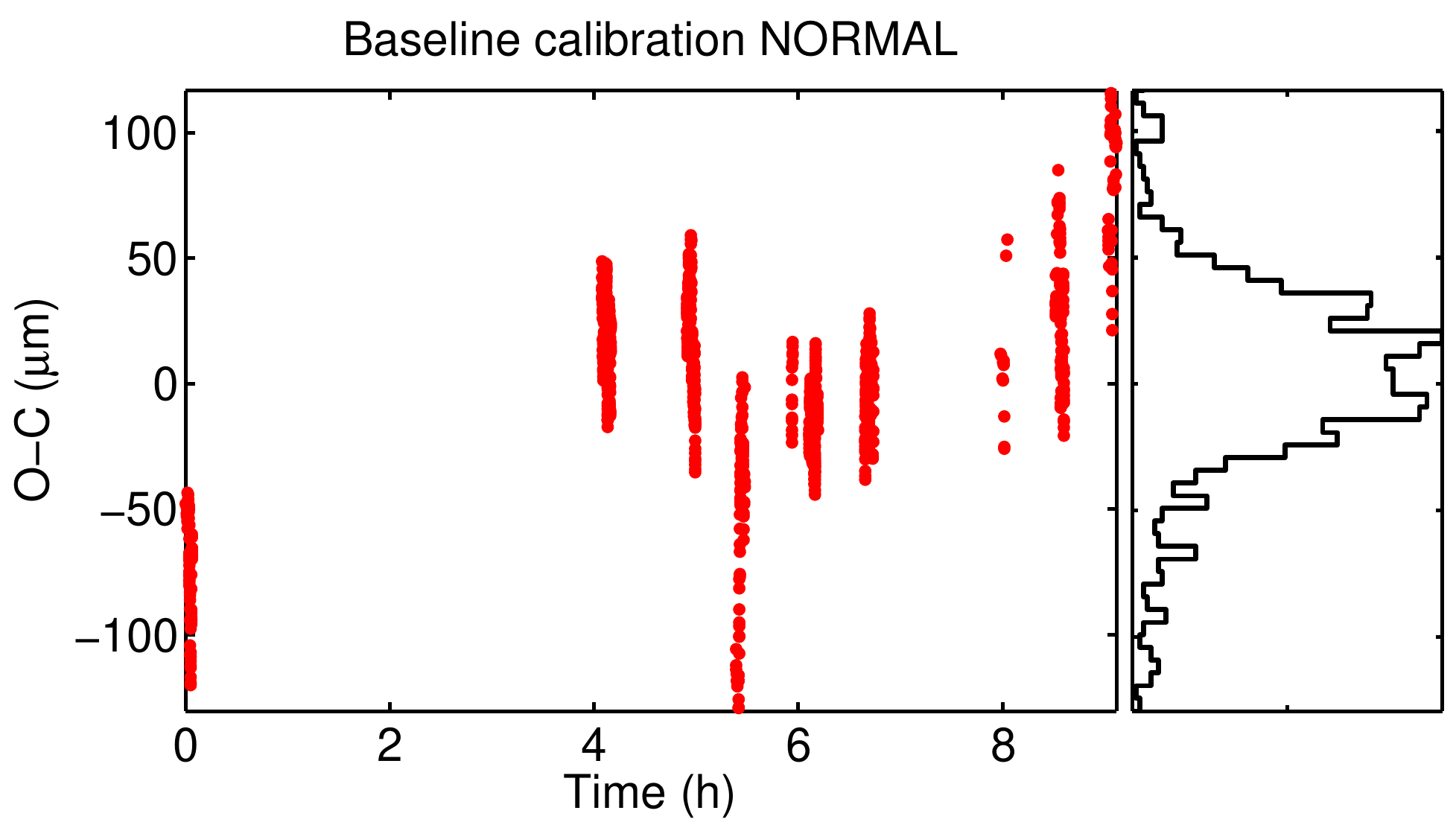}
\includegraphics[width=0.7\linewidth]{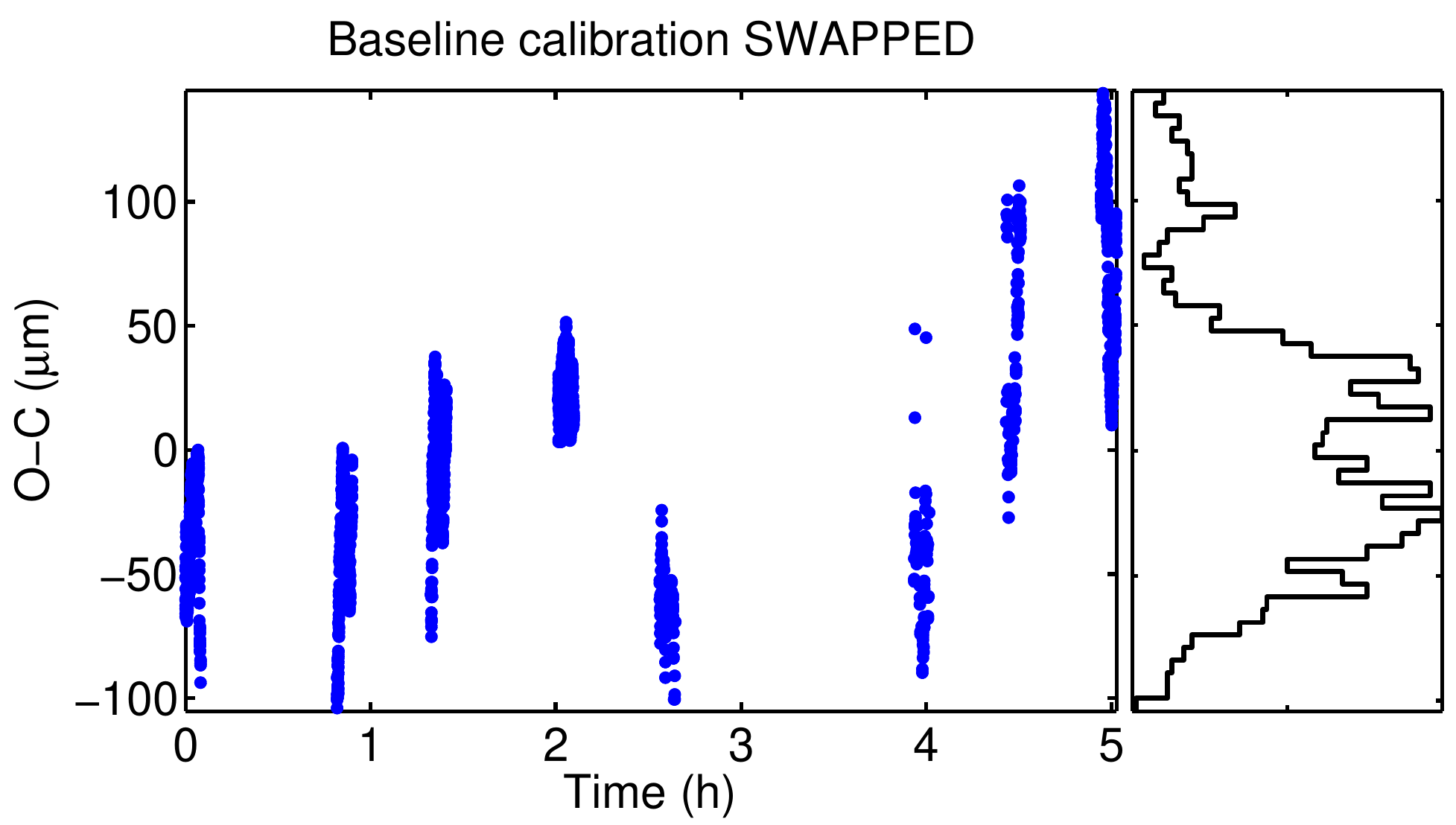}
\caption{\emph{Top}: Residuals of the baseline fit to normal mode data only. The RMS is 38 $\mu$m. \emph{Bottom}: The same for swapped mode only. The RMS is 50 $\mu$m.}\label{fig:residuals2}
\end{center}
\end{figure}

\noindent 
During the night of November 21, 2011, the observations for a baseline model were collected. The stars were chosen from the FK6 catalogue \citep{Wielen:1999lr} based on their $K$-magnitude, observability, and sky distribution, which is constrained by the delay line limits as shown in Fig.~\ref{fig:vlti}. Since these stars usually are single, a fake secondary star at +8\arcsec~in right ascension was considered for each observation. The telescopes thus pointed at 4\arcsec~from the target. To investigate the baseline in normal and swapped mode, each target was observed in both modes. Note that the measurements in normal mode yield the wide-angle baseline $\vec B_\mathrm{B,Wide}$ of the FSUB feed, whereas the swapped mode data determine a modified wide-angle baseline $\vec B'_\mathrm{A,Wide}$ of the FSUA feed, because the derotator is not in the nominal position. Nine stars were observed and are listed in Table~\ref{tab:ALL}. The accurate instantaneous coordinates were computed accounting for proper motion correction,  precession, and nutation, using the parameters of the FK6 catalogue.\\
Three different datasets were modelled. In the first case all data in both modes were considered, whereas in the second and third case only data in normal or swapped mode were used, respectively. When doing the combined model, the model function Eq.~\ref{eq:wab01} has to account for an additional parameter which is the internal DOPD $\Delta_\mathrm{int}$ which offsets the delay between normal and swapped states (see Sect.~\ref{sec:MDLfit}):
  \begin{equation}\label{eq:wab05}
\frac{\Omega_\mathrm{DL}}{ n' } \,n_{\mathrm{FSU}} = \vec B_\mathrm{Wide} \cdot \vec s_0 + B_0 + \mathcal{H}_S \,\, \Delta_{int}, 
\end{equation} 
where $\Omega_\mathrm{DL}$ is the delay line metrology measurement, $n'$ and $n_{\mathrm{FSU}}$ are the wavelength-dependent refraction terms introduced in the previous section, $\vec B_\mathrm{Wide}$ in the 'average' wide-angle baseline of both normal and swapped configuration, $B_0$ is a constant internal delay, and $\mathcal{H}_S$ is the Heaviside-type function
\begin{equation}\label{eq:HS}
 \mathcal{H}_S = \left\{ \begin{array}{rcr}
 0 & \mathrm{in} & \mathrm{normal~mode}\\
 1 & \mathrm{in} & \mathrm{swapped~mode}
 \end{array}\right.
\end{equation}
for the swapped mode. Thus the model for the baseline fit of all data combined has 5 parameters: $B_X$, $B_Y$, $B_Z$, $B_0$, and $\Delta_\mathrm{int}$, where the offsets $B_0$ and $\Delta_\mathrm{int}$ are assumed constant over the night. For the actual data, the refraction terms $n'-1$ and $n_{\mathrm{FSU}}-1$ were of the order of $(1 \pm0.4) \cdot 10^{-6}$ and $(1.997\pm0.007) \cdot 10^{-4}$, respectively. The result of the global fit to all data is shown in Fig.~\ref{fig:BCALresiduals}. Over the 9-hour timespan of the observations, the atmosphere appears to introduce an OPD variation of $\sim 250\,\mu$m peak-to-valley. The residual's RMS after the 5-parameter fit is 46~$\mu$m. At first sight, no obvious systematics or correlations with elevation, azimuth angle, hour angle, or stellar coordinates can be identified. It rather appears like the apparent structure is caused by slowly varying atmospheric parameters with the effect that the residual's distribution does not resemble a Gaussian curve. Normal and swapped data were also separately adjusted with a 4-parameter model. The resulting graphs are shown in Fig.~\ref{fig:residuals2} and reveal that the fit quality is comparably poor in both cases, presumably due to atmospheric turbulence.
\begin{table}[h!]
\caption{Baseline fit results. Numbers in brackets indicate the Monte Carlo derived errors in the last significant digit (here $\mu$m).}
\label{tab:ALLresults}  \centering  
\small
\begin{tabular}{l r@{\,}l   r@{\,}l  r@{\,} l}
\hline\hline %
 & \multicolumn{2}{r}{ALL} & \multicolumn{2}{r}{NORMAL} & \multicolumn{2}{r}{SWAPPED}\\ 
 & \multicolumn{2}{r}{(m)} &  \multicolumn{2}{r}{(m)}          & \multicolumn{2}{r}{(m)}\\ 
\hline 
 & \multicolumn{2}{c}{$\vec B_\mathrm{Wide}$}  & \multicolumn{2}{c}{$\vec B_\mathrm{B,Wide}$} & \multicolumn{2}{c}{$\vec B_\mathrm{A,Wide}'$} \\[1pt] 
$B_0$ & $-60.920729$ & (59) &  $-60.920707$ & (59) & $-60.920536$ & (98) \\  
$B_x$ & $76.387823$ & (42) &  $76.387792$ & (35) & $76.387839$ & (70) \\ 
$B_y$ & $-49.877600$ & (24) &  $-49.877595$ & (12) & $-49.877606$ & (52) \\ 
$B_z$ & $-0.017542$ & (62) &  $-0.017527$ & (62) & $-0.017533$ & (105) \\ 
$\Delta_\mathrm{int}$ & $0.000193 $&(15) &   \ldots & & \ldots &  \\
RMS & $0.000045$    &         & $0.000038$&           & $0.000050$& \\ 
\hline
\end{tabular}
\end{table}

\noindent
Tables~\ref{tab:ALLresults} and \ref{tab:ALLdiff} summarise the results. Formal error bars were obtained from Monte Carlo simulations and some parameters are strongly correlated (e.g. $B_0$ and $B_z$). The internal DOPD appears here in the difference between $B_0$ in normal and swapped. We conclude that the differences between the average, the normal, and the swapped baseline are not significant and that the formal uncertainties are compatible with the requirement in Sect.~\ref{sec:designgoal}. We will use the baseline $\vec B_\mathrm{Wide}$ obtained from the combined data in all of the following analysis, i.e.\ as wide-angle baseline and as approximation of the narrow-angle baseline, which is only valid as zeroth order approximation.
\begin{table}[h!]
\caption{Baseline fit results: differences between normal and swapped mode data.}
\label{tab:ALLdiff}  \centering  
\begin{tabular}{l r@{\,$\pm$\,}l   r@{\,$\pm$\,}l  r@{\,$\pm$\,} l}
\hline\hline %
 & \multicolumn{2}{c}{NORM -- SWAP} & \multicolumn{2}{c}{ALL -- NORM} & \multicolumn{2}{c}{ALL -- SWAP}\\ 
 & \multicolumn{2}{c}{($\mu$m)} &  \multicolumn{2}{c}{($\mu$m)}          & \multicolumn{2}{c}{($\mu$m)}\\ 
\hline 
$B_0$ & $-173$ & 115 & $-22$ & 85& $-195$& 116\\  
$B_x$ & $-47$ & 78 & $ 30$ & 55& $-16$& 82\\ 
$B_y$ & $ 11$ & 53 & $ -5$ & 27& $  6$& 57\\ 
$B_z$ & $  4$ & 121 & $-13$ & 88& $ -8$& 123\\ 
\hline
\end{tabular}
\end{table}

\subsection{Fitting the main delay line feedback}\label{sec:DLfitting} 
\begin{figure}\begin{center}
\includegraphics[width = 0.8\linewidth,  trim = 0cm 0cm 0cm 0cm, clip=true]{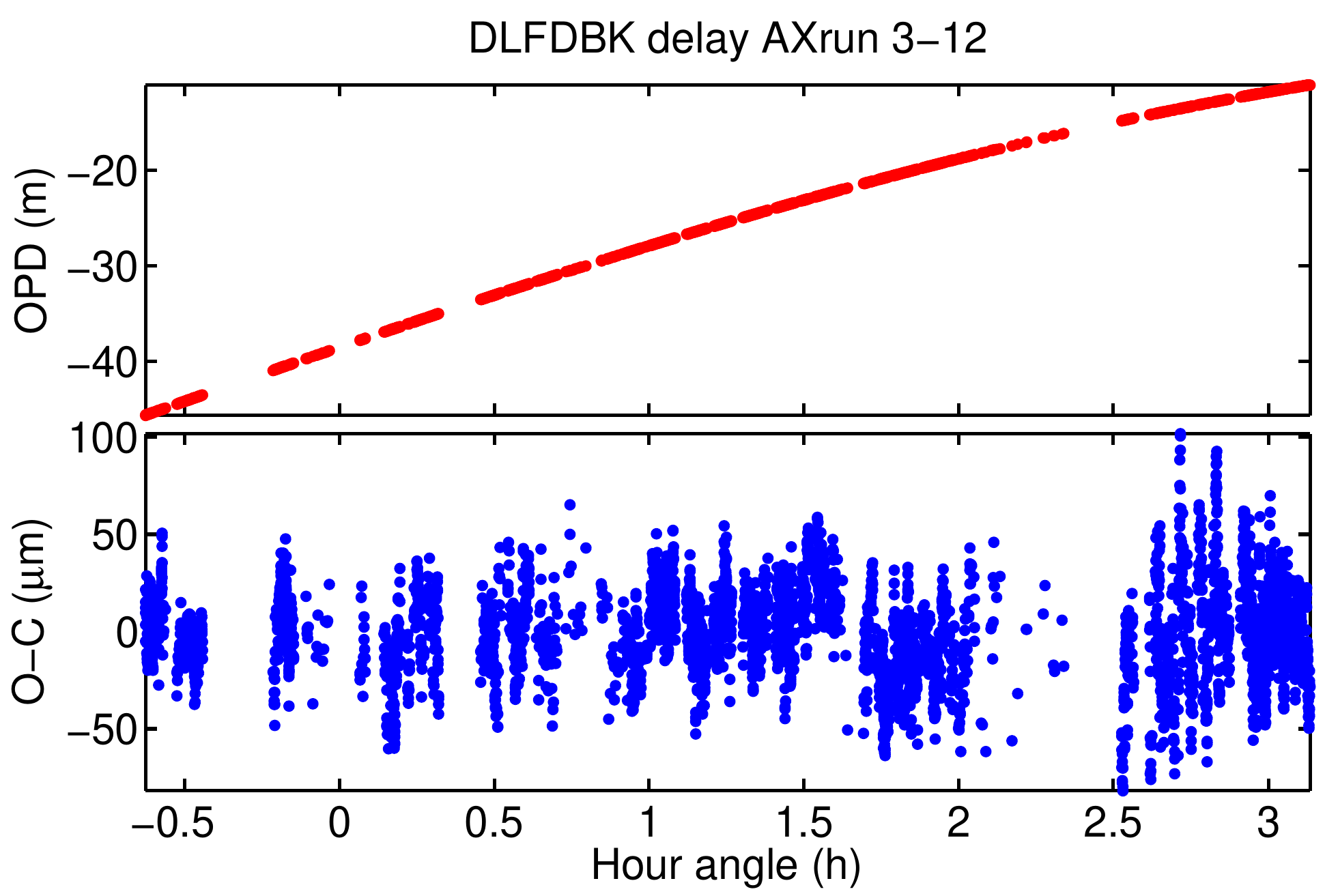}
\caption{\emph{Top}: Main delay measured with the DL metrology during the demonstration run. \emph{Bottom}: Residuals after adjusting the model Eq.~\ref{eq:DLmodel} including offsets to the primary coordinates. The atmospheric perturbations are visible and have a total RMS of $24.0\,\mu$m.}\label{fig:MDLfit}\end{center}
\end{figure}

Assuming the baseline is known, the main delay $\Omega_{DL}$ measured with the main delay line internal metrology during an astrometric observation can be used to determine the position of the primary star\footnote{We neglect any effect related to the off-axis pointing by $\rho/2$ of the telescope.}. The corresponding model has four free parameters: an offset $\gamma_{DL}$, the primary star coordinates in RA $\alpha_P$ and DEC $\delta_P$, and an internal DOPD $\Delta_{int}$. The internal DOPD corresponds to the difference of zero-OPD positions of FSUA and FSUB, which in principle is arbitrary, because the VLTI does not provide the instruments with a common OPD reference. We have to take it into account, because we model combined normal and swapped observations and $\Delta_{int}$ is added to the DL position in one mode but not in the other. We assume $\Delta_{int}$ is constant in time and arbitrarily assigned it to the swapped observation. 
\begin{equation}\label{eq:DLmodel}
\Omega'_{DL} = \gamma_{DL} + w(\alpha_P, \delta_P) + \mathcal{H}_S \,\, \Delta_{int}, 
\end{equation} 
where $w$ models the OPD (Eq.~\ref{eq:OPDform}). The baseline is assumed to be $\vec B_\mathrm{Wide}$ determined in Sect.~\ref{sec:opdmodeldet}. The fit result is shown in Fig.~\ref{fig:MDLfit} and summarised in Table~\ref{tab:maindelyfits}. The table lists coordinate offsets $\Delta \alpha_P$ and $\Delta \delta_P$ instead of absolute coordinates. These offsets are computed relative to the instantaneous target position, after proper motion and precession correction. The fit residuals show the atmospheric piston effect, which introduces variations of typically $\pm50\,\mu$m with an RMS of $24.0\,\mu$m. The larger variations of $\pm90\,\mu$m at large hour angle $>2.5$\,h correspond to observations at high airmass of $1.35-1.45$. Atmospheric conditions were stable with an average coherence time of $\tau_0\!\sim\!1.8$\,ms and seeing of $\sim$1.2\arcsec, as reported by the observatory monitoring system measuring in the optical.

\begin{figure}\begin{center}
\includegraphics[width = 0.8\linewidth,  trim = 0cm 0cm 0cm 0cm, clip=true]{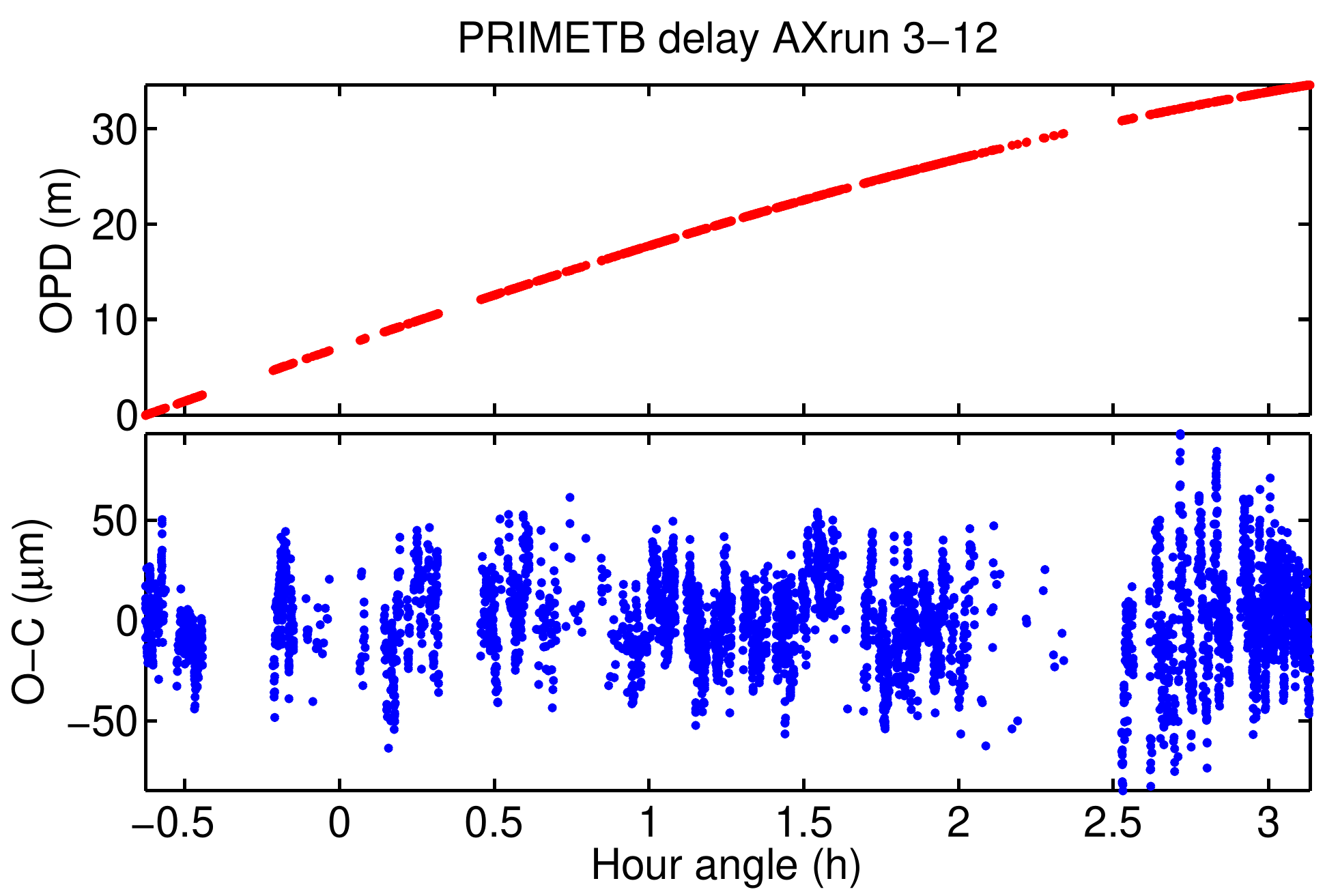}
\caption{\emph{Top}:  Change of {\small PRIMETB} during the observation for Run 9. \emph{Bottom}: Residuals after adjusting the model Eq.~\ref{eq:PRIMETBmodel}. The residual RMS is $23.1\,\mu$m.}\label{fig:MDLfit2}\end{center}
\end{figure}
\begin{table}
\caption{Results of main delay fitting}
\label{tab:maindelyfits}  \centering  
\small
\begin{tabular}{l r c c} 	
\hline\hline %
Parameter & Unit & DL feedback & {\small PRIMETB} \\
 \hline
$\gamma_\mathrm{DL}$ & (m)&$-12.36546 \pm 0.00002$& \ldots\\
$\gamma_\mathrm{PB}$ & (m)&\ldots& $32.27081\pm 0.00002$\\
$\Delta \alpha_P$ & (\arcsec) & $-0.42 \pm 0.05$ & $-1.25 \pm 0.09$\\
$\Delta \delta_P$ & (\arcsec) & $-0.40 \pm 0.02$& $-0.20 \pm 0.03$\\
$\Delta_{int}$ & (mm) &$-0.123 \pm 0.001$ &\ldots\\
$\varrho$ & (\arcsec) & \ldots & $11.445 \pm 0.013$\\
$\Phi$ & (deg) & \ldots & $188.00 \pm 0.04$\\
\hline 
\end{tabular} 
\end{table}

\subsection{Fitting the PRIMETB main delay}
The {\small PRIMETB} delay $\Delta L_B$ is modelled similarly to the main delay line OPD, but the model function is slightly different. The internal DOPD $\Delta_{int}$ does not have to be modelled, because it is transparent to {\small PRIMETB}, see Fig.~\ref{fig:labsketch}: in normal mode, $\Delta_{int}$ is compensated by the differential delay actuator DDL1, which is not seen by {\small PRIMETB}. In swapped mode, $\Delta_{int}$ is introduced by both the tracking DL (seen by {\small PRIMETB}) and by the differential delay actuator DDL2 (also seen by {\small PRIMETB}) and the net effect on {\small PRIMETB} is therefore zero. However, {\small PRIMETB} measures the differential delay introduced by DDL2 in swapped mode to compensate for the external differential delay. Thus, we can use {\small PRIMETB} to obtain a first estimate of the pair separation $\varrho$ and position angle $\Phi$. The complete model function is 
\begin{equation}\label{eq:PRIMETBmodel}
\Delta L'_B = \gamma_{PB} + w(\alpha_P, \delta_P) + \mathcal{H}_S \,\, \Delta w(\alpha_P, \delta_P, \varrho,\Phi), 
\end{equation} 
where $\gamma_{PB}$ is an offset and $\Delta w$ models the differential delay (Eq.~\ref{eq:dopdmodel}). The results are summarised in Table~\ref{tab:maindelyfits} and Fig.~\ref{fig:MDLfit2} shows the fit residuals. The best fit coordinate offsets agree with the results of the main delay line fit in the previous section within 0.8\arcsec~in RA and within 0.2\arcsec~in DEC. The separation fit yields first realistic estimates of $\varrho$ and $\Phi$.

\subsubsection{Comparison of the residuals when fitting PRIMETB and main delay}
\begin{figure}[h!]\begin{center}
\includegraphics[width = 0.8\linewidth,  trim = 0cm 0cm 0cm 0cm, clip=true]{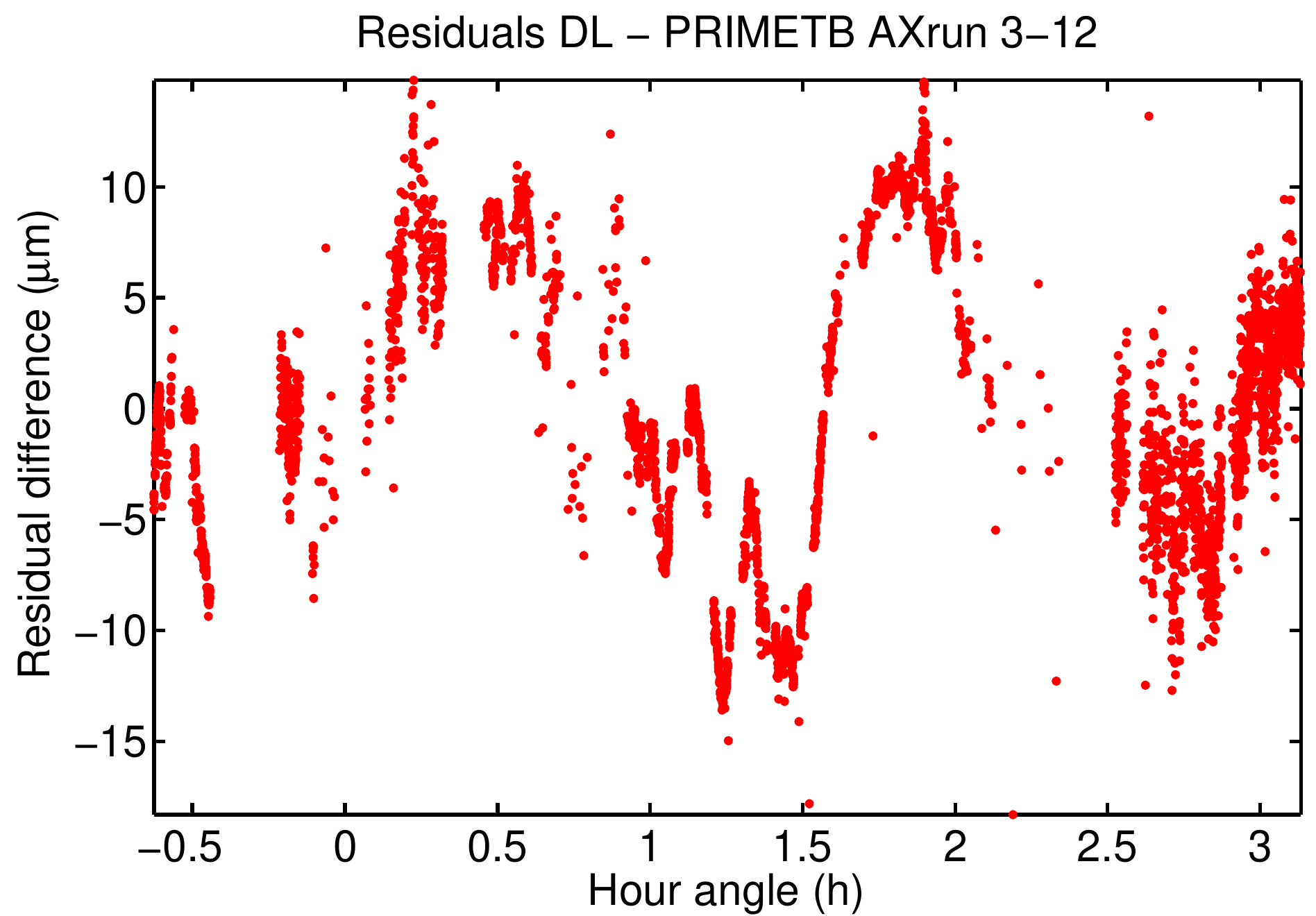}
\caption{Difference between the residuals measured in Fig.~\ref{fig:MDLfit}. The residual RMS is $4.2\,\mu$m.}\label{fig:resdiff}\end{center}
\end{figure}

Ideally, the residuals shown in Fig.~\ref{fig:MDLfit} and \ref{fig:MDLfit2} should be identical, but in practice we expect differences because the monitored optical path is not the same: {\small PRIMETB} propagates in the central obscuration of the stellar beams from the beam combiner to the retroreflector at the telescope, whereas the delay line metrology measures only within the DL tunnel. Additionally, the laser wavelengths are different with {\small PRIMETB} in the near infrared ($\sim1.3\,\mu$m) and the delay line in the visible ($\sim0.6\,\mu$m). Figure~\ref{fig:resdiff} shows the residual difference and reveals variations of the order of 20~$\mu$m PTV with an RMS of 6.4~$\mu$m. This residual difference could be caused by the atmospheric turbulence present in the beam-path not common to the two metrology systems, i.e. in the beam combination laboratory or between the delay line tunnel and the telescopes. Fluctuations in temperature, pressure, and humidity during the observation and the associated chromatic variations of the refractive index of air can also contribute to the observed turbulence.

\subsection{Astrometry fit of a separation vector}
The astrometric observable $AX_{obs}$ is a linear combination of the differential delay measured with the {\small PRIMA} metrology and the two fringe positions of primary and secondary object measured with the two fringe sensors. The model function for the relative astrometry fit is
\begin{equation}\label{eq:axfit}
\begin{split}
AX_{obs} &= \Delta L - GD_A - GD_B\\
&= c + \mathcal{H}_N \,\, \Delta w(\varrho,\Phi, \vec B_\mathrm{AX}) + \mathcal{H}_S \,\, \Delta w(\varrho,\Phi, -\vec B_\mathrm{AX}),
\end{split}
\end{equation}
where $\vec B_\mathrm{AX}$ is the narrow-angle astrometric baseline vector and $\mathcal{H}_N$ is the Heaviside-type function
\begin{equation}\label{heavisideN}
 \mathcal{H}_N = \left\{ \begin{array}{rcr}
 1 & \mathrm{in} & \mathrm{normal~mode}\\
 0 & \mathrm{in} & \mathrm{swapped~mode}
 \end{array}\right.
\end{equation} 
for the normal mode. The constant $c$ is the {\small PRIMA} metrology zero-point, which depends on the interferometer configuration at the last metrology fringe counter reset. In this description, the swapped observation corresponds to an observation with inverted baseline orientation compared to the normal mode. We initially make the coarse approximation that $\vec B_\mathrm{AX} = \vec B_\mathrm{Wide}$, thus the three free parameters in the model Eq.~\ref{eq:axfit} are the metrology zero point $c$, the target separation $\varrho$, and the position angle $\Phi$ as defined in Fig.~\ref{fig:myPair}. 
\begin{table}
\caption{Astrometry fit results for the demonstration run on HD\,10360.}
\label{tab:axfits}  \centering  
\begin{tabular}{l r r} 	
\hline\hline %
Parameter & Unit &  Value \\
 \hline
$c$ & (mm) & $-1.34971 \pm 0.00002 $\\
$\varrho$ & (mas) & $11434.53 \pm 0.15$\\
$\Phi$ & (\degr) & $-172.1640 \pm 0.0007$\\
$\sigma_{AX}$& ($\mu$m) & 1.025\\
\hline
\end{tabular} 
\end{table}
\begin{figure*}\begin{center}
\sidecaption
\includegraphics[width = 12cm,  trim = 0cm 0cm 0cm 0cm, clip=true]{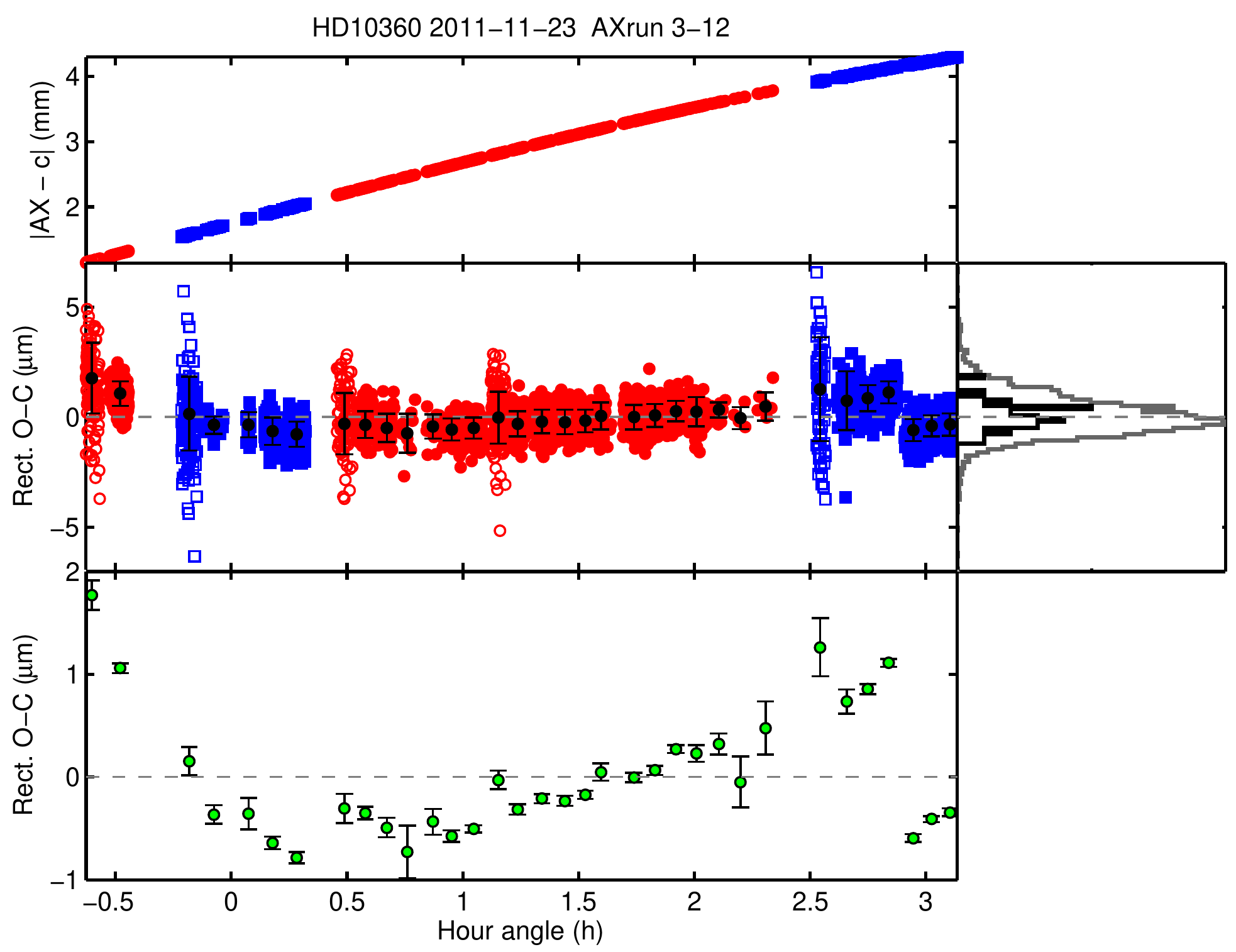}
\caption{\emph{Top}: Absolute value of the measured differential delay $|AX_{obs} - c|$ for the demonstration run as a function of hour angle. Normal mode data are shown in red and swapped mode data in blue. \emph{Middle}: Rectified residuals of the astrometric fit. Open and filled symbols indicate scanning and tracking data, respectively. File-average bins are shown in black. \emph{Bottom}: Rectified and binned residuals with error bars that rely on Gaussian statistics (error of the mean).} 
\label{fig:axres1}\end{center}\end{figure*}
Figure~\ref{fig:axres1} shows the result of the astrometry fit and the corresponding best-fit parameter values are listed in Table~\ref{tab:axfits}. The errors given in Table~\ref{tab:axfits} reflect the precision of the measurement, but not its accuracy. As is shown in Fig.~\ref{fig:axres1}, the model Eq.~\ref{eq:axfit} does not reproduce the data accurately, because the fit residuals exhibit strong systematics. The accuracy is therefore not limited by the measurement precision on the astrometric observable, but by a systematic effect that is not taken into account in the model. On this target with $\sim$11\arcsec~ separation, the RMS of the residuals is of the order of 1 $\mu$m and corresponds to $\sim$2~mas on the sky for a 100~m baseline.

\subsection{Auxiliary data of critical elements}
The anticipated astrometric accuracy sets tight requirements on the alignment and stability of the VLTI-{\small PRIMA} optical train. Whereas thermal drifts of passive mirrors are difficult to monitor, the information about the instantaneous configuration of controllable mirrors can be recorded during the observation for posterior inspection. Basic information about moving systems is included in the file header keywords, i.e.\ at relatively low cadence. For the commissioning, a low-level recording tool was devised to collect data at higher cadence and continuously. As an example, the motion of the derotator is discussed below, but a detailed presentation of other available data and their characteristics is given in \cite{Sahlmann2012PhD}.\\
Figure~\ref{fig:derot20} shows the position angle of the derotators in the star separator modules of AT3 and AT4 during the demonstration run. The angle $\omega_{rot}$ is the only degree of freedom of the derotator implemented with a relative encoder in the sense that $370\degr  = 10\degr$. Fast changes by 90\degr~correspond to swap/unswap motions, because the field rotates by twice the derotator angle. The derotator mechanical range is limited to 0-180\degr~and both derotators execute the swap procedure in equal fashion. In Fig.~\ref{fig:derot20} the derotator of AT3 reaches the software limit at an hour angle of $\sim$2.8~h and wraps by 180\degr.
\section{PRIMA astrometry precision and accuracy as of February 2012}\label{sec:prec}
Characterising the precision and accuracy of the {\small PRIMA} astrometric measurements is not straight-forward because of the large number of external effects to be considered and the resulting large parameter space to be explored. For the initial analysis, we apply the simplest three-parameter model and assume that the astrometric baseline is given by the wide-angle baseline determined in November 2011. This model is applied to all astrometric sequences collected so far with the aim of obtaining a general picture of the performances.  

\subsection{Amplitude and structure of the residuals}
\begin{figure}[h!]\begin{center}
\includegraphics[width = 0.75\linewidth]{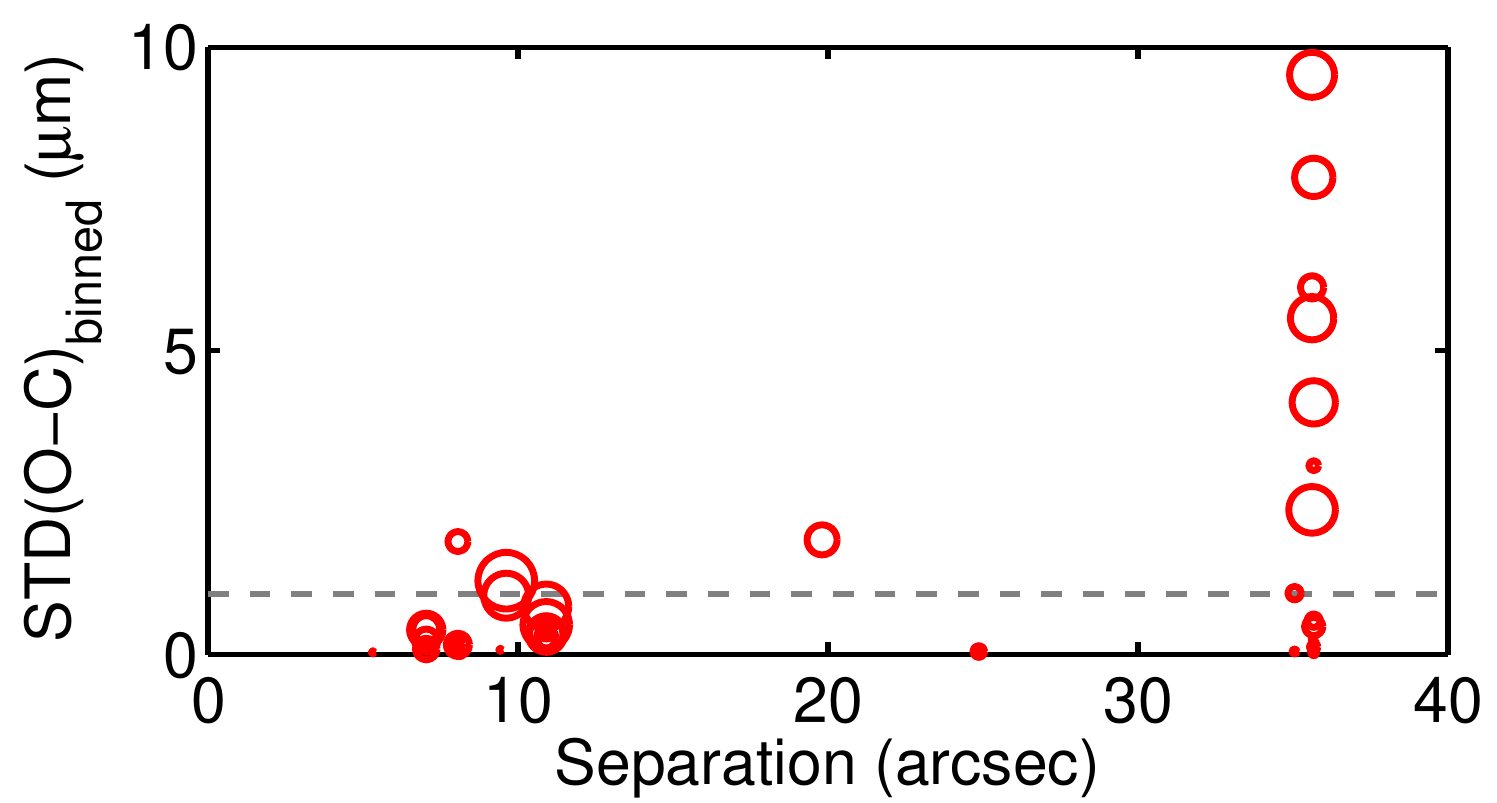}
\caption{Dispersion of the binned residuals for 48 astrometric runs obtained on 10 different target pairs as a function of target separation. The symbol size is proportional to the observation time span. The dashed line is at $1~\mu$m.}
\label{fig:axstatomc}\end{center}\end{figure}

Depending on the target separation, the time coverage of the observation sequence, and the amount of collected data, the amplitude and structure of the residuals about the best astrometry fit changes dramatically. For wide separation binaries the residual structure is orders of magnitude larger than the measurement noise and can reach a peak-to-valley amplitude of $\sim\!30~\mu$m. For small separation targets, the levels of systematic and measurement noise can become comparable and the residual's dispersion can fall below the $\mu$m level. Figure~\ref{fig:axstatomc} shows the residual amplitude for different targets and sequence lengths. To distinguish between measurement noise and systematic biases, the dispersion of the binned residuals are shown (one bin per file). We see that short sequences tend to have small residuals and the residual level at separations of $\sim\!10\arcsec$~is much lower than at $\sim\!35\arcsec$. The intermediate separation range is poorly covered with observations. One has to be cautious when using solely the residuals to characterise the quality of an observation. The fact that short sequences have smaller residuals does not necessarily mean that those are of better quality. Instead, it is the consequence of the fitting procedure which determines the best-fit parameters for the respective sequence. For instance, if the sequence consists of only one normal and one swapped observation, any systematic error between these will be absorbed by the metrology zero-point which is a free fit parameter, thus biasing the other free parameters separation and position angle.\\
Already when analysing the first light data, we found hints for measurement biases that seemed to correlate with the field rotation, which appear as a wavelike structure in the residuals. Since the field rotation is not seen by the {\small PRIMA} metrology which provides us with the astrometric observable, biases that are related to field rotation are not calibrated with the swap procedure but conversely are amplified and appear with opposite signs in normal and swapped mode. This led to the idea to 'rectify' the observed residuals to ease their visual interpretation, as is illustrated in Fig.~\ref{fig:rectified}. Whenever the residuals were rectified, this is indicated by the axis labelling. The dependency of the residual's amplitude on target separation is also illustrated in Fig.~\ref{fig:rectified2}, which shows that the wavelike pattern is similar for different objects with an amplitude that depends on the separation.

\subsection{Astrometric precision and the atmospheric limit}
\begin{figure}\begin{center}
\includegraphics[height = 0.46\linewidth,  trim = 1cm 4.4cm 2cm 0cm, clip=true]{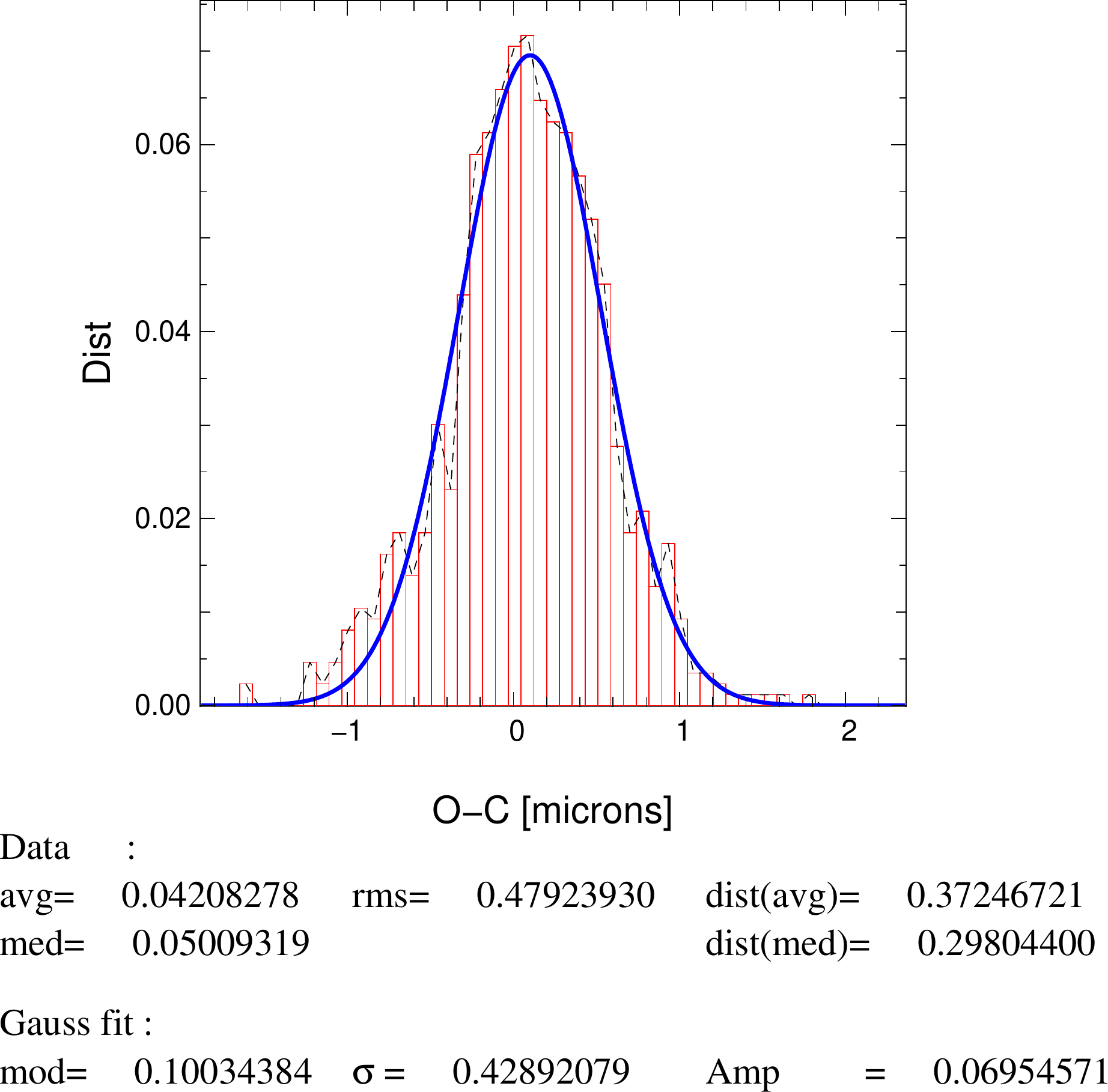}
\includegraphics[height = 0.47\linewidth,  trim = 0cm 0cm 0cm 0cm, clip=true]{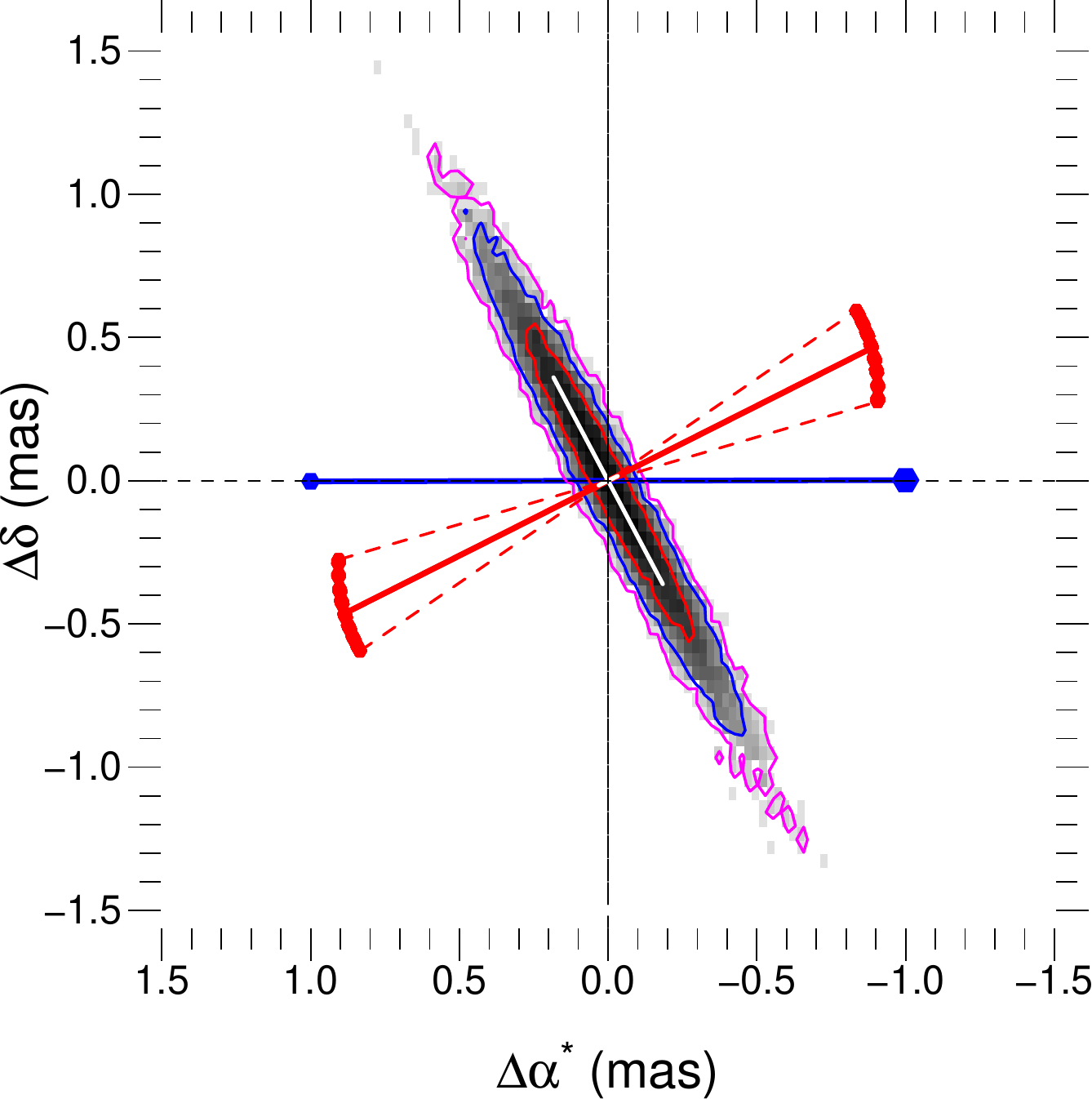}
\caption{\emph{Left:} Residual's histogram of the HD\,202730 observation shown in Fig.~\ref{fig:rectified2} and the best Gaussian fit to it. \emph{Right:} The two-dimensional distribution of the relative secondary's position in the sky for 10\,000 Monte Carlo simulations is shown in grey-shading. Contour lines indicate 1,2,3-$\sigma$ confidence intervals. The blue line indicates the position angle of the binary, the red circles show the equivalent $u$-$v$-coordinates of the observations, and the solid red line illustrates the average projected baseline orientation. }
\label{fig:precisionHD202730}\end{center}
\end{figure}

\begin{figure}
\begin{center}
\includegraphics[width = 0.9\linewidth]{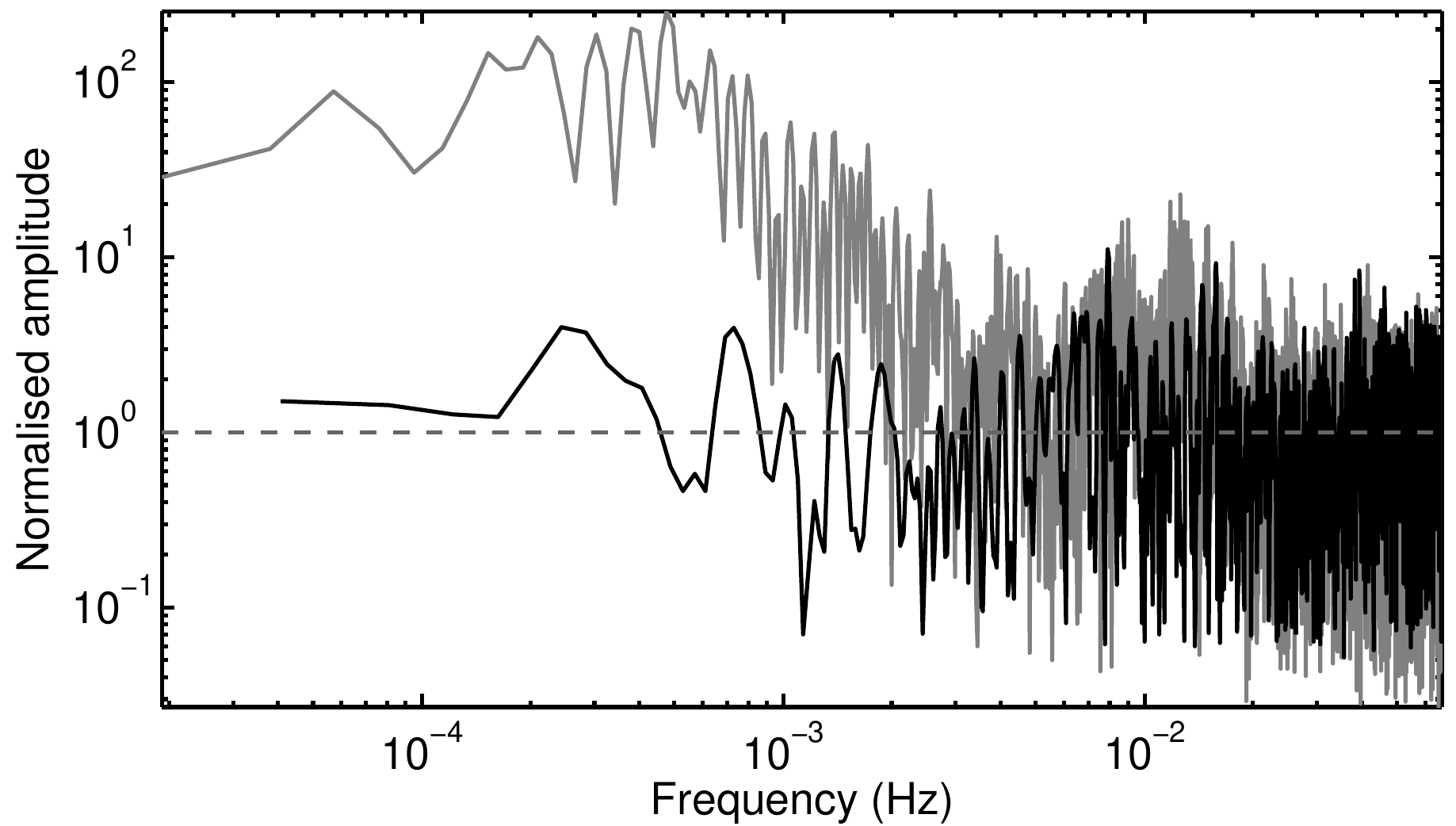}
\caption{Lomb Scargle periodogram \citep{Press:1986eu} of the 1-second average residuals for the observations of HD\,202730 (black curve, see Figs.~\ref{fig:precisionHD202730} and \ref{fig:rectified2}) and of HD\,10360 (grey curve) shown in Fig.~\ref{fig:axres1}.}
\label{fig:lomb}\end{center}\end{figure}

The residuals for the tight binary HD\,202730 shown in Fig.~\ref{fig:rectified2} yield an approximately normal distribution shown in Fig.~\ref{fig:precisionHD202730}. It is therefore reasonable to assume that the systematic errors for this specific observation are smaller than the measurement noise, thus hidden by it. 
Inspection of the residuals' periodogram in Fig.~\ref{fig:lomb} confirms that the noise of this observation is approximately white down to the lowest accessible frequency. This is in strong contrast to the observation of HD\,10360 shown in Fig.~\ref{fig:axres1}, where the systematic errors cause a large excess of low-frequency noise\footnote{For this comparison, only tracking data was considered and scanning data was discarded.}. In the case of HD\,202730, we can therefore assume Gaussian statistics and determine the formal errors on the fit parameters by Monte Carlo simulations. The distribution of the secondary star position is also displayed in Fig.~\ref{fig:precisionHD202730} and shows that the separation vector is best constrained along the projected baseline orientation, as expected. The observation time span was $\sim$1.6~h during which the projected baseline rotated, resulting in a less tight but still considerable constraint across the principal baseline orientation. The average projected baseline length during this observation was $B_\mathrm{p}\simeq86~m$ and the fit residual's dispersion is $\sigma_\mathrm{O-C} = 0.48~\mu$m for a total of $N_\mathrm{mes}=865$ data points, representing the 1~s averages of an effective observation time of 0.24~h. Under the assumption of Gaussian statistics, the expected astrometric precision $\sigma_\mathrm{AX}$ can be estimated as
\begin{equation}\label{xx1}
\sigma_\mathrm{AX} = \frac{\sigma_\mathrm{O-C}}{\sqrt{N_\mathrm{mes}}\,B_\mathrm{p}} ~\mathrm{rad} \simeq 0.039~\mathrm{mas}.
\end{equation}
\noindent
A principal component analysis of the uncertainy ellipse in the sky shown in Fig.~\ref{fig:precisionHD202730} yields a dispersion of 0.436~mas across the baseline orientation and of 0.038~mas along the baseline orientation. Thus both approaches of estimating the measurement precision along the principal direction yield a comparable result, which is of the order of 40~micro-arcseconds. Using the relationships of \cite{Shao1992}, the expected astrometric error caused by atmospheric turbulence alone for this observation is 0.018~mas, where we neglected the influence of the actual atmospheric conditions and the difference between the atmospheric profiles at Mauna Kea and Paranal and, which is justified considering the measurements obtained at Cerro Armazones, a neighbour of Cerro Paranal at 22 km (e.g. \citealt{Schock:2009uq}). Thus, the astrometric precision achievable with {\small PRIMA} in the described configuration and with the above data reduction and analysis techniques is approximately two times worse than the expected atmospheric limit. This performance can potentially be improved by refining the data reduction, in particular the estimation of the astrometric observable Eq.~\ref{eq:axobs}. On the other hand, the excess noise presumably can be attributed to the systematic errors of the measurement, which stand out clearly on targets with a wider separation.\\
The above observation is not one with the best precision achieved. For instance, the observation of HD\,10360 on August 26, 2011, taken over $\sim$1.3~h and presented in \cite{Sahlmann:2012uq} yield a residual dispersion of $\sigma_\mathrm{O-C} = 0.40~\mu$m for $N_\mathrm{mes}=937$ data points ($\sim$0.26~h effective observation time) for an average baseline length of 90~m. The corresponding error ellipse has a 1-$\sigma$ dispersion of 0.032~mas along the baseline, which is comparable to the estimation of 0.030~mas through Eq.~\ref{xx1} and to the theoretical limit of 0.027~mas \citep{Shao1992}. Some observations of HD\,10360 discussed in Sect.~\ref{sec:closer} yield even better precision with 1-$\sigma$ uncertainties of 0.022--0.025~mas along the baseline.\\
We stress that precision is not equal to accuracy and the astrometric precision derived above reflects how well the fringes can be tracked and how precisely the internal differential delay can be measured, i.e.\ the stability of the differential delay measurements on short time scales. In this case, a high precision does not ensure that the best-fit separation is equally accurate, i.e.\ corresponds to reality. We estimate the astrometric accuracy of {\small PRIMA} in Section~\ref{sec:accuracy}.

\subsection{Combination of multiple epochs}
\begin{figure}\begin{center}
\includegraphics[width = 0.65\linewidth,  trim = 0cm 0cm 4.5cm 0cm, clip=true]{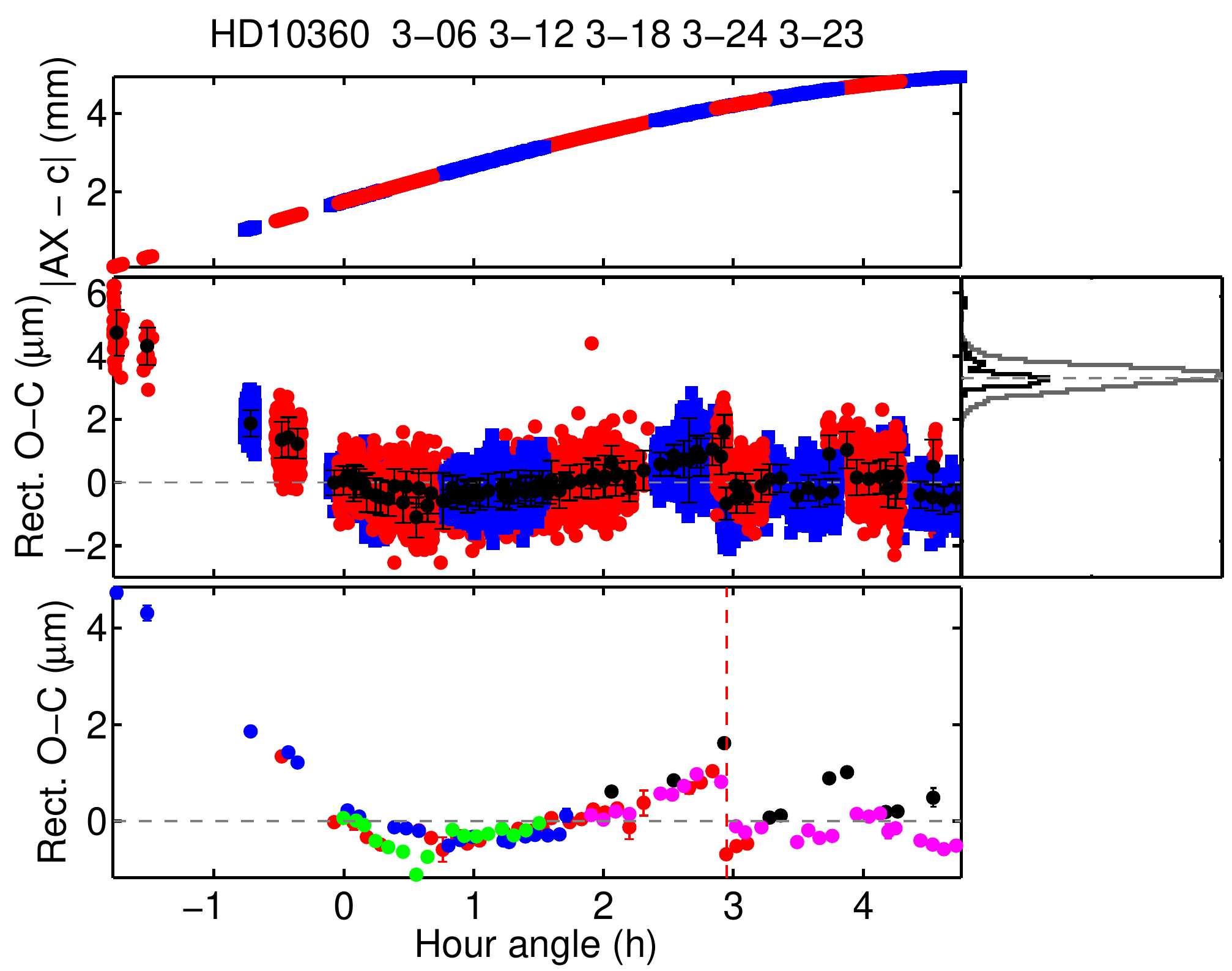}
\caption{Result of the combined fit of five epochs spanning over 5 days for HD\,10360. The colour coding is like in Fig.~\ref{fig:rectifiedmulti1}. In the bottom panel, the vertical dashed line indicates the STS-AT3 derotator wrapping.}
\label{fig:rectifiedmulti2}\end{center}
\end{figure}

Since several targets have been observed multiple times, the astrometric fit can be applied to the combination of multi-epoch data. The simplest case applies when the targets' relative position change can be neglected and the model function only has to account for the different metrology zero-points
\begin{equation}\label{eq:axfitmulti}
AX_\mathrm{obs} = \mathcal{H}_N \,\, \Delta w(\varrho,\Phi, \vec{B}_\mathrm{AX}) + \mathcal{H}_S \,\, \Delta w(\varrho,\Phi, -\vec{B}_\mathrm{AX}) + \sum_j c_j \mathcal{H}_j,
\end{equation}
where, the constant $c_j$ is the {\small PRIMA} metrology zero point at epoch $j$, which is applied via the Heaviside-type function
\begin{equation}\label{xx2}
 \mathcal{H}_j = \left\{ \begin{array}{rlr}
 1 & \mathrm{if} & \mathrm{data\; taken\; in\; epoch}\: j\\
 0 & \mathrm{else}. & \\
  \end{array}\right.
\end{equation} 
The model has thus $2+N_{j}$ free parameters, where $N_{j}$ is the number of epochs. The relative motion of HD\,66598 is uncertain but expected to be $\lesssim10$~mas\,yr$^{-1}$, because the separation change over 100 years recorded in the WDS catalogue \citep{Mason:2001fk} is approximately 1\arcsec. The largest expected motion between November 2011 and January 2012 is thus about 1~mas and much smaller than the measurement uncertainty due to the large systematic residuals of this target. Thus the simple combined fit is justified and Fig.~\ref{fig:rectifiedmulti1} shows the resulting residuals, which reveal that their principal structure is repeatable over several nights and occurs on hour-angle scales of several hours.\\
A similar analysis for the tight binary HD\,10360 of five epochs spanning 5 days is shown in in Fig.~\ref{fig:rectifiedmulti2} and confirms the previous findings, however with a down-scaled amplitude. The residuals of several epochs align with each other and show a structure on hour-long timescales. The discontinuity at hour angle $\sim$3~h coincides with a 180\degr~rotation of the telescope derotator in AT3, due to the software limit as shown in Fig.~\ref{fig:derot20}. Although the field rotates by 360\degr~and the system's configuration should thus be identical before and after this event, the step change in the residuals proves that it is not the case. The observed step change of $\sim$2~$\mu$m is a strong hint that the systematic errors are at least partially related to the derotator's position angle.

\subsection{Astrometric accuracy and plate scale}\label{sec:accuracy}
The first step to determine the astrometric accuracy is to examine the repeatability of the measurement results. In the case of the commissioning data for {\small PRIMA} astrometry, the observation epochs are typically separated by days or weeks. The measurement precision is in the range of 0.1~mas, which sets the stability requirement for the relative separation of an ideal calibration target. It turns out that most of the observed targets show relative motions of this order of magnitude \emph{per day}, i.e.\ these targets have to be characterised in detail before they can be used as reference for the {\small PRIMA} astrometry. The following sections present two of those targets and the corresponding {\small PRIMA} measurements.

\subsubsection{Relative motion of HD\,202730}
\begin{figure}[h!]\begin{center}
\includegraphics[width = \linewidth,  trim = 0cm 0cm 0cm 0cm, clip=true]{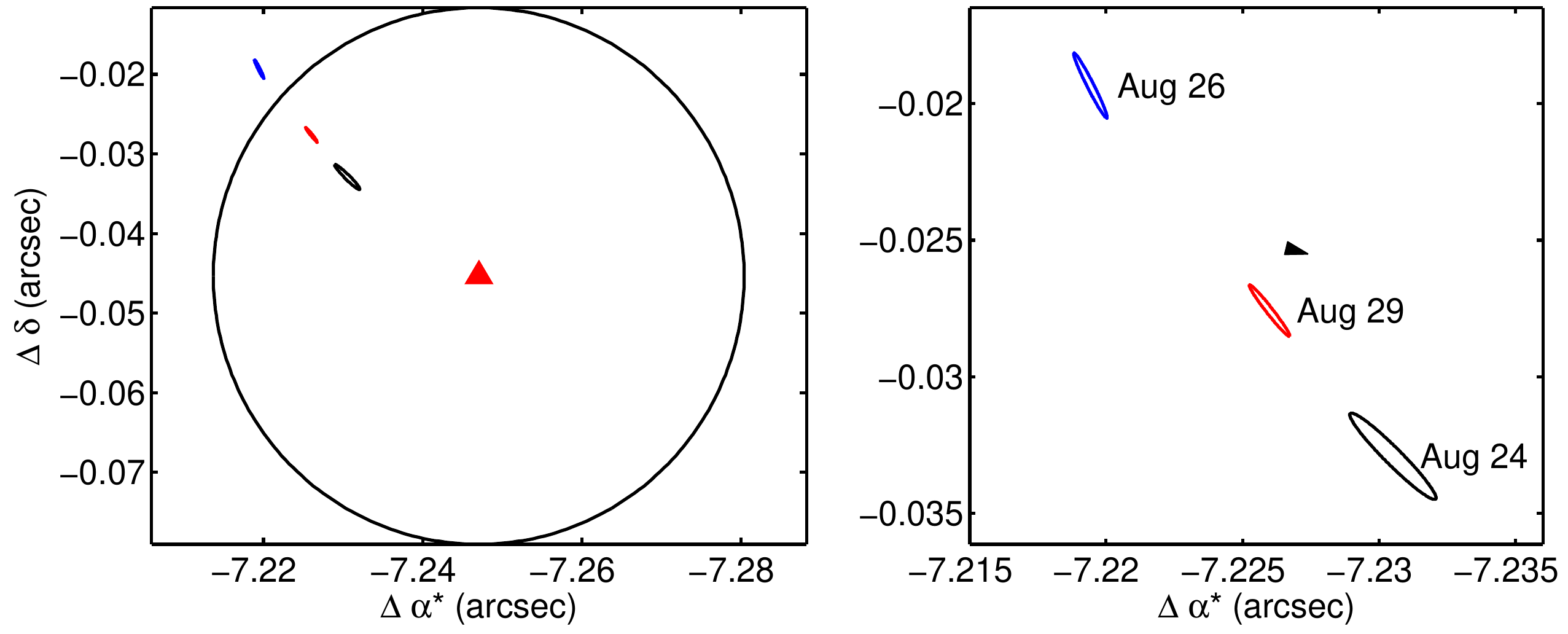}
\caption{\emph{Left:} The expected separation of HD\,202730 at the end of August 2011 (red triangle) and its error ellipse (black curve). The 3-$\sigma$ formal error ellipses of {\small PRIMA} measurements are visible at the upper left. \emph{Right:} Close-up view showing the large scatter relative to the expected motion of $\sim$0.35~mas during the 5 day timespan shown by the black arrow.}
\label{fig:HD202730ellipses}\end{center}
\end{figure}

To characterise the relative motion of the binary HD\,202730 - GJ\,9733\,B, we used the measurements of the Washington double star catalogue \citep{Mason:2001fk}. It shows considerable linear motion over the 160 year time base and rectilinear elements are given in the respective catalogue\footnote{\url{http://www.usno.navy.mil/USNO/astrometry/optical-IR-prod/wds/lin1}}. Because of the need for accurate error estimation, we used the raw WDS data and determined the linear elements in the following way. The model function is
\begin{equation}\label{eq:xx3}
\begin{split}
x &= R\,\sin \Theta  = x_0 + \mu_x (t-<t>)\\
y &= R\,\cos \Theta  = y_0 + \mu_y (t-<t>),
\end{split}
\end{equation}
where the separation $R$ and the position angle $\Theta$ are given by the WDS. We used uniform weighting after an initial sigma-clipping that resulted in discarding 19 of 76 valid measurements. The global linear fit yields the elements listed in Table~\ref{tab:HD202730lin}, which are reasonably close to the catalogue values and imply a separation change of 0.070 and 0.027 mas per day in RA and DEC respectively. The data, the best fit motion, and the residuals are shown in Fig.~\ref{fig:HD202730fit}. Using the obtained parameters, the expected separation at the time of {\small PRIMA} observations can be estimated, which is shown in Fig.~\ref{fig:HD202730ellipses}. The axes of the {\small PRIMA} measurement ellipses correspond to the 3-$\sigma$ dispersion along the respective principal orientations defined by the baseline orientation (cf. Fig~\ref{fig:precisionHD202730}).\\ 
The relative positions measured with {\small PRIMA} are in agreement with the predicted position within the admittedly large error of 33 mas coming from the WDS extrapolation. However, the scatter of the {\small PRIMA} measurements of $\sim$10~mas is much larger than their precision and can not be explained by the expected separation change, which is an order of magnitude smaller. Instead, we note that the hour-angle coverage is not identical for these observations and the Aug. 26 epoch is covering the meridian passage. 
\begin{table}[h]
\caption{Linear elements of the HD\,202730 separation change described by Eq.~\ref{eq:xx3}}
\label{tab:HD202730lin}  \centering  
\begin{tabular}{l r r} 	
\hline\hline %
Parameter & Unit & Value\\
 \hline
$x_0$ & (\arcsec) & $-5.4991 \pm 0.0174$ \\
$y_0$ & (\arcsec) & $+0.6341 \pm 0.0174$ \\
$\mu_x$ & (\arcsec \,yr$^{-1}$) & $-0.0256 \pm 0.0004$ \\
$\mu_y$ & (\arcsec \,yr$^{-1}$) & $-0.0100 \pm 0.0004$ \\
$\sigma_{O-C}$ & (\arcsec) & $0.130$  \\
\hline 
\end{tabular} 
\end{table}

\subsubsection{Orbital motion of HD\,10360 - HD\,10361}
\begin{figure*}
\begin{center}
\includegraphics[width = 0.8\linewidth,  trim = 0cm 0cm 0cm 0cm, clip=true]{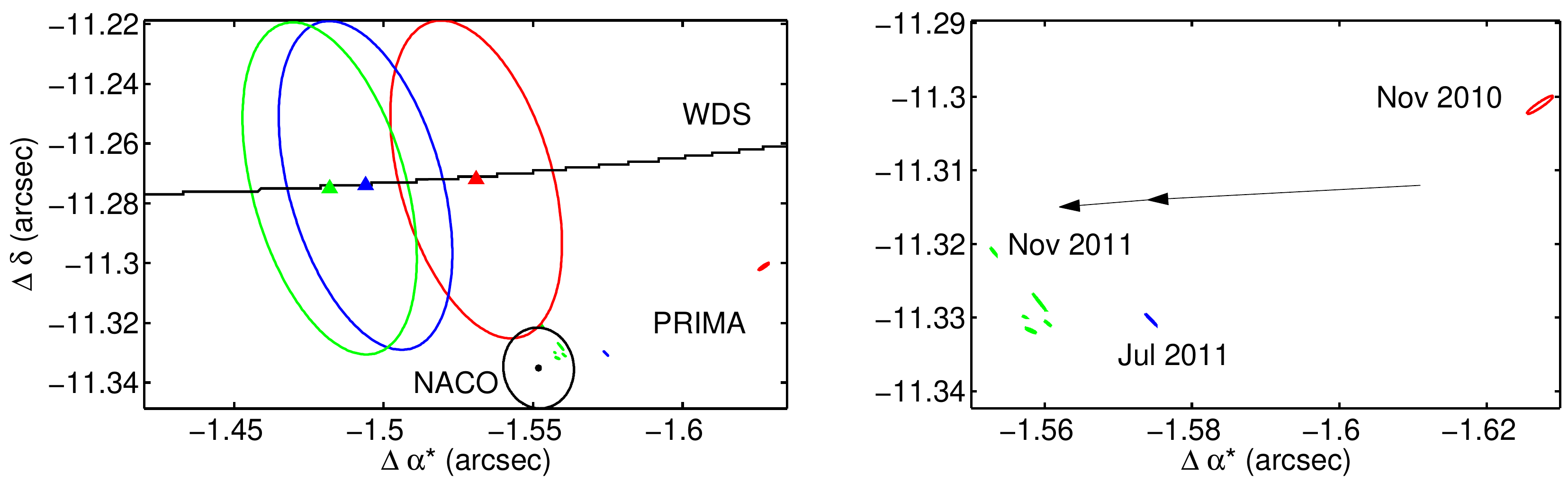}
\caption{\emph{Left:} Close up view of the updated HD\,10360 model orbit (black curve) showing the predicted positions (triangles) and their 1-$\sigma$ error ellipses for Nov. 2010, July 2011, and Nov. 2011 in red, blue, and green, respectively. The {\small PRIMA} error ellipses are visible towards to lower right of the panel. The 1-$\sigma$ error ellipse of the {\small NACO} measurement taken in Nov. 2011 is shown in black. \emph{Right:} The 3-$\sigma$ error ellipses of the {\small PRIMA} measurements for the single epochs in Nov. 2010 and Jul. 2011 and the five epochs in Nov. 2011. The black arrows correspond to the expected separation change between the epochs.}
\label{fig:HD10360fit2}
\end{center}
\end{figure*}

HD\,10360 is a very bright visual binary and a solution for the relative orbit has been published \citep{van-Albada:1957lr}. The WDS lists 162 valid measurements and the comparison between the published orbit and the more recent data from the WDS showed significant disagreement. Thus, we fitted an updated orbit considering uniform weighting and the WDS data remaining after an initial sigma-clipping, which discarded 9 measurements. The result is shown in Fig.~\ref{fig:HD10360fit} and Table~\ref{tab:HD10360Orb}. The new orbit is comparable in angular size, but has a considerably shorter orbital period, a higher eccentricity, and a smaller inclination. For the purpose of comparison with {\small PRIMA} astrometry, it is important that the model motion fits well the recent WDS measurements, which is the case. The separation change in 2011 and 2012 corresponding to this orbit is $129\pm5~\mu$as/day eastwards.
\begin{table}[h]
\caption{Updated orbital elements of HD\,10360}
\label{tab:HD10360Orb}  \centering  
\begin{tabular}{l r r r} 	
\hline\hline %
Parameter & Unit & \cite{van-Albada:1957lr} & This work\\
 \hline
$P$ & (Yr) & $483.7$ & $312.4 \pm 13.3$ \\ 
$e$ &        & $0.53$ & $0.74 \pm 0.03$ \\ 
$T_0$ & (JDB) & $-16575$ & $-14024 \pm 322$ \\ 
$\omega$ & (\degr) & $18.37$ & $-49 \pm 5$ \\ 
$\Omega$ & (\degr) & $13.12$ & $-20 \pm 2$ \\ 
$i$ & (\degr) & $142.8$ & $130.3 \pm 3.3$ \\ 
$a_{rel}$ & (\arcsec) & $7.82$ & $7.77 \pm 0.27$ \\ 
$\sigma_{O-C}$ & (\arcsec) & \ldots &$0.137$  \\
 \hline 
\end{tabular} 
\end{table}\\
HD\,10360 was extensively observed with {\small PRIMA} for test purposes already before the astrometry facility became operational. Using dual-feed data taken in November 2010, we could recover astrometric observations by using the internal metrology signal of the DDL to construct the astrometric observable, thus those measurements are intrinsically less accurate. Then HD\,10360 was observed once in July 2011 and five times in November 2011. Figure~\ref{fig:HD10360fit2} shows that the {\small PRIMA} measurements do not coincide with the predicted separation based on the updated orbit within the latter's error ellipses of $25-55$~mas size, but they are located within $\sim$3-$\sigma$. 
Considering the incomplete orbital coverage and the uncertain error bars of the WDS measurements, the updated orbit solution may be subject to systematic errors, which could explain the disagreement between the orbit model and the {\small PRIMA} measurements. With {\small PRIMA} we systematically measure a $\sim$\,60~mas larger separation and a larger position angle, which indicates that these measurements are coherent. This is also enforced by the observation that the separation change of HD\,10360 over 1 year is measured with the correct order of magnitude, cf. Fig.~\ref{fig:HD10360fit2} right.\\ 
In addition these results were obtained under the assumption that the astrometric baseline is the same for all epochs, which is certainly not justified at the required level of precision. Over the 1 year time base, the telescopes and the STS were moved, aligned, and possibly modified. Also we did not consider any effect related to the fit residual structure and the November 2010 epoch was recovered from non-astrometric observations. Thus the shown results are affected by unquantified systematic errors, which cause the observed inconsistencies. But these measurements provide us with a starting point for the necessary improvement by several orders of magnitude.

\begin{figure}[h!]\begin{center}
\includegraphics[width = 0.48\linewidth,  trim = 0cm 0cm 4.7cm 0cm, clip=true]{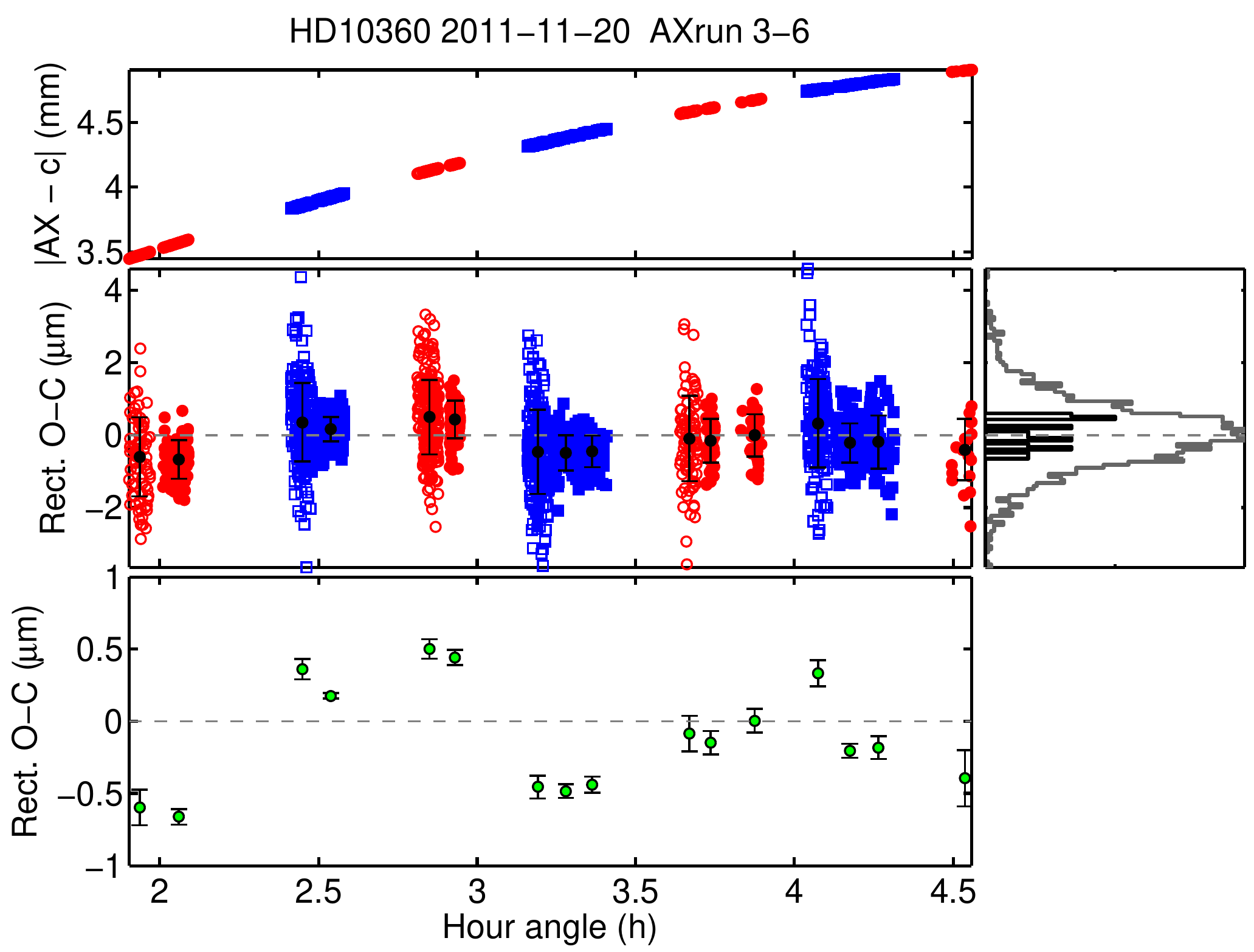}\hspace{2mm}
\includegraphics[width = 0.47\linewidth]{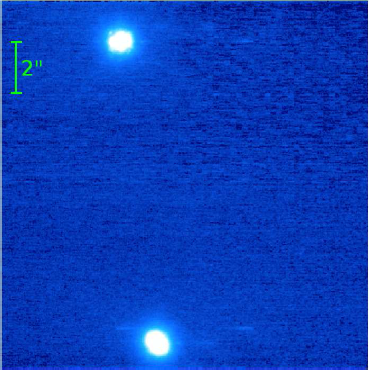}
\caption{Simultaneous observation of HD\,10360 with {\small PRIMA} and {\small NACO}. \emph{Left:} Result of the {\small PRIMA} astrometry fit. \emph{Right:} {\small NACO} $K$-band image of HD\,10360 (top) and HD\,10361 (bottom).}
\label{fig:HD10360naco}\end{center}
\end{figure}

\begin{figure}\begin{center}
\includegraphics[width = 0.65\linewidth,  trim = 0cm 0cm 0cm 0cm, clip=true]{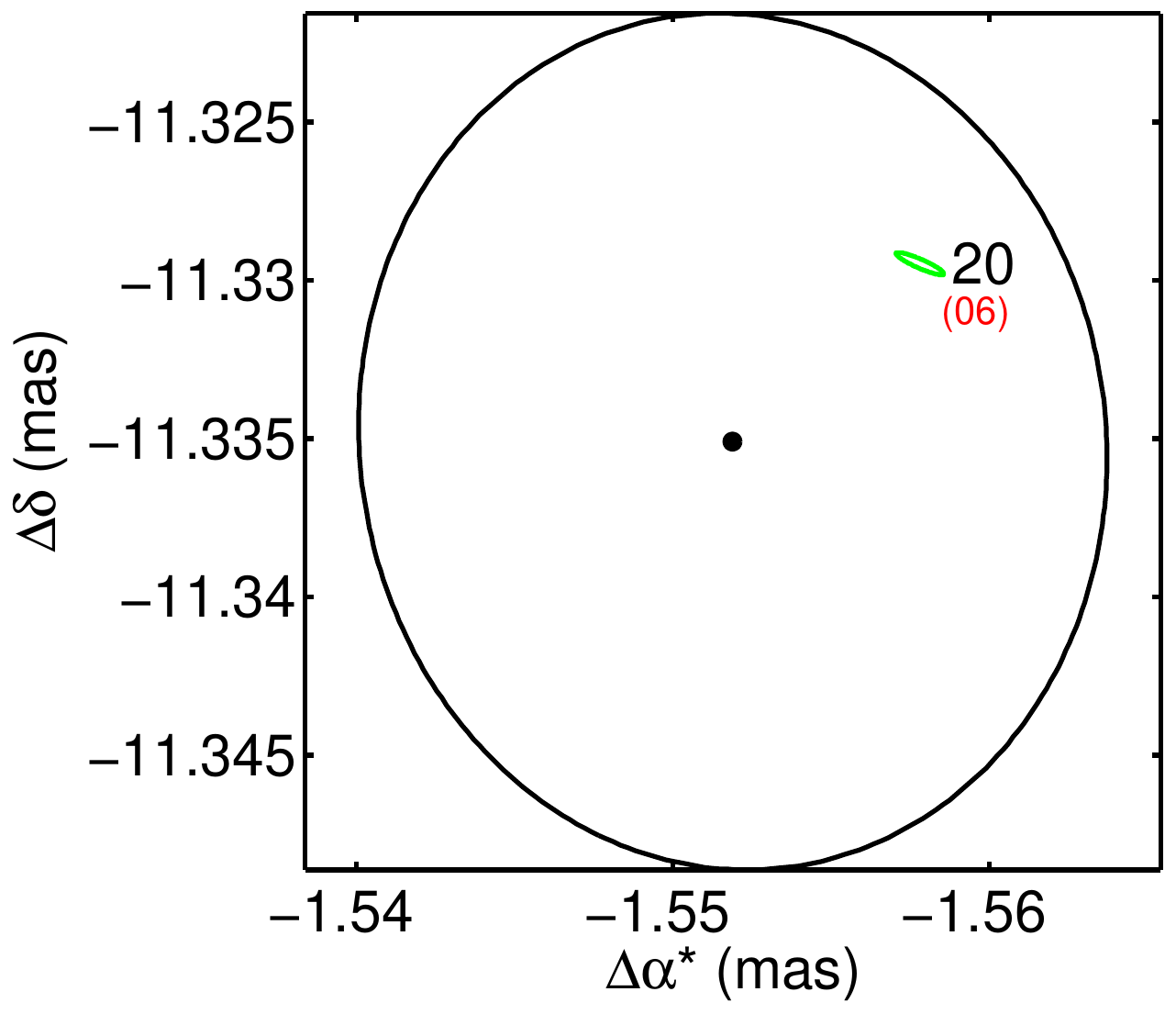}
\caption{The separation of HD\,10360 on November 20, 2011 measured with {\small NACO} (black) and {\small PRIMA} (green) are consistent within the former's error ellipse.}
\label{fig:HD10360naco2}\end{center}
\end{figure}

\subsubsection{Adaptive optics observation with NACO}
As shown above, the quality assessment of {\small PRIMA} measurements of HD\,10360 is limited by the knowledge of the pair's orbital motion, which is not going to improve significantly in the near future because high precision measurements over long time-spans will be necessary. Thus we used an independent method to obtain a high-precision reference measurement for this target and obtained technical observation time with NACO, the adaptive optics assisted infrared camera at the VLT \citep{Lenzen:2003vn, Rousset:2003ys}. In the night of November 20, 2011, HD\,10360 was observed simultaneously with {\small PRIMA} and {\small NACO} (Fig.~\ref{fig:HD10360naco}). We used the 2.17 $\mu$m Br$_\gamma$ filter of the {\small NACO} S27 camera to obtain a total of 218 s exposure time in 2000 individual frames. To calibrate the {\small NACO} plate scale, we subsequently observed an area of the Trapezium cluster and used the astrometry of selected cluster stars published by \cite{Close:2012fk}, and the measurement was corrected for differential refraction. Because of the target separation of $\sim$11\arcsec, the plate scale calibration limits the {\small NACO} astrometric accuracy to 13 mas. The formal solution of the {\small PRIMA} measurement is a separation of $\rho = (11.4361 \pm 0.0001)$\arcsec~and a position angle of $\Phi =(187.829 \pm 0.001)$\degr. Figure~\ref{fig:HD10360naco2} shows that the {\small NACO} and {\small PRIMA} measurements agree within the {\small NACO} error bar of $\sim13$~mas, which confirms that the orbit solution based on WDS data alone is biased towards smaller separations and a larger position angle, cf. Fig~\ref{fig:HD10360fit2}. The overlap of both independent measurements and the robust astrometric calibration of {\small NACO} allow us to conclude that the plate scale of {\small PRIMA} is correct at the $\sim$13~mas level for a 11.4\arcsec~binary, i.e.\ exhibits a relative error smaller than $1.1\cdot10^{-3}$ \footnote{Since PRIMA is not a field imaging instrument, its \emph{plate scale} refers to the absolute scale fidelity, which for instance can be affected by an error in the astrometric baseline or in the metrology wavelength.}.

\subsection{A closer look at the November 2011 epochs}\label{sec:closer}
\begin{figure}[h!]\begin{center}
\includegraphics[width = \linewidth,  trim = 0cm 0cm 0cm 0cm, clip=true]{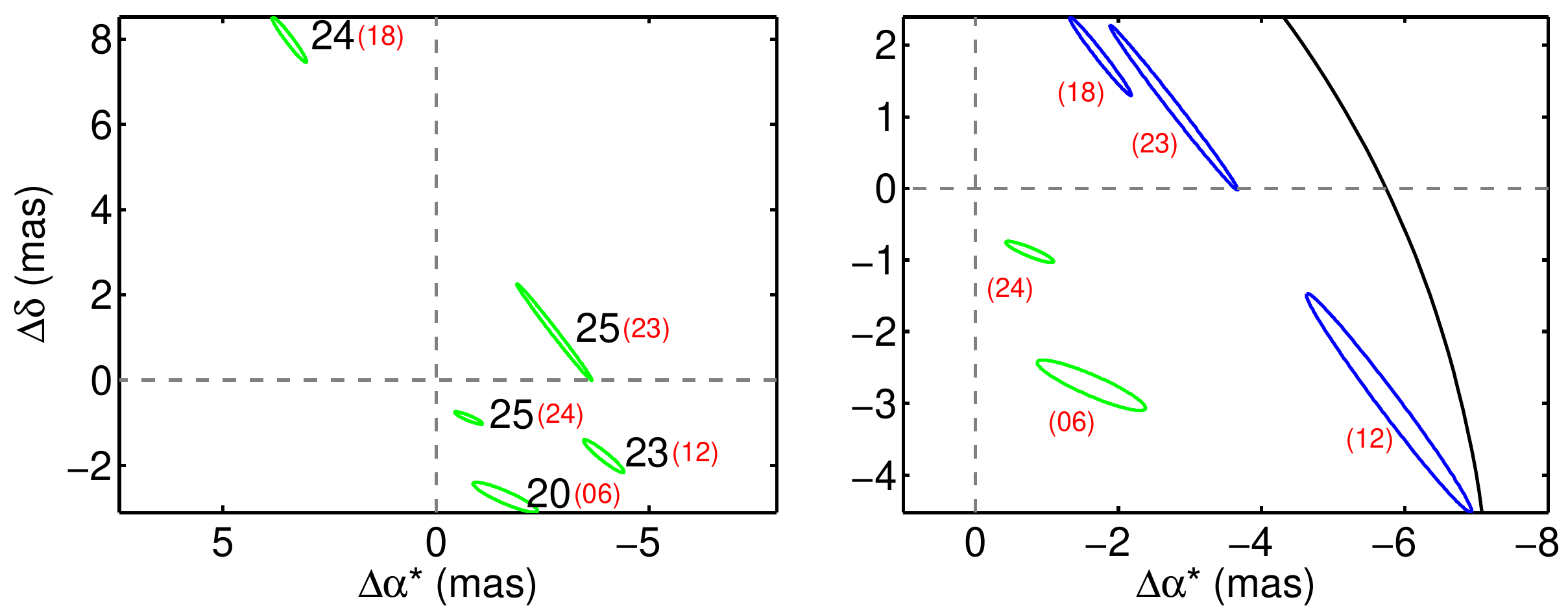}
\caption{\emph{Left:} {\small PRIMA} error ellipses from the analysis of the raw data on HD\,10360 for the five epochs in November 2011. The offsets to a constant vector are shown. Numbers indicate the datum in November 2011 and numbers in brackets show the run number of the respective measurement. \emph{Right:} The same, but after limiting the hour angle range of three epochs (blue). The black curve is the error ellipse section of {\small NACO} shown in Fig.~\ref{fig:HD10360fit2}.}
\label{fig:HD10360fit3}\end{center}
\end{figure}

The five epochs on HD\,10360 taken over a time base of five days in November 2011 offer the opportunity to study the short term stability of the measurement. The raw {\small PRIMA} measurements show a scatter of the order of $\pm5$~mas as shown in Fig.~\ref{fig:HD10360fit3}.
For the following discussion, it is necessary to list the sequence of observations and their conditions.
\begin{description}
  \item[November 20, 2011:] Simultaneous observations with {\small NACO} and {\small PRIMA} are obtained and the latter cover a hour angle range of 2 - 4.5 (run 06). 
  \item[November 21, 2011:] Observations to determine the baseline model are made, which are discussed in Sect.~\ref{sec:opdmodeldet}.
  \item[November 22, 2011:] Failure of the technical CCD of AT4 at the beginning of the night prevents observations. Consequently AT4 was moved and the star separator module was taken out of the telescope station to give access to the telescope electronics. The system was then reassembled and thus may have shifted within the mechanical tolerances.
  \item[November 23, 2011:]  {\small PRIMA} observations in the hour angle range of -0.5 - 3 (run 12).
  \item[November 24, 2011:]  {\small PRIMA} observations in the hour angle range of -1.5 - 1.5 (run 18).
  \item[November 25, 2011:]  {\small PRIMA} observations in the hour angle range of 0 - 1.5 (run 23) and 2 - 4.5 (run 24).  
  \end{description}
Three runs are affected by the AT3 derotator unwrapping at hour angle $\sim$2.8. In Fig.~\ref{fig:HD10360fit3} we see that the observations obtained at large hour angle (runs 6 and 24) yield similar separation and position angle, whereas the observations at small hour angles (runs 12, 18, and 23) result in a systematically higher position angle (westward shift) and a higher spread in separation. Run 18 is offset relative to the other runs, which may be explicable by the large hour angle coverage symmetric about meridian passage. In an attempt to limit the effects of unequal hour angle range, we excluded data of those three runs outside $HA = [-0.2,1.6]$ and the results are shown in Fig.~\ref{fig:HD10360fit3} (right). The error ellipse of run 18 is now close to the other measurements at small hour angles.\\
From the analysis of the HD\,10360 data several conclusions can be drawn: (a) The measurement dispersion is of order $\pm3$~mas and is dominated by hour-angle dependent biases which are much larger than the formal uncertainties. (b) The expected eastward separation change of $\sim$130\,$\mu$as/day can therefore not be detected. (c) The effect of the telescope and star separator relocation on the astrometric measurement was smaller than $\sim$2 mas. (d) The uncontrolled biases appear to have a stronger influence on the position angle than on the separation.

\subsection{Identifying the sources of systematic errors}\label{sec:sster}
\begin{figure}\begin{center}
\includegraphics[width = 0.8\linewidth,  trim = 0cm 0cm 0cm 0cm, clip=true]{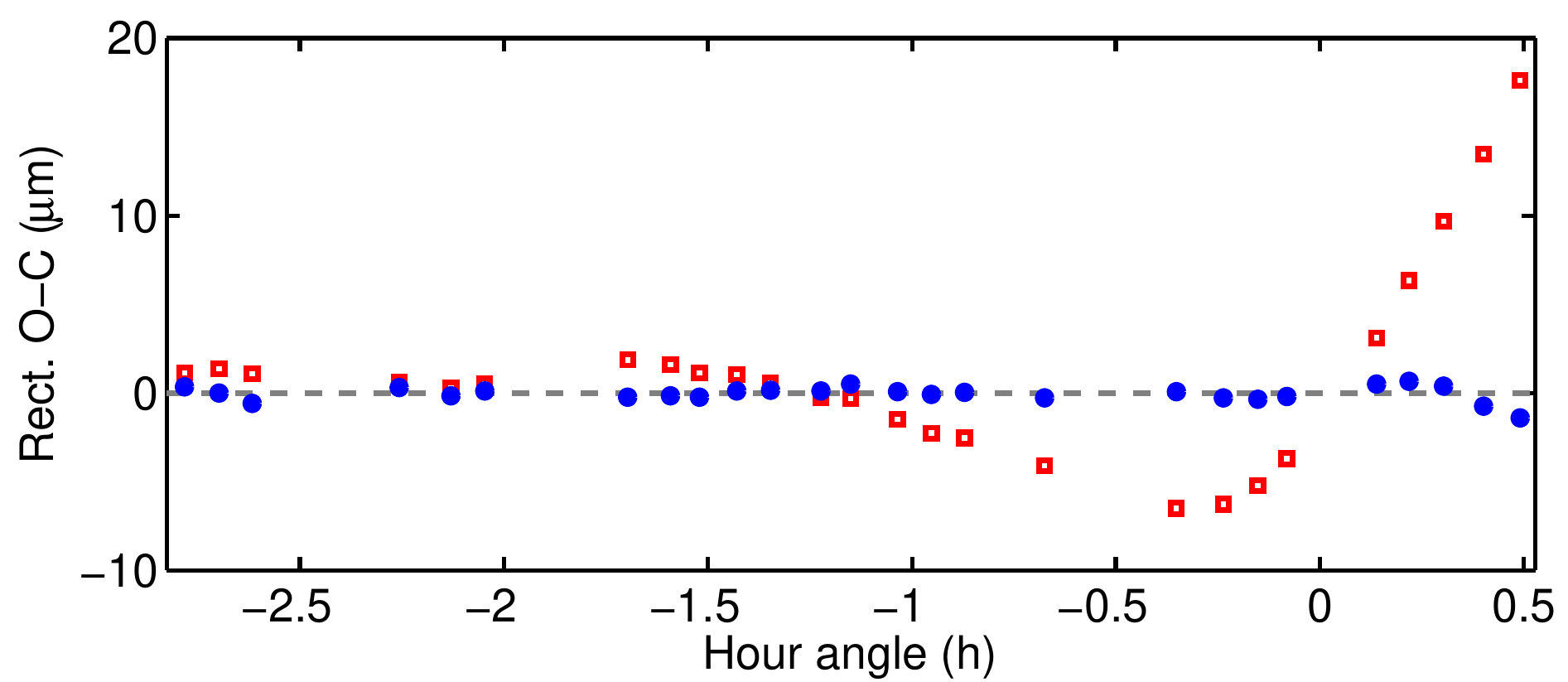}
\caption[Comparison of residuals with the standard and modified model function.]{Rectified and binned residuals of HD\,66598 observations after adjustment of the standard model (red open squares, $5.53~\mu$m RMS) and with the modifed, 5-parameter model (blue circles, $0.43~\mu$m RMS).}
\label{fig:3132}\end{center}
\end{figure}

Recalling Eq.~\ref{eq:nab04}
\begin{equation*}
\label{eq:nab44}
\Delta w(t) + \Delta \epsilon(t) = \left[ \vec B_\mathrm{AX} - \vec \mu_{1}(t) + \vec \mu_{2}(t)\right] \cdot \vec \Delta s(t) + c_\Delta,
\end{equation*}
we note that there are two terms capable of introducing biases that reach an amplitude of up to tens of micro-metres, i.e.\ far larger than one fringe packet and close to the coherence length of the {\small PRIMA} bandpass. One is the non-common path error $\Delta \epsilon(t)$ between metrology and stellar beams, and the other is the time-dependent relative position change $\vec \mu_{i}(t)$ between the metrology end-point and the telescope pivot point. A possibility to distinguish between the different error terms may be to compare the differential delay measured with the {\small PRIMA} metrology to the differential delay line position, which is discussed in Appendix~\ref{sec:diffMET}. To make accurate astrometry at $10\,\mu$as level possible, those terms have to be quantified, measured, and calibrated at the 5~nm level for the non-common path error and at the $100\,\mu$m level for $\vec \mu_{i}(t)$.\\
The fit residuals shown in Fig.~\ref{fig:rectifiedmulti1} and \ref{fig:rectifiedmulti2} indicate that they are correlated with the field rotation $\omega$. To verify this hypothesis, we added a bias term to the astrometric model function Eq.~\ref{eq:axfit}:
\begin{equation}\label{eq:axfit2}
\begin{split}
 AX_{obs} = c \;+ \;&\mathcal{H}_N \,\, \Delta w(\varrho,\Phi, \vec{B}) + \mathcal{H}_S \,\, \Delta w(\varrho,\Phi, -\vec{B})\\
  + \;&sign(\mathrm{mode}) \cdot  (r_1 + r_2\,\omega),
 \end{split}
\end{equation}
where $r_1$ is an offset and $r_2$ models a linear dependency on the field angle, and $sign(\mathrm{mode})$ is $+1$ in normal mode and $-1$ in swapped mode. Instead of computing the field rotation angle, we obtained it from the auxiliary data of the derotator angles $\omega_{rot,3}$ and $\omega_{rot,4}$ of STS3 and STS4, respectively, from the relation $\omega = (\omega_{rot,3}+\omega_{rot,4})/2$. When applying the modified model function to a suitable test sequence on the wide binary HD\,66598 taken on November 23, 2011, the binned residual dispersion $\sigma_{O-C, binned}$ decreases by more than a factor of 10 as is shown in Fig.~\ref{fig:3132} and Table~\ref{tab:313}. This is a compelling evidence that the dominant systematic errors in the {\small PRIMA} astrometry data are correlated with field rotation.
\begin{table}
\caption{Fit results with standard and modified model function.}
\label{tab:313}  \centering
\small  
\begin{tabular}{l c c c} 	
\hline\hline %
Parameter & Unit & Standard model & Model Eq.~\ref{eq:axfit2}\\
 \hline
$c$ & (mm) & $-10.08724 \pm 0.00009$ & $-10.0860 \pm 0.0002$ \\ 
$\rho$ &  (\arcsec)      & $35.7789 \pm 0.0007$ & $36.162 \pm 0.024$ \\ 
$\Phi$ & (\degr) & $134.677 \pm 0.003$ & $134.449 \pm 0.024$ \\ 
$r_1$ & ($\mu$m) & \ldots & $-644 \pm 37$ \\ 
$r_2$ & ($\mu$m/\degr) & \ldots & $3.9 \pm 0.2$ \\ 
$\sigma_{O-C, binned}$ & ($\mu$m) & 5.53 &$0.43$  \\
 \hline 
\end{tabular} 
\end{table}

\begin{figure}
\begin{center} 
\includegraphics[width= \linewidth]{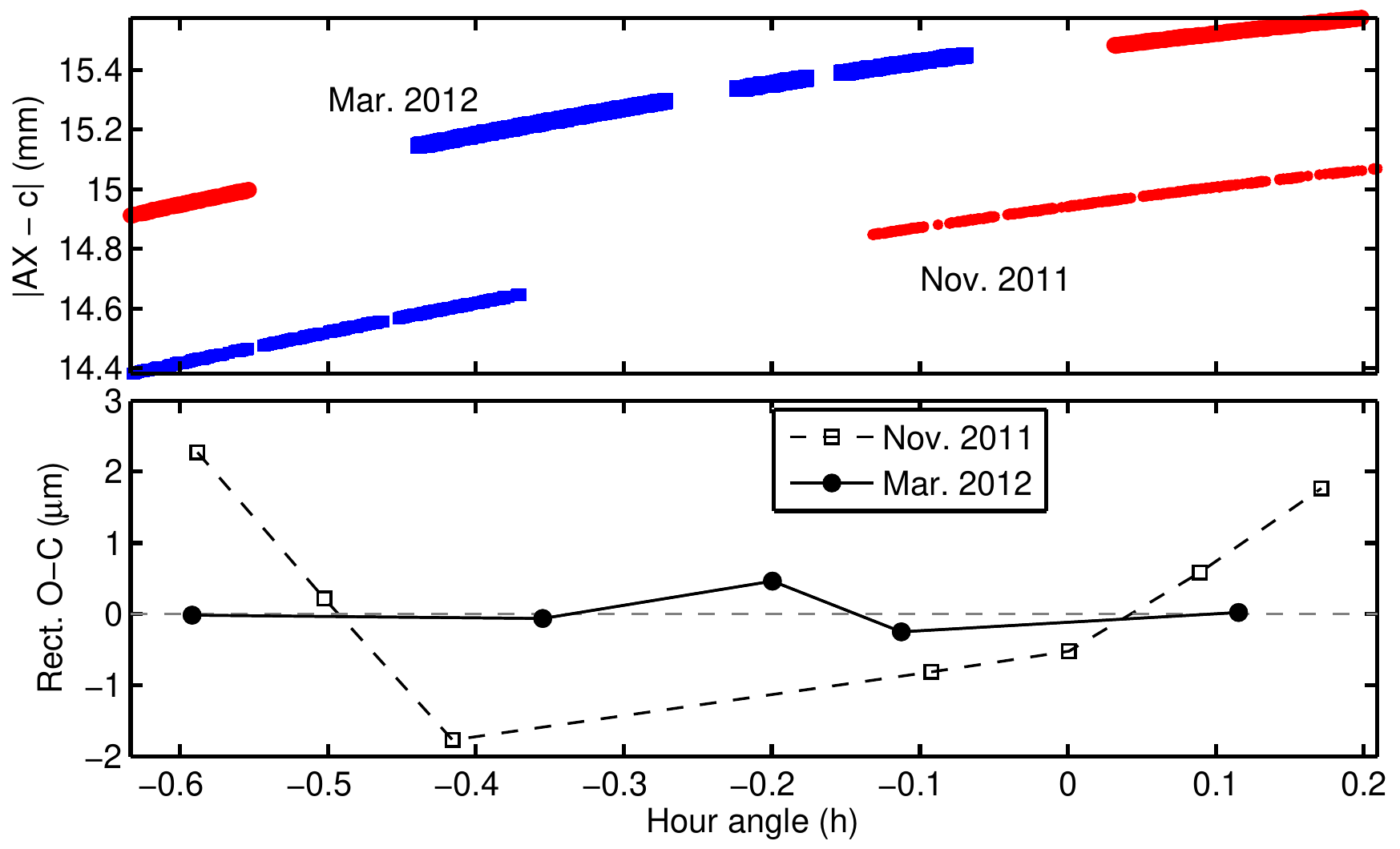} 
\caption{Observations of HD\,66598 before and after telescope re-alignment. \emph{Top:} Differential delay of both sequences. For better readability, the Nov. 2011 sequence was shifted by -0.5 mm. \emph{Bottom:} Binned residuals before (dashed line and open squares) and after (solid line and filled circles) the alignment intervention. The binned residual dispersion decreases from 1.46\,$\mu$m to 0.27\,$\mu$m. Error bars are smaller than the symbol size.}
\label{fig:comp}
 \end{center} 
\end{figure}

\noindent
The physical cause was discovered in January 2012, when important and pointing-dependent obscurations of the stellar pupils and a significant lateral pupil run-out as a function of the telescope azimuth angle of both telescopes were detected, which points at insufficient alignment of the {\small PRIMA-VLTI} optics.
The alignment of the auxiliary telescopes (primarily of M4) used for PRIMA was improved in March 2012, which resulted in a significant reduction of residual amplitude. Figure~\ref{fig:comp} shows the effect with the help of observations of HD\,66598, which were selected to lie within the same hour-angle range and have a comparable amount of collected data.
Before the intervention, the residuals show a u-shaped pattern with several $\mu$m amplitude typical for observations of a wide-separation binary close to meridian. After the intervention, the remaining residuals have decreased by more than a factor of five in RMS. This significant decrease of the systematic error supports the idea that it is introduced in the part of the optical beamtrain that is not monitored by the {\small PRIMA} metrology, i.e.\ above M9 in the derotator and the telescope.
\section{Discussion and conclusions}\label{sec:concl}
The concept of narrow-angle astrometry with a dual-feed interferometer has been developed two decades ago and {\small PRIMA} at the {\small VLTI} is the first large scale facility to implement this mode of operation with the goal of offering it to the astronomical community. {\small PRIMA} was integrated in the {\small VLTI} and several technical runs and three astrometric commissionings have taken place in 2011/12 with the purpose of establishing the instrument's astrometric capabilities. The technical challenges of operating {\small PRIMA} have been overcome and allowed us to collect data of good quality to test the system and obtain first estimates of the achievable astrometric precision and accuracy.\\ 
As a critical instrument hardware contribution, the {\small ESPRI} consortium has successfully built, installed, and tested differential delay lines for {\small PRIMA} according to the technical requirements set by ESO. The differential delay lines are now permanently installed at {\small VLTI} and integrated in the interferometer infrastructure. They implement robust and fast piston actuators for four beams with a possible extension to eight beams, and are available for {\small PRIMA} astrometry and for the use in conjunction with other and future {\small VLTI} instruments.\\
The astrometric precision of {\small PRIMA} observations is limited by the atmosphere at a level close to the theoretical expectations and a value of $\sim$0.04~mas was obtained for 0.24~h effective observation time on a bright and small-separation binary. This confirms that the anticipated scientific programme, requiring uncertainties in the 0.01 milli-arcsecond range, is feasible with this facility if a long-term stability at the same level can be achieved, which is not the case with the present setup:\\ 
For wide-separation binaries ($\gtrsim$10\arcsec) and long ($>$2 h) observing sequences, large systematic errors that are correlated with the field rotation are observed. Those errors can amount to several tens of micro-metres in differential delay corresponding to tens of milli-arcseconds in astrometry. The astrometric accuracy obtained so far is not sufficient and several equivalent {\small PRIMA} measurements exhibit a scatter much larger than the individual measurement uncertainty. On a bright 11\arcsec~binary, for instance, the measurement repeatability over five days is of order 3~milli-arcsec. The dominant systematic errors originate in the beam train above M9 that is not monitored by the {\small PRIMA} metrology and can therefore be attributed to the non-common path term $\Delta \epsilon(t)$ of the model function Eq.~\ref{eq:nab04}. So far, the large amplitude of this term has prevented us from quantifying the contributions of the other error terms and potential biases occurring at the fringe position measurement level. An extension of the metrology endpoint up to M2 or beyond would mitigate a large fraction of the observed systematic errors and allow us to significantly improve the astrometric accuracy. The necessary modification to the telescope and star separator systems are under study at ESO.\\   
To verify the {\small PRIMA} astrometric measurement independently, we obtained calibrated {\small NACO} astrometry of a 11\arcsec~binary simultaneously with a {\small PRIMA} observation. Both measurement agree within the {\small NACO} uncertainty of 13\,mas, which allows us to conclude that the {\small PRIMA} astrometric plate scale is correct at the $\sim$0.1\% level. Similar measurements will be helpful to confirm the {\small PRIMA} accuracy, which has to reach the $10^{-5}-10^{-6}$ level to meet the {\small ESPRI} science requirement.\\
In conclusion, the present astrometric performance of {\small PRIMA} is not sufficient to start the {\small ESPRI} astrometric search for extrasolar planets. The factors that limit the astrometric accuracy are being identified and corrected, which will be an iterative process in which the true error sources are eliminated in descending order of the introduced error magnitude. The dominant contributor has been identified to be the unmonitored beamtrain between M9 and M1.\\
In this paper, we have described the interferometer infrastructure and the setup of the {\small PRIMA} astrometric instrument. The operation principles and the data reduction and analysis strategies were introduced. Our findings show that modifications to the metrology system are necessary to monitor a larger part of the stellar beamtrain. Despite the changes in the setup provoked by those improvements in the future, the described {\small PRIMA} observation principles retain their validity.

\subsection{Outlook}
The immediate future of the {\small PRIMA} and {\small ESPRI} projects will be focussed on establishing the astrometric performance necessary for the {\small ESPRI} science programme. Once this is accomplished, we will use  {\small PRIMA} astrometry to constrain the mass distribution of giant planets around Sun-like stars and to advance our knowledge on long-period planets around young stars. This research will be pioneering the era of the {\small GAIA} astrometry mission (e.g. \citealt{Lindegren:2010kx}), which in the intermediate future will have grand influence on the exoplanet field \citep{Casertano:2008th} and will bring the application of high-precision astrometry for exoplanet research to maturity.\\
Because the future of the {\small ASTRA} project \citep{Woillez:2010rt} is uncertain due to ceasing operations at the Keck Interferometer, {\small PRIMA} is the only dual-feed interferometer facility worldwide. The future {\small GRAVITY} instrument \citep{Eisenhauer:2011fr} is based on similar operational principles and will take the realisation of high-precision interferometric astrometry one step further.

\begin{acknowledgements}
For the research presented in this paper, we have made use of the databases at the Centre de Donn\'ees astronomiques de Strasbourg (\href{http://cds.u-strasbg.fr/}{CDS}) and of NASA's Astrophysics Data System Service (\href{http://adsabs.harvard.edu/abstract_service.html}{ADS}). This research has made use of the Washington Double Star Catalog (\href{http://ad.usno.navy.mil/wds/}{WDS}) maintained at the U.S. Naval Observatory. J.S. thanks J. Woillez for sharing his astrometric baseline description and S. Lacour for insightful discussions on the interferometer modelling. J.S. kindly acknowledges support as a visitor at the Centro de Astrobiolog\'ia in Villanueva de la Ca\~nada (Madrid), where parts of this paper have been written. 
All authors are grateful to the technicians, engineers, and scientists of the PRIMA, VLT/VLTI, and ESPRI teams that made the deployment of the PRIMA facility possible. We thank M. Accardo, J. Alonso, H. Bonnet, P. Bourget, A. {Cortes}, R. {Frahm}, B. {Gilli}, P. Gitton, N. Gomes, S. Guisard, P. Haguenauer, C. Hummel, L. Jocou, A. Jost, C. {Maire}, S. {M\'enardi}, S. {Morel}, J. Ott, R. {Palsa},  E. Pedretti, I. {Percheron}, A. {Pino}, D. Popovic, E. Pozna, F. Puech, A. Ramirez, F. Somboli, I. {Stilz}, G. {Valdes}, and many others. We thank A. Kaufer for granting us NACO technical time and D. Mawet and L. Tacconi-Garman for kindly preparing and executing the NACO observations. 
\end{acknowledgements}

\bibliography{/users/sahlmann/astro/papers} 

\clearpage
\begin{appendix}
\section{Corrections to PRIMA metrology measurements}\label{sec:PRIMETcorr}
\subsection{Correcting PRIMA metrology fringe counter overflows}
The {\small PRIMA} metrology measurements of differential delay ($\Delta L$, {\small PRIMET}) and internal delay of the FSUB feed ($\Delta L_B$, {\small PRIMETB}) rely on fringe counters which are stored internally in a memory location using $N_\mathrm{bit,FC}$ bits. The value of $N_\mathrm{bit,FC}$ defines the maximum measurement range. The {\small PRIMA} metrology fringe counter overflows are flagged by the respective {\small PACMAN} file header keywords and the counter size is $N_{bit,FC}=31$ bits. The laser frequency has been calibrated with a laser frequency comb system and is stabilised to $\lambda_P = 1319.1762$~nm during operation. The value of the offset caused by an overflow is
\begin{equation} \label{eq:deltaFC2}
\Delta_{FC}= \frac{\lambda_P }{2} \,\frac{\Delta \nu_P}{\nu_P} \, 2^{N_{bit,FC}} \simeq 486.783\,\,\mu\mathrm{m},
\end{equation}
where $\nu_P$ is the {\small PRIMA} metrology laser frequency and $ \Delta \nu_P$ is the frequency shift given in Table~\ref{tab:PRIMETdef}. The factor $\Delta \nu_P/\nu_P$ originates in the heterodyne detection principle of the system and the necessary phase error compensation. Two counters are involved and the corresponding overflows occur approximately every 80 min, provoking an upward offset by $+\Delta_{FC}$ followed by a downward offset of $-\Delta_{FC}$ a few seconds later, because of the Doppler-shift due to the moving delay line. During the data reduction, those offsets are detected and corrected. 

\subsection{Correcting PRIMETB wrapping}
The {\small PRIMA} metrology system {\small PRIMETB} measures the internal delay $\Delta L_B$ of the FSUB feed. By design, {\small PRIMETB} has a wrapping period of $L_{PM}$ caused by fringe counter overflows in the phasemeter:
\begin{equation}\label{eq:LPM2}
L_{PM} =  \frac{\lambda_P }{2}  2^{N_{bit,PM}} \simeq 345.814121~\mathrm{mm},
\end{equation}
where $c_P$ is the vacuum speed of light and $N_{bit,PM}=19$ is the number of bits allocated for this counter. Before using the {\small PRIMETB} data for analysis, it has to be uwrapped accurately. More details about the corrections can be found in \cite{Sahlmann2012PhD}.

\section{Difference between differential OPD measured with PRIMA metrology and DDL feedback}\label{sec:diffMET}
During the observation, the differential delay is introduced with the DDL and measured with the {\small PRIMA} metrology. Ideally, the measurements of the internal DDL laser metrology $\delta \Omega_{DDL}$ ($\lambda_{DDL} = 632.991\, \mu$m) and the {\small PRIMA} metrology ($\Delta L$) should be equal except for a constant offset. In reality, their difference varies in time and has offsets between normal and swapped states. 
\begin{figure}[h]
\begin{center}
\includegraphics[width = 0.8\linewidth]{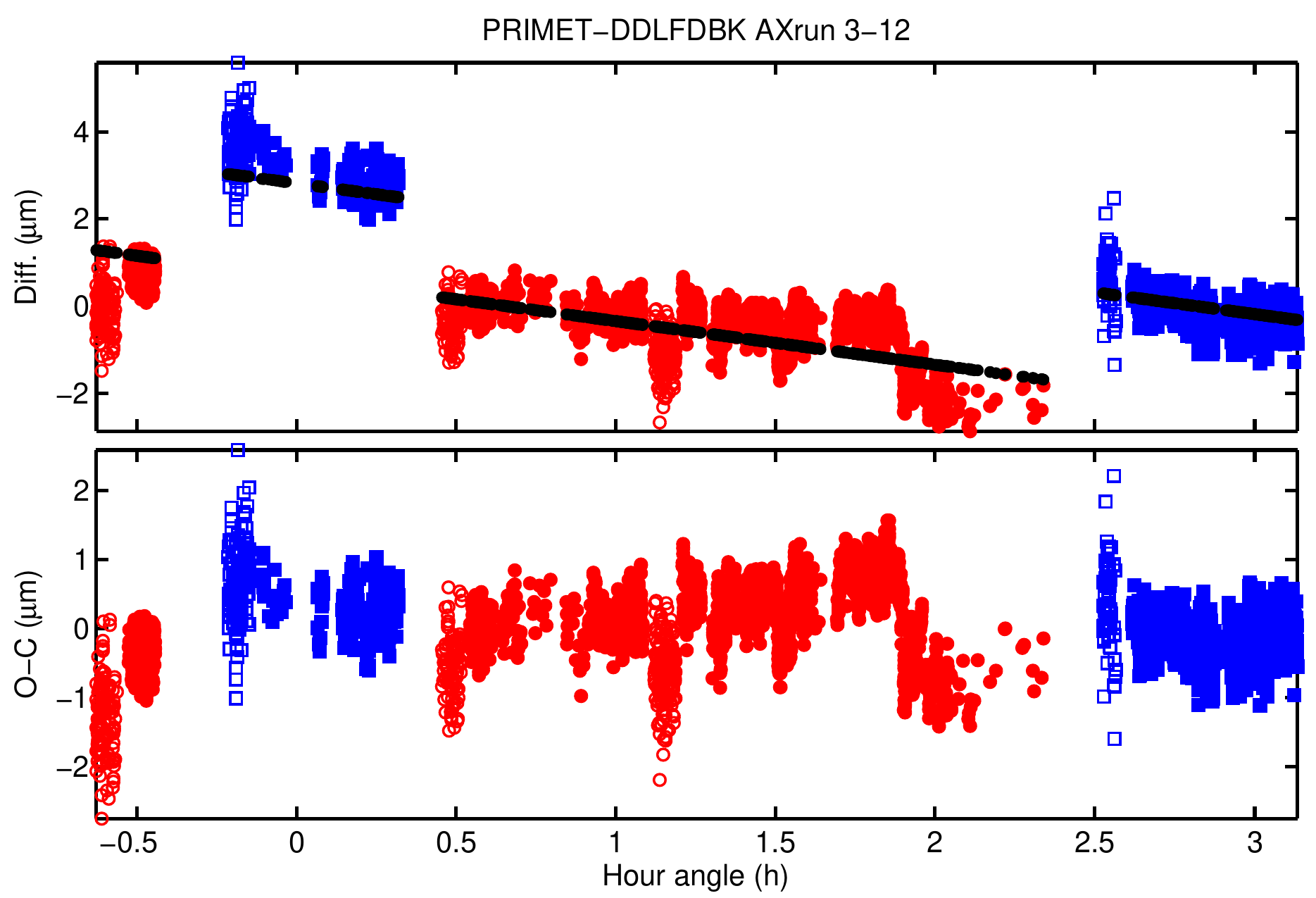}
\caption{\emph{Top}: {\small PRIMA} metrology minus DDL feedback during the observations of run 3-12. The black dots show the best linear fit with a first order polynomial. \emph{Bottom}: Residuals after adjusting the model Eq.~\ref{eq:diffmodel} with three parameters (offset, slope, step change amplitude).}
\label{fig:PRIMETDDL}
\end{center}
\end{figure}
Figure~\ref{fig:PRIMETDDL} shows the {\small PRIMA} metrology reading minus the DDL feedback as a function of time. The sequence starts with normal mode observations, where the DDL feedback corresponds to DDL2. In swapped mode, the DDL feedback corresponds to DDL1. Without assuming an underlying physical model, we fitted a linear dependence in time to quantify the apparent variations. The model function is
\begin{equation}\label{eq:diffmodel}
\Delta L - \delta \Omega_{DDL} = c_{DDL} + m_{DDL} \,t + \mathcal{H}_S \,\,\gamma_{DDL},
\end{equation}
where $c_{DDL}$ is an offset, $m_{DDL}$ is the observed slope, $t$ is the time, and $\gamma_{DDL}$ is the additional offset visible in swapped mode. The difference shows a linear drift of $\!-1.004 \,\mu$m h$^{-1}$ and a step change at each normal-swapped cycle with an amplitude of $\sim\! 2.163 \, \mu$m. Table~\ref{tab:difffits} shows the fit results, where $\sigma_{DDL}$ is the RMS of the fit residuals. A look at the residuals reveals that this simple model does not account for all systematics apparent in the data. 
\begin{table}[h]
\caption{Results of PRIMET -- DDL modelling}
\label{tab:difffits}  \centering  
\begin{tabular}{l r r r} 	
\hline\hline %
Parameter & Unit & Run 3-12 \\
 \hline
$c_{DDL}$ & (mm) & $-1.48732 \pm 0.00002$ \\
$m_{DDL}$ & ($ \mu $m h$^{-1}$) &$-1.00368101 \pm 0.00000001$\\
$\gamma_{DDL}$ & ($ \mu $m) &     $2.16259646 \pm 0.00000002$\\
$\sigma_{DDL}$& ($\mu$m) & 0.591\\
\hline 
\end{tabular} 
\end{table}\\
Since the DDL are controlled in closed loop by the fringe sensors combining the stellar beams and the {\small PRIMA} metrology is a passive system measuring the differential delay up to its endpoints and thus does not cover the complete internal stellar beam-train, the difference shown in Fig.~\ref{fig:PRIMETDDL} may carry essential information about the non-common path error of the {\small PRIMA} system and the interpretation of the astrometric observable and the astrometric baselines. This motivates a closer investigation of the observed drift and the following hypotheses are proposed:
\begin{itemize}
\item Refraction: Differential chromatic refraction between the metrologies of {\small PRIMA} and DDL and between the two feeds are suppressed because the DDL are kept under vacuum. However, second order differential effects due to gradients along the {\small VLTI} beam train are possible and remain to be quantified.
\item One or several of the non-tracking DDL introduce the drift and offsets: The internal metrologies of all four DDL have been examined and only the tracking one shows values different from zero. The passive DDL are controlled internally on the basis of their respective independent metrology system. A large drift of the internal metrology wavelength during the observation could explain the observed drift, because it would not appear in the data, but the required amplitude appears unreasonable.
  \item An error in the {\small PRIMA} metrology or DDL metrology laser wavelength: In Sect~\ref{sec:MDLfit}, we have shown that the {\small PRIMA} wavelength appears to be accurate, thus the DDL wavelength can be suspected. A DDL wavelength error would result in a linear drift as a function of DDL position and should not result in normal-swap offsets. The corresponding model fit yields a value for the drift of about $-1.1\,\mu$m/m, i.e. an error of order $10^{-3}$, which is very large and does not explain the normal-swap offsets. A DDL wavelength error is therefore an unlikely cause.
\item An error introduced by the delay line variable curvature mirror, which is located in an image plane: The delay line VCM curvature depends linearly on the DL position. The effect of the DL VCM curvature on the PRIMET measurement and on the DOPD between the two interferometer feeds has to be studied and understood, thus remains a possible explanation. 
\end{itemize}
We can also examine the problem using the baseline formalism developed in Sect.~\ref{sec:principles}. When neglecting both constant terms and error terms, the delays measured with delay line ($\Omega_\mathrm{DL}$) and differential delay line ($\delta \Omega_\mathrm{DDL}$) metrologies are related to the wide-angle baselines. In normal mode, we have
\begin{equation}\label{eq:wab10}
\begin{split}
\Omega_\mathrm{DL}+\delta \Omega_\mathrm{DDL1}&= \vec B_\mathrm{A, Wide} \cdot \vec s_A \\
\Omega_\mathrm{DL}&= \vec B_\mathrm{B, Wide} \cdot \vec s_B,
\end{split}
\end{equation}
because it is DDL1 that introduces the differential delay and it follows
\begin{equation}\label{eq:wab11}
\delta \Omega_\mathrm{DDL1}= \vec B_\mathrm{A, Wide} \cdot \vec s_A -\vec B_\mathrm{B, Wide} \cdot \vec s_B
\end{equation}
When considering non-common path errors of the delay line metrologies $\epsilon_\mathrm{DL}$, the correction terms relating the metrology terms to the pivot points $\vec \mu_\mathrm{DL}$, and biases introduced by the fringe position measurements of the FSU $\eta_\mathrm{FSU}$ we obtain
\begin{equation}\label{eq:wab12}
\begin{split}
\delta \Omega_\mathrm{DDL1} + \epsilon_{N,\mathrm{DL}}= &\left(\vec B_\mathrm{A, Wide} -\vec \mu_\mathrm{A1,DL}+\vec \mu_\mathrm{A2,DL}\right) \cdot \vec s_A + \eta_\mathrm{FSUA}\\
 -&\left(\vec B_\mathrm{B, Wide} -\vec \mu_\mathrm{B1,DL}+\vec \mu_\mathrm{B2,DL}\right) \cdot \vec s_B + \eta_\mathrm{FSUB}
\end{split}
\end{equation}
and the equivalent relation in swapped mode, where the primed quantities stand for the potential baseline modification by turning the derotators:
\begin{equation}\label{eq:wab13}
\begin{split}
\delta \Omega_\mathrm{DDL2} + \epsilon_{S,\mathrm{DL}}= &\left(\vec B'_\mathrm{A, Wide} -\vec \mu'_\mathrm{A1,DL}+\vec \mu'_\mathrm{A2,DL}\right) \cdot \vec s_B + \eta'_\mathrm{FSUA}\\
 -&\left(\vec B'_\mathrm{B, Wide} -\vec \mu'_\mathrm{B1,DL}+\vec \mu'_\mathrm{B2,DL}\right) \cdot \vec s_A + \eta'_\mathrm{FSUB}.
\end{split}
\end{equation}
Note that in swapped mode, the secondary star ($\vec s_A$) is in the B feed and observed with FSUB. The equation for the {\small PRIMA} metrology measurement is (cf. Eq.~\ref{eq:nab03})
\begin{equation}\label{eq:wab14}
\begin{split}
\mathrm{Normal:}\hspace{0.3cm}\Delta L + \epsilon_{\Delta L}= &\left(\vec B_\mathrm{A, Wide} -\vec \mu_{\mathrm{A1},\Delta L}+\vec \mu_{\mathrm{A2},\Delta L} \right) \cdot \vec s_A + \eta_\mathrm{FSUA}\\
 -&\left(\vec B_\mathrm{B, Wide} -\vec \mu_{\mathrm{B1},\Delta L}+\vec \mu_{\mathrm{B2},\Delta L}\right) \cdot \vec s_B + \eta_\mathrm{FSUB}
\end{split}
\end{equation}
\begin{equation}\label{eq:wab15}
\begin{split}
\mathrm{Swapped:}\hspace{0.1cm}\Delta L + \epsilon_{\Delta L}= &\left(\vec B'_\mathrm{A, Wide} -\vec \mu'_{\mathrm{A1},\Delta L}+\vec \mu'_{\mathrm{A2},\Delta L} \right) \cdot \vec s_B + \eta'_\mathrm{FSUA}\\
 -&\left(\vec B'_\mathrm{B, Wide} -\vec \mu'_{\mathrm{B1},\Delta L}+\vec \mu'_{\mathrm{B2},\Delta L}\right) \cdot \vec s_A + \eta'_\mathrm{FSUB},
\end{split}
\end{equation}
where $\epsilon_{\Delta L}$ is the 4-beam non-common path error of the {\small PRIMA} metrology, and the $\vec \mu_{\Delta L}$ model the offsets from the respective endpoints to the pivot points. Subtraction of Eq.~\ref{eq:wab12} from Eq.~\ref{eq:wab14} yields the observable plotted in Fig.~\ref{fig:PRIMETDDL}:
 \begin{equation}\label{eq:wab16}
\begin{split}
\mathrm{Normal:}\\
\Delta L - \delta \Omega_\mathrm{DDL1} = &\; \epsilon_{\Delta L}-\epsilon_{N,\mathrm{DL}}\\
&+ \left(-\vec \mu_{\mathrm{A1},\Delta L}+\vec \mu_{\mathrm{A2},\Delta L} +  \vec \mu_\mathrm{A1,DL}-\vec \mu_\mathrm{A2,DL}\right) \cdot \vec s_A\\
&- \left(-\vec \mu_{\mathrm{B1},\Delta L}+\vec \mu_{\mathrm{B2},\Delta L} +  \vec \mu_\mathrm{B1,DL}-\vec \mu_\mathrm{B2,DL}\right) \cdot \vec s_B \\
\mathrm{Swapped:}\\
\Delta L - \delta \Omega_\mathrm{DDL2} = &\; \epsilon_{\Delta L}-\epsilon_{S,\mathrm{DL}}\\
&+ \left(-\vec \mu'_{\mathrm{A1},\Delta L}+\vec \mu'_{\mathrm{A2},\Delta L} +  \vec \mu'_\mathrm{A1,DL}-\vec \mu'_\mathrm{A2,DL}\right) \cdot \vec s_B\\
&- \left(-\vec \mu'_{\mathrm{B1},\Delta L}+\vec \mu'_{\mathrm{B2},\Delta L} +  \vec \mu'_\mathrm{B1,DL}-\vec \mu'_\mathrm{B2,DL}\right) \cdot \vec s_A \\
\end{split}
\end{equation}\\
There are thus several terms that can alter the difference $\Delta L - \delta \Omega_\mathrm{DDL}$ and they are either attributable to non-common path errors or to changes in the relative positions of metrology endpoints and telescope pivot points. At the actual state of the analysis, we are unable to conclude on the dominant effect. 

\section{Tables and figures}
 \begin{figure}[h]
 \begin{center}
\includegraphics[width = 0.7\linewidth,  trim = 0cm 0cm 0cm 0cm, clip=true]{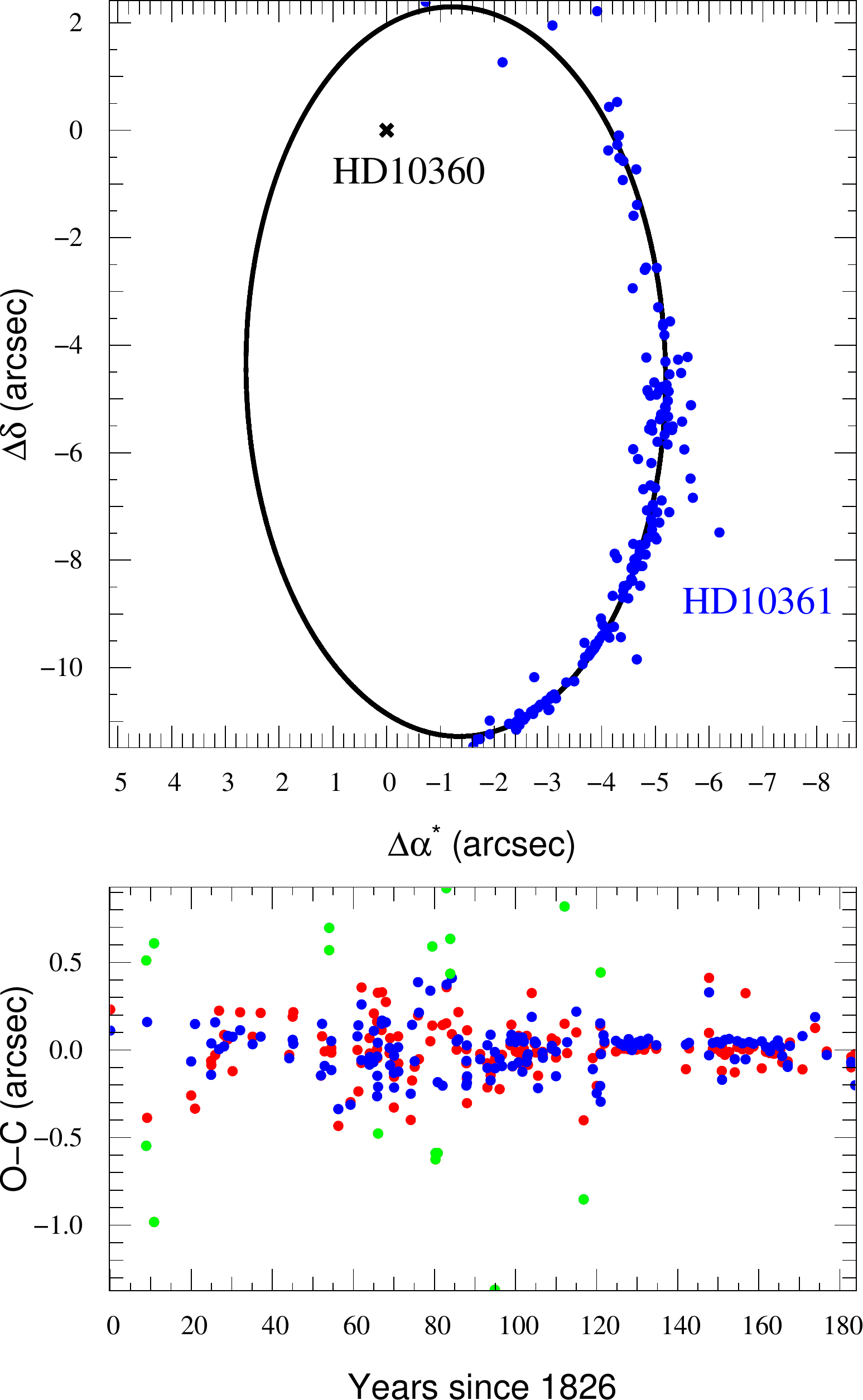}
\caption{\emph{Top:} The relative motion of HD\,10361 (blue circles) and HD\,10360 (black cross) in the sky and the updated orbital fit (solid curve). \emph{Bottom:} The fit residuals in RA (red) and DEC (blue). Data marked with green symbols was not used for the fit.}
\label{fig:HD10360fit}\end{center}
\end{figure}

\begin{table*}[h]
\caption{Number of reflections before injection into the FSU fibre when using ATs.}
\label{tab:reflection}  \centering  
\begin{tabular}{l | l r r r r r r r r r r } 	
\hline\hline  
Mode &Element     & AT & STS & M12 & DL & M16 &BC\tablefootmark{a} & SY\tablefootmark{b} & DDL & FSU\tablefootmark{c}  & All\\
\hline
Dual &Reflections& 8 & 8  &1 & 5& 1&3&1&5& 6 & 38\\
Single &Reflections& 11 & -  &1 & 5& 1&-&1&-& 6 & 25\\
\hline 
\end{tabular} 
\tablefoot{\tablefoottext{a} {Beam Compressor.} \tablefoottext{b} {Switchyard.} \tablefoottext{c} {This includes the folding mirror before the FSU and excludes the IRIS $H$/$K$-dichroic passed in transmission.} }
\end{table*}
\begin{table*}[h!]
\caption{Metrology wavelengths and calculated index of refraction for the demonstration run.}
\label{tab:DLDDLwavelength}  \centering  
\begin{tabular}{l r r r r} 	
\hline\hline %
 &  & Applied value\tablefootmark{a} & Vac. value & Refract. index \\ 
 \hline
$\lambda_\mathrm{DL}$ & (nm) & 632.863000 & 632.991354 & $n_\mathrm{DL}-1=2.025 \pm 0.003 \cdot 10^{-04}$\\  
$\lambda_\mathrm{DDL}$ & (nm) & 632.991000& 632.991372 & 0 \\  
$\lambda_P$ & (nm)      & 1\,319.176183 & 1\,319.176183& $n_\mathrm{P}-1=2.007\pm0.008 \cdot 10^{-04}$\\
$\lambda_{FSU}$ & (nm)      & $\sim$2\,250 & \ldots& $n_\mathrm{FSU}-1=2.003 \pm 0.008\cdot 10^{-04}$\\[5pt]
Pressure & (mbar) &\multicolumn{3}{c}{$745.4 \pm 0.3$ }\\
Rel. humidity &(\%) &\multicolumn{3}{c}{$6.8 \pm 0.6 $}\\
Avg. temp. & (\degr C) &\multicolumn{3}{c}{$15.6 \pm 1.1$}\\
\hline 
\end{tabular} 
\tablefoot{\tablefoottext{a} {The applied value of wavelength multiplies the fringe count to yield the length reported by the real-time system. The refraction correction applied to this length is described in the text.}}
\end{table*}

\begin{table*}[h]
\caption{Characteristics of selected targets.}
\label{tab:targets}  \centering  
\begin{tabular}{l c r r |r r |r r } 	
\hline\hline %
\multicolumn{2}{l}{System} & \multicolumn{2}{c}{HD10360}& \multicolumn{2}{c}{HD202730}& \multicolumn{2}{c}{HD66598}\\
\multicolumn{2}{l}{Name}  &HD\,10360 &HD\,10361 &HD\,202730 &GJ\,9733\,B &HD\,66598 &HD\,66598\,B  \\
\multicolumn{2}{l}{Sp. Type}  &K2V &K0V &A5V &A7V &K2.5III &\ldots  \\
$m_V$ & (mag) &5.96 &5.07 &4.40 &7.20 &5.82 &8.81  \\
$m_K$ & (mag) &3.56 &3.51 &4.10 &5.42 &3.04 &4.63  \\
$m_H$ & (mag) &6.68 &\ldots &5.51 &\dots &3.17 &4.92  \\
$\mu_{\alpha^\star}$ & (mas/yr) &282 &302 &107 &97 &-12 &-1 \\
$\mu_{\delta}$ & (mas/yr) &22 &-14 &-66 &-70 &8 &8  \\[5pt] 
$\alpha$ & (h:m:s) &\multicolumn{2}{r|}{01:39:47.55\tablefootmark{a}} &\multicolumn{2}{r|}{21:19:51.98} &\multicolumn{2}{r}{08:03:04.16}  \\
$\delta$ & (d:m:s) &\multicolumn{2}{r|}{-56:11:36.16\tablefootmark{a}} &\multicolumn{2}{r|}{-53:26:57.93} &\multicolumn{2}{r}{-32:27:48.78}  \\
$\varpi$ & (mas) &\multicolumn{2}{r|}{148}&\multicolumn{2}{r|}{33.0}&\multicolumn{2}{r}{6.9} \\
$\rho$ & (\arcsec) & \multicolumn{2}{r|}{10.941}& \multicolumn{2}{r|}{ 7.067}& \multicolumn{2}{r}{35.630} \\ 
$\Phi$ & (\degr) &  \multicolumn{2}{r|}{180.656}&  \multicolumn{2}{r|}{267.014}&  \multicolumn{2}{r}{134.242} \\ 
$\Delta \alpha^\star$ & (\arcsec) &  \multicolumn{2}{r|}{-0.125}& \multicolumn{2}{r|}{-7.057}&   \multicolumn{2}{r}{25.527} \\ 
$\Delta \delta$ & (\arcsec) &  \multicolumn{2}{r|}{-10.940}&  \multicolumn{2}{r|}{-0.368}&  \multicolumn{2}{r}{-24.859} \\ 
\hline 
\end{tabular}
\tablefoot{\tablefoottext{a} {The coordinates of HD\,10360 originate from the Tycho catalogue \citep{Hog:2000kx}, which lists this target with zero proper motion. This makes a proper motion correction from the Tycho epoch to ICRS necessary.} }
\end{table*}

 \begin{figure*}[h]
 \sidecaption
\includegraphics[width = 12cm,  trim = 0cm 0cm 0cm 0cm, clip=true]{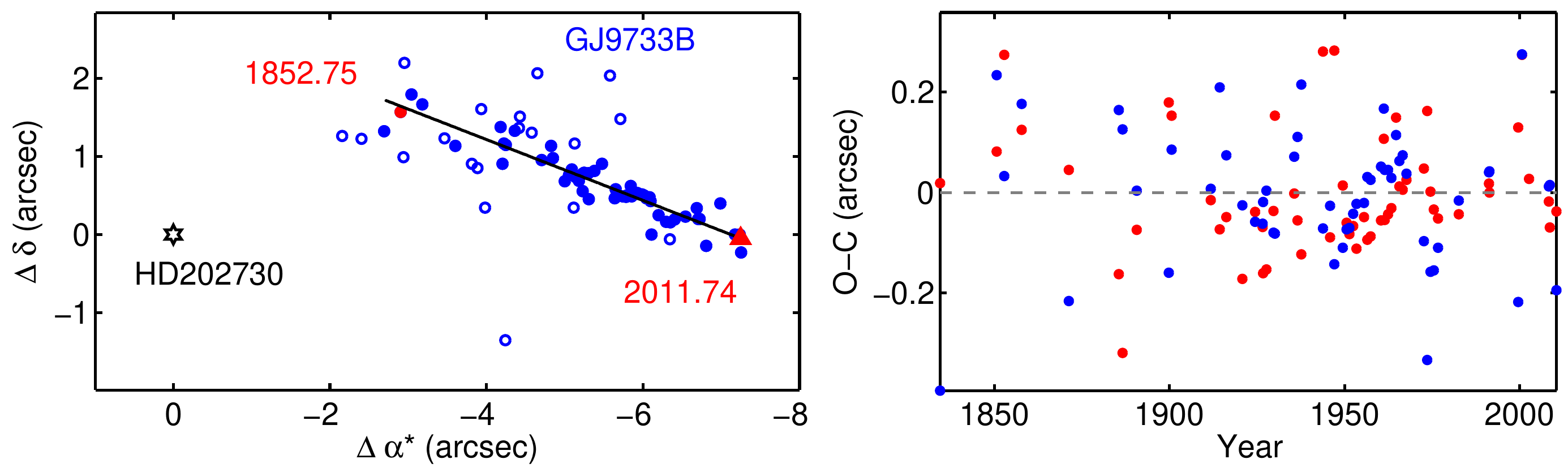}
\caption{Linear fit to the relative motion of HD\,202730. \emph{Left:} The location of primary (black star) and secondary (blue circles) and the best linear fit (black line) are shown. Open circles identify measurements not considered for the fit. The expected separation in August 2011 is shown with the red triangle. \emph{Right:} The fit residuals in RA (red) and DEC (blue).}
\label{fig:HD202730fit}
\end{figure*}

\begin{table*}[h]
\caption{PRIMA-VLTI control loops acting on the stellar beams.}
\label{tab:controlloops}  \centering  
\begin{tabular}{l r r r c c c} 	
\hline\hline %
Function & Sensor & Actuator  & Bandwidth\tablefootmark{f} & FB\tablefootmark{a} & Band &\#\tablefootmark{e}\\
&  &   &(Hz) &  \\
 \hline
Telescope tracking (Alt.+Az.)& Encoders & Torque motors  & $\sim5$ & 1 &\ldots&2\\
Telescope guiding (Alt.+Az.) & CCD & Torque motors & $\sim5$ & 1 &$V$&2\\
Image stabilisation telescope& STRAP\tablefootmark{b}  & M6  & $\sim$ 5-10 & 1 &$V$&2\\
Image stabilisation laboratory & IRIS  & STS-FSM  & $\sim1$ & 1 &$H$&4\\
Lateral pupil stabilisation\tablefootmark{c} & Quadcell & STS-VCM &$\sim1$ & 1& $1.3\,\mu$m&4\\
Primary fringe tracking & FSUB/FSUA & One delay line &$\sim10$ & 1&$K$&1\\
Secondary fringe tracking & FSUA/FSUB & One DDL &$\sim$ 5-10 & 1&$K$&1\\
Longitudinal pupil position\tablefootmark{d} & \ldots & DL-VCM &\ldots & 0 &\ldots&1\\
Image position on STS-M10 & \ldots & Derotator &\ldots& 0 &\ldots&2\\
Telescope focus\tablefootmark{g} (manual)  & IRIS & M2 & \ldots & 0 & $H$ & 2\\
Angle tracking\tablefootmark{h} (optional) & IRIS  & FSU-ACU  & $\sim5$ & 0 &$H$&4\\
\hline 
\end{tabular} 
\tablefoot{\tablefoottext{a} {The flag indicates whether it is a feedback loop.} \tablefoottext{b} {Quadcell sensor based on avalanche photo diodes.} \tablefoottext{c} {The prima metrology beams are stabilised.} \tablefoottext{d} {A linear model depending on the carriage position is used to blindly adjust the mirror curvature.} \tablefoottext{e} {Number of instances.} \tablefoottext{f} {Lowest frequency of -3 dB gain loss.} \tablefoottext{g} {The telescope focussing is done manually by the operator.} \tablefoottext{h} {The angle tracking (IFG) consists of offloading the residual image offsets measured with IRIS onto the image actuators of the FSU \citep{Sahlmann:2009kx}. It is usually not used during astrometric observations.}}
\end{table*}

\begin{figure*}[h]
\begin{center}
\includegraphics[width = 0.8\textwidth,  trim = 0cm 0cm 0cm 0cm, clip=true]{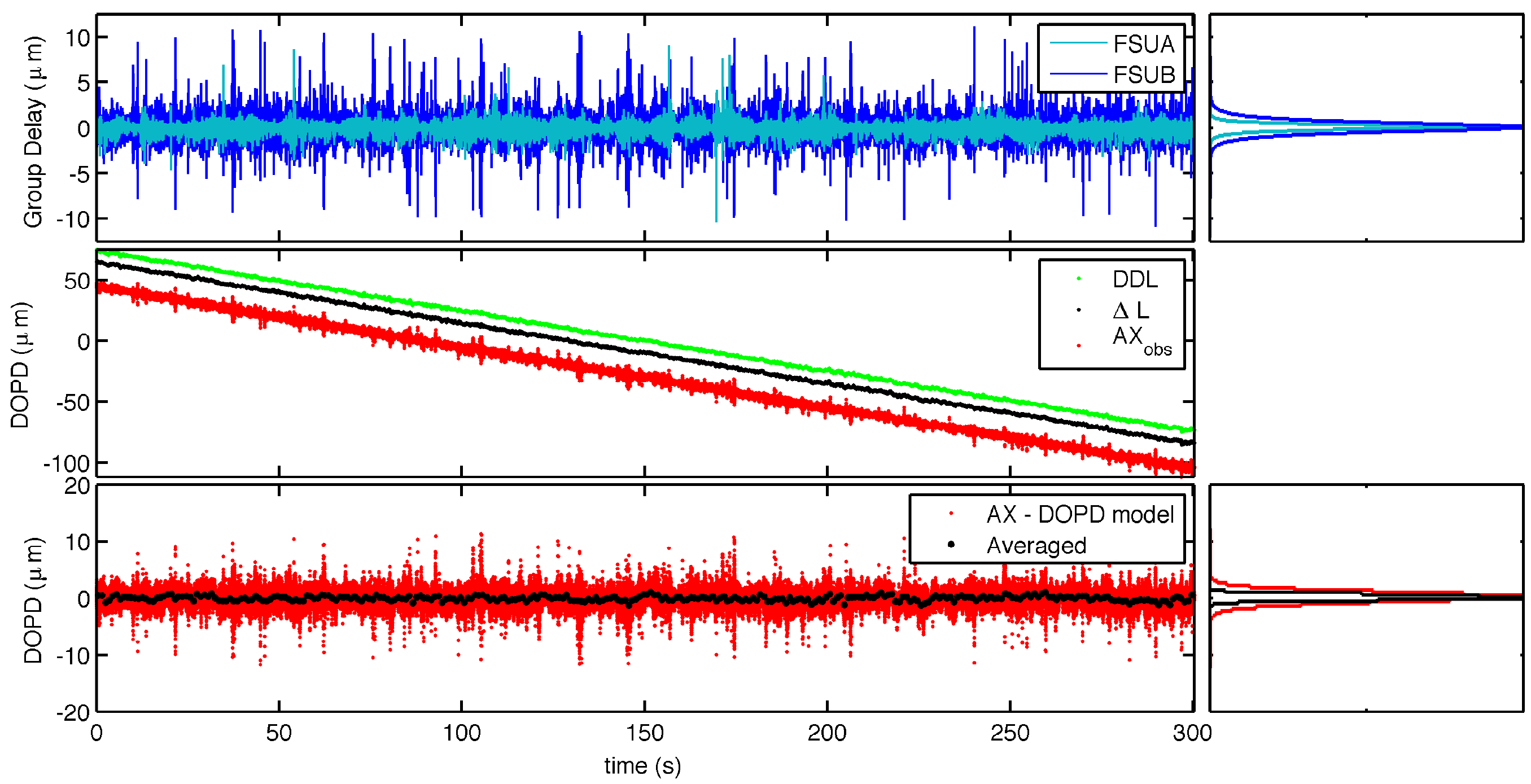}
\caption{Summary graphic of a fringe tracking file. The top panel shows the raw group delays, the middle panel shows the change of differential delay measured with the DDL metrology, the {\small PRIMA} metrology, and the astrometric observable. The raw model-subtracted astrometric observable is shown in red in the lowermost panel and the 1-second averages are shown as solid black circles.}
\label{fig:summary}\end{center}
\end{figure*}

\begin{figure}[h]
\begin{center}
\includegraphics[width = 0.9\linewidth,  trim = 0cm 0cm 0cm 0cm, clip=true]{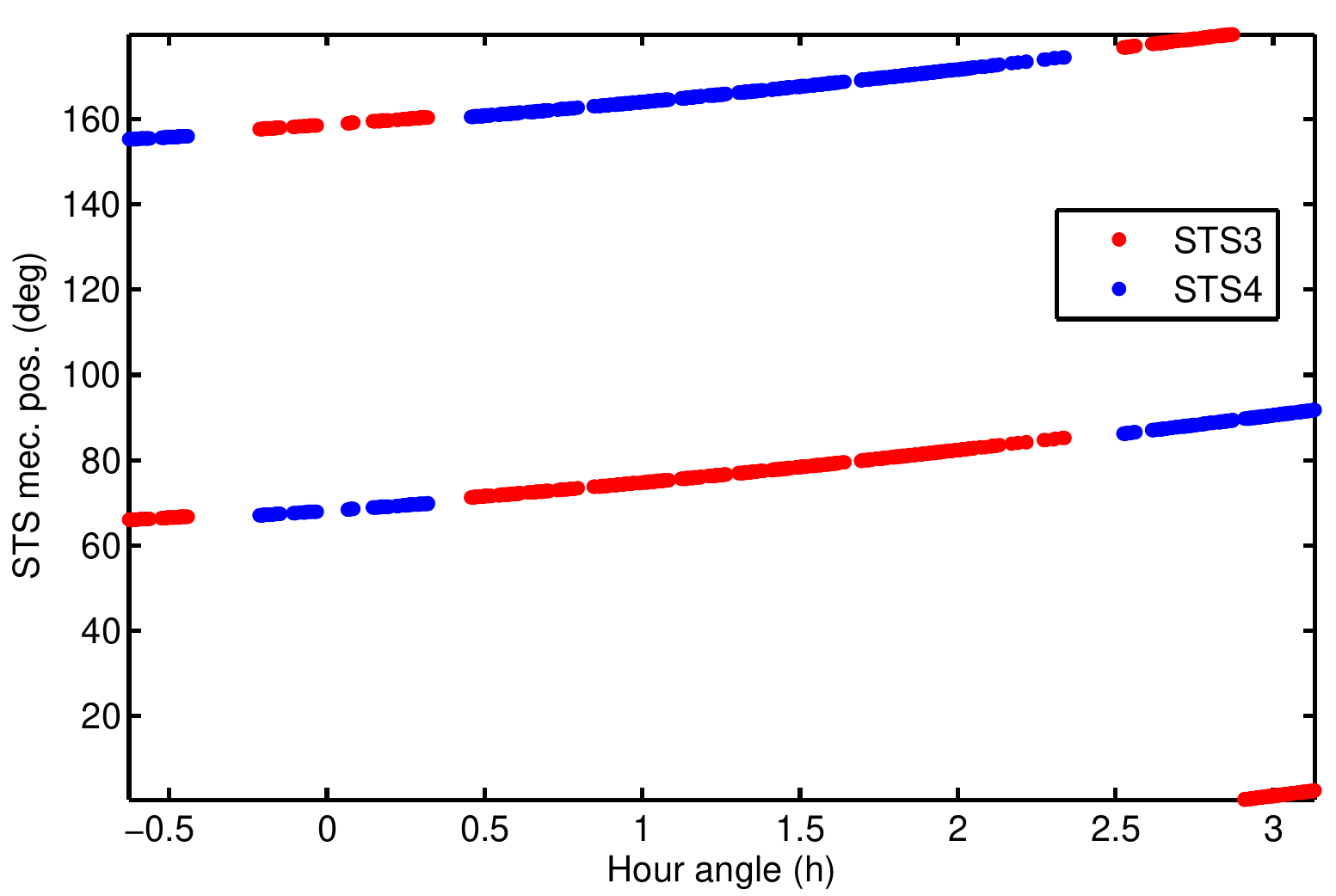}
\caption{Derotator angle of STS3 (red) and STS4 (blue) during the demonstration run. The data is available continuously but only the interpolation on the astrometry data points is shown.}\label{fig:derot20}\end{center}
\end{figure}

 \begin{figure}[h]
 \begin{center}
\includegraphics[width = 0.7\linewidth,  trim = 0cm 0cm 4.7cm 0cm, clip=true]{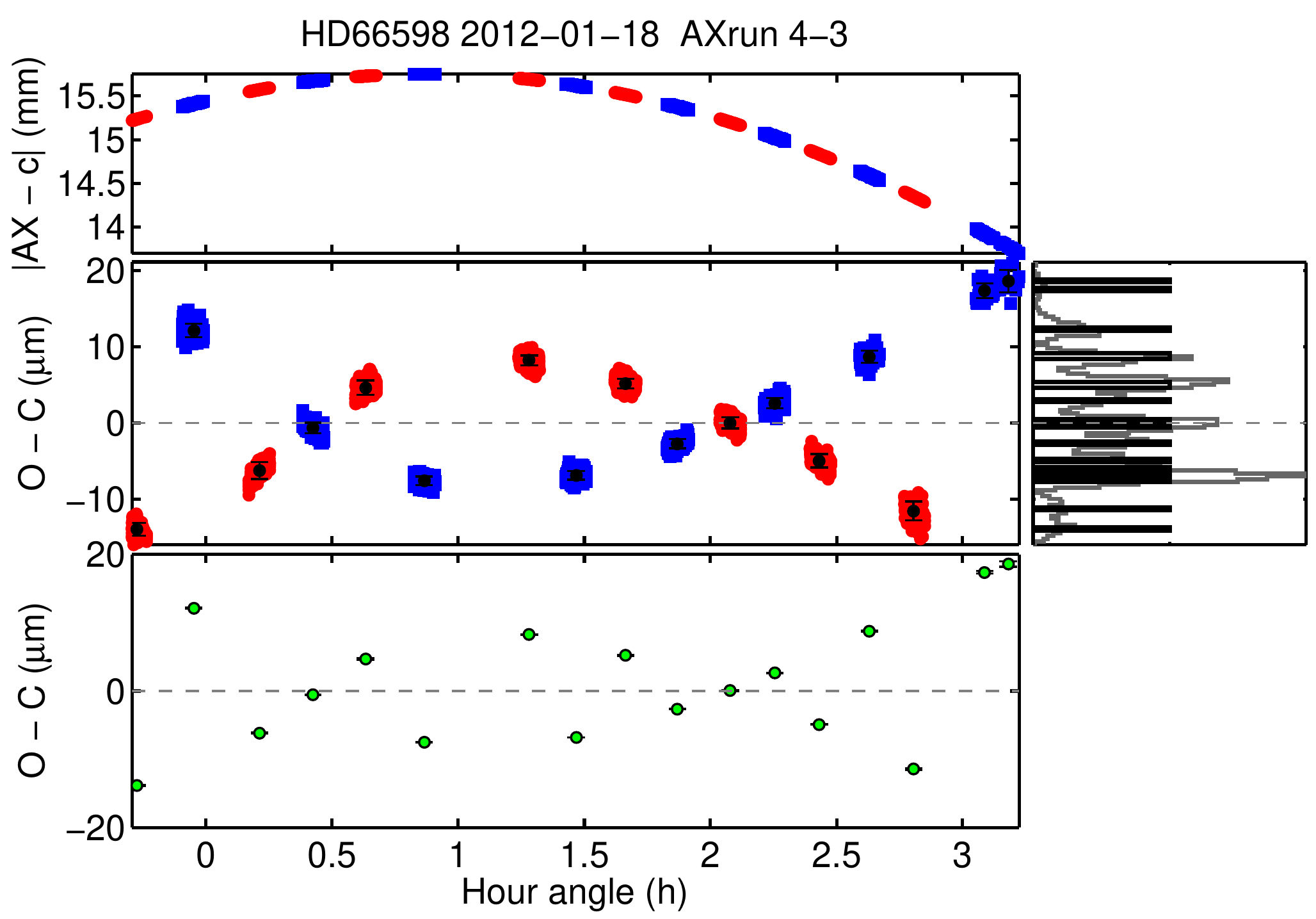}
\includegraphics[width = 0.7\linewidth,  trim = 0cm 0cm 4.7cm 0cm, clip=true]{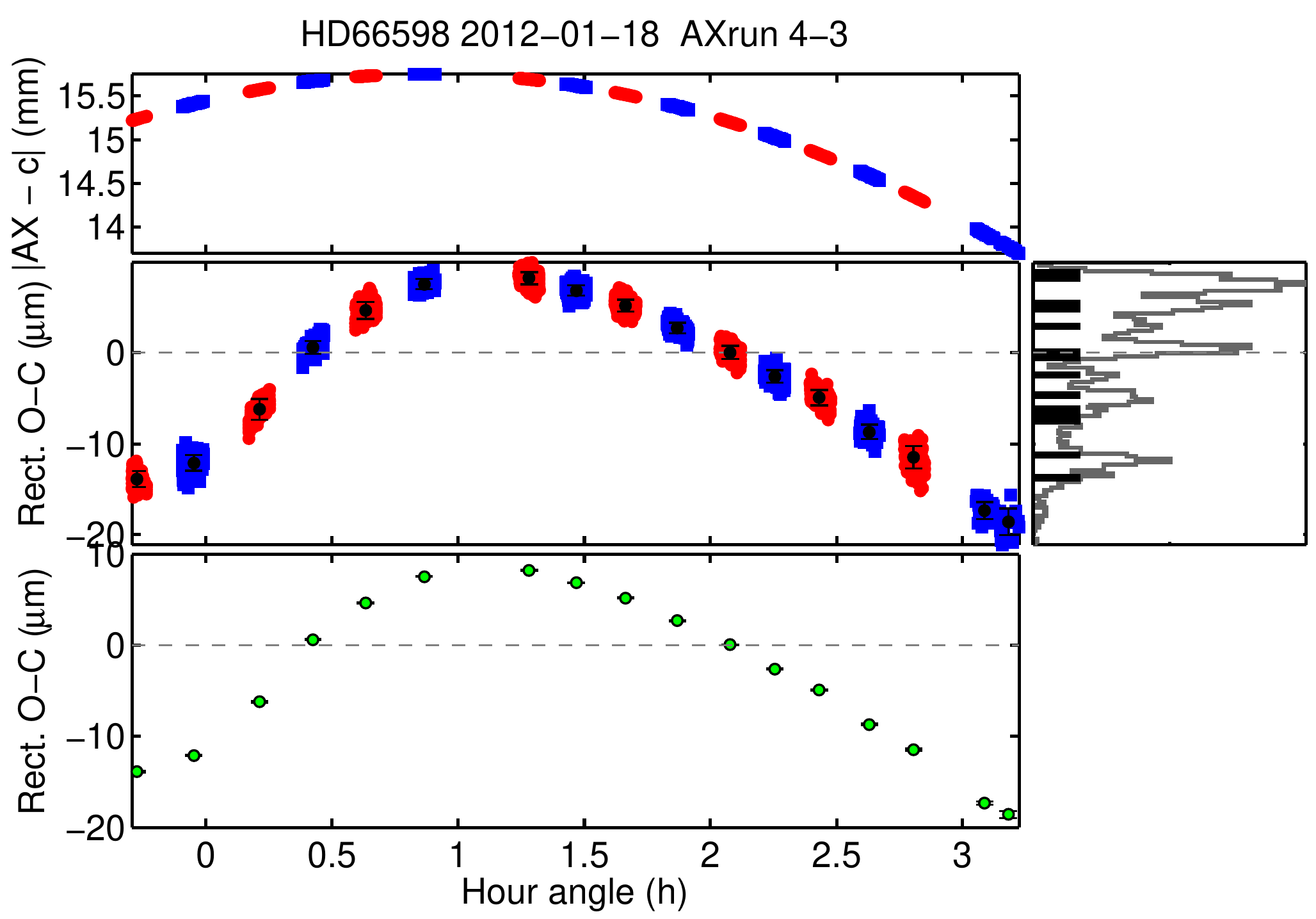}
\caption{Illustration of the fit residuals for HD\,66598 ($\rho\sim36\arcsec$). Colour coding and panel definition as in Fig.~\ref{fig:axres1}. \emph{Top}: Nominal residuals. \emph{Bottom}: Rectified residuals: the O-C of swapped mode data are multiplied with $-1$.}
\label{fig:rectified}\end{center}
\end{figure}

\begin{figure}[h]
\begin{center}
\includegraphics[width = 0.7\linewidth,  trim = 0cm 0cm 4.5cm 0cm, clip=true]{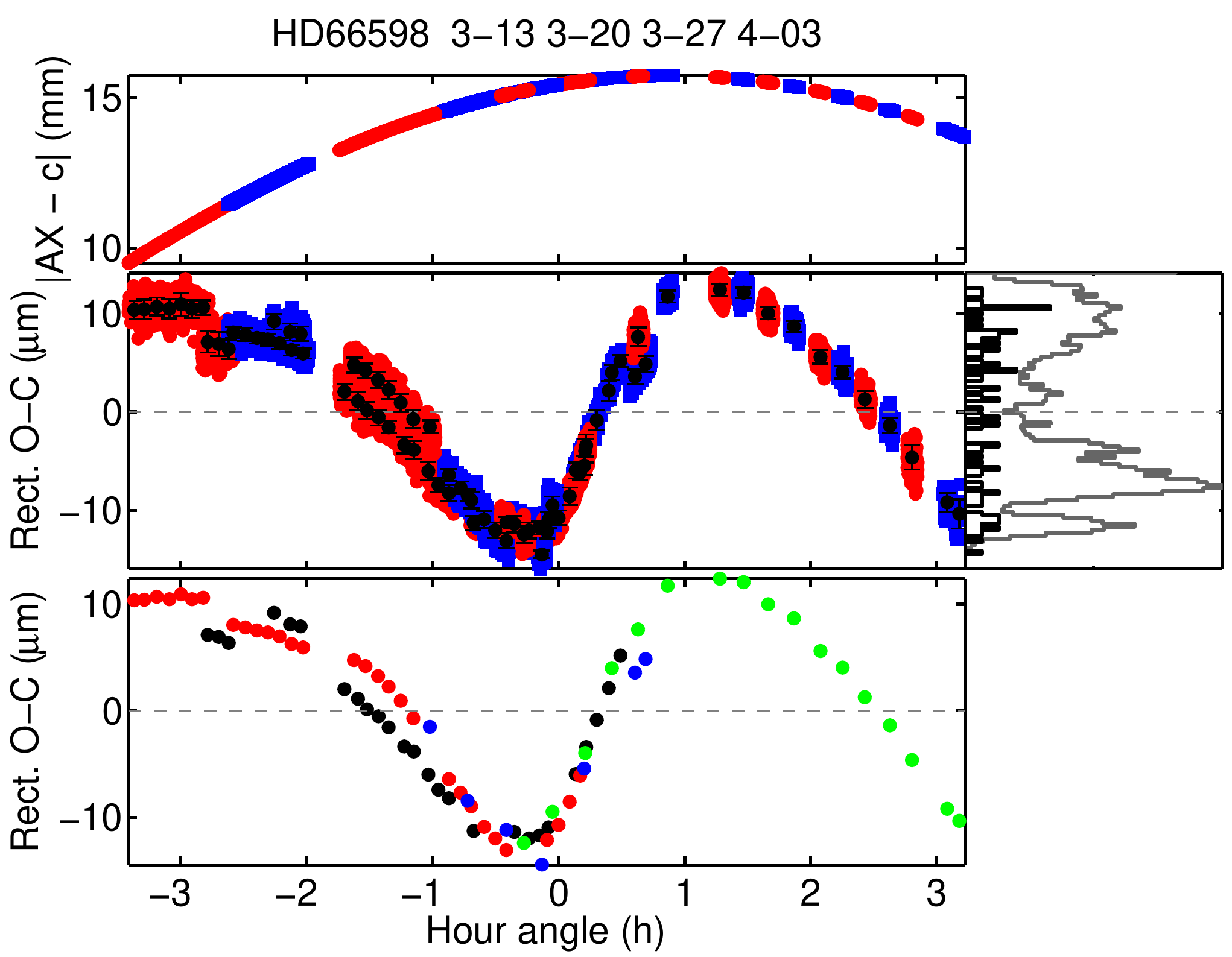}
\caption{Result of the combined fit of four epochs spanning over 56 days for HD\,66598. The colour coding and panel definition is similar to Fig.~\ref{fig:axres1}, except that each colour in the bottom panel marks the data of one epoch.}
\label{fig:rectifiedmulti1}\end{center}
\end{figure}

\begin{figure*}[h]
\begin{center}
\includegraphics[width = 6cm,  trim = 0cm 0cm 4.7cm 0cm, clip=true]{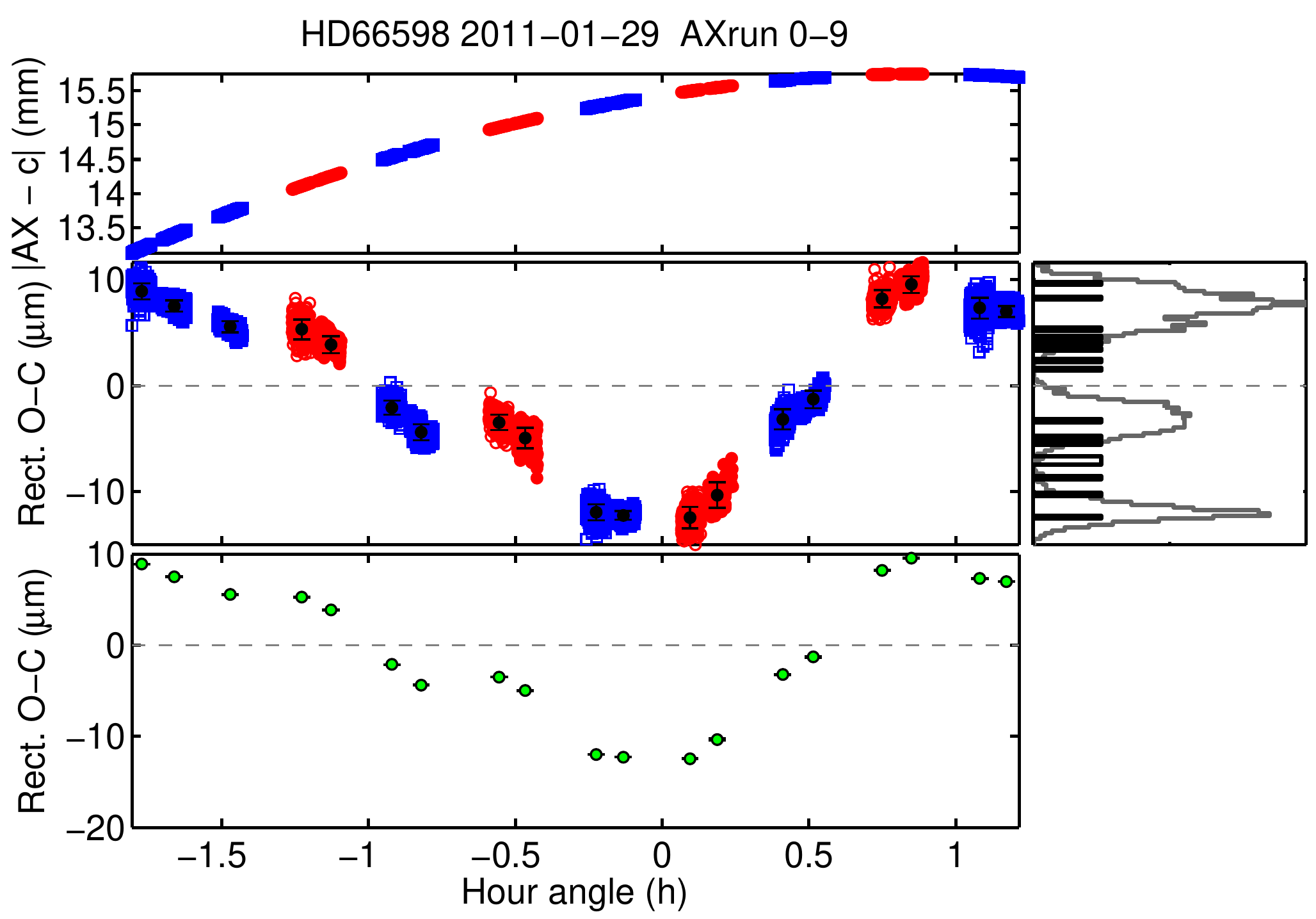}\hspace{6mm}
\includegraphics[width = 6cm,  trim = 0cm 0cm 4.7cm 0cm, clip=true]{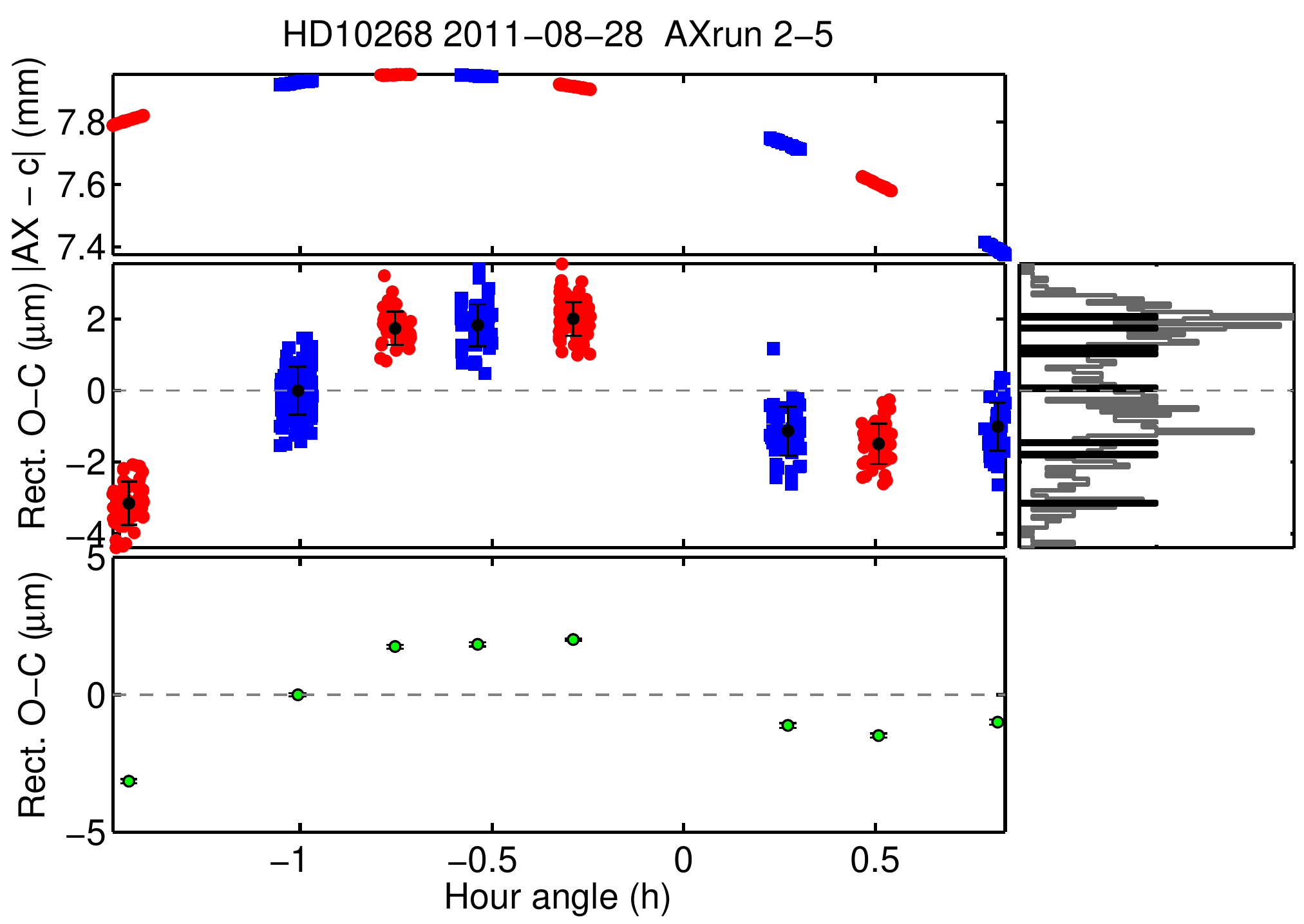}\hspace{4mm}
\includegraphics[width = 6cm,  trim = 0cm 0cm 4.7cm -0.5cm, clip=true]{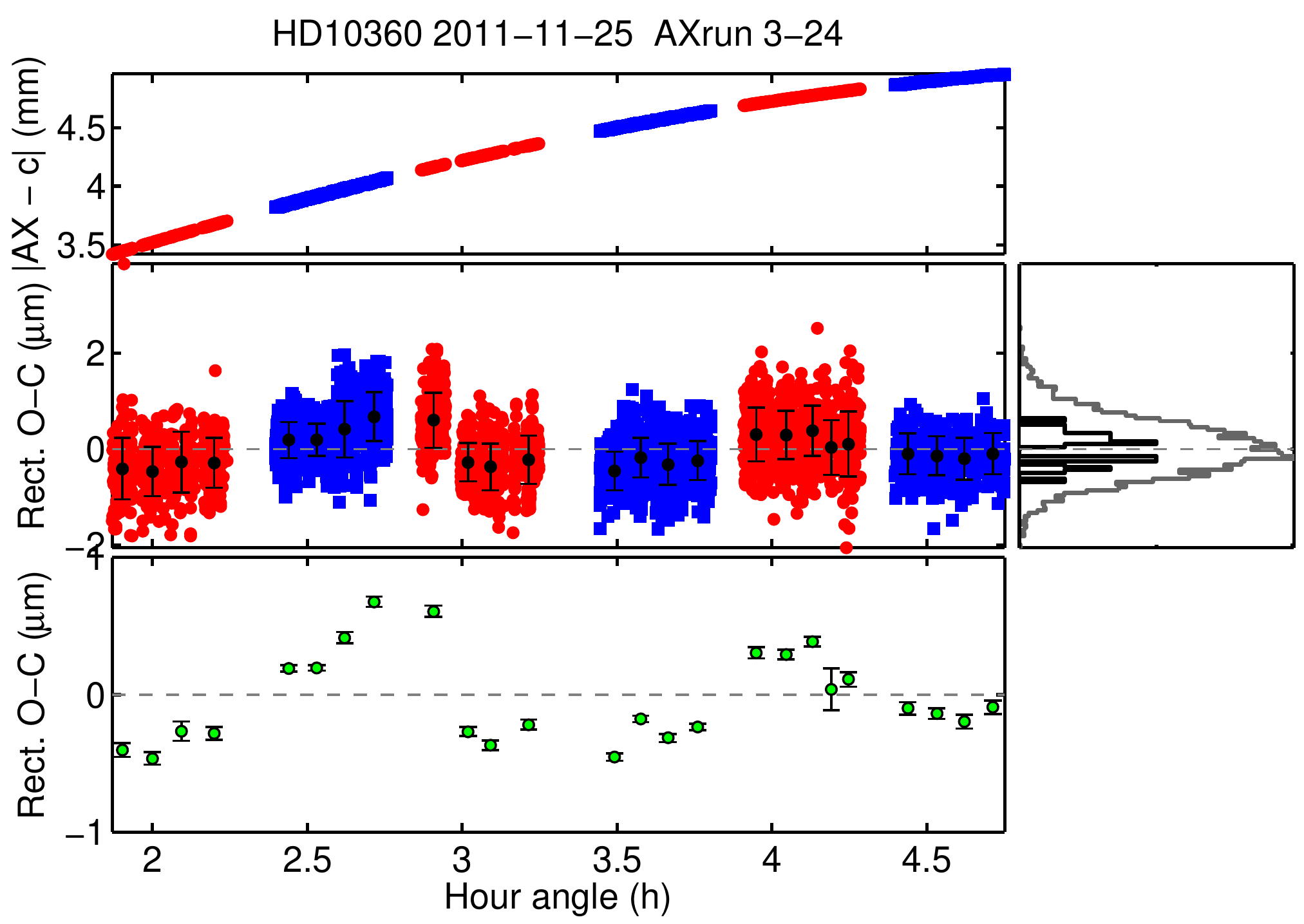}\hspace{4mm}
\includegraphics[width = 6.2cm,  trim = 0cm 0cm 4.55cm -0.5cm, clip=true]{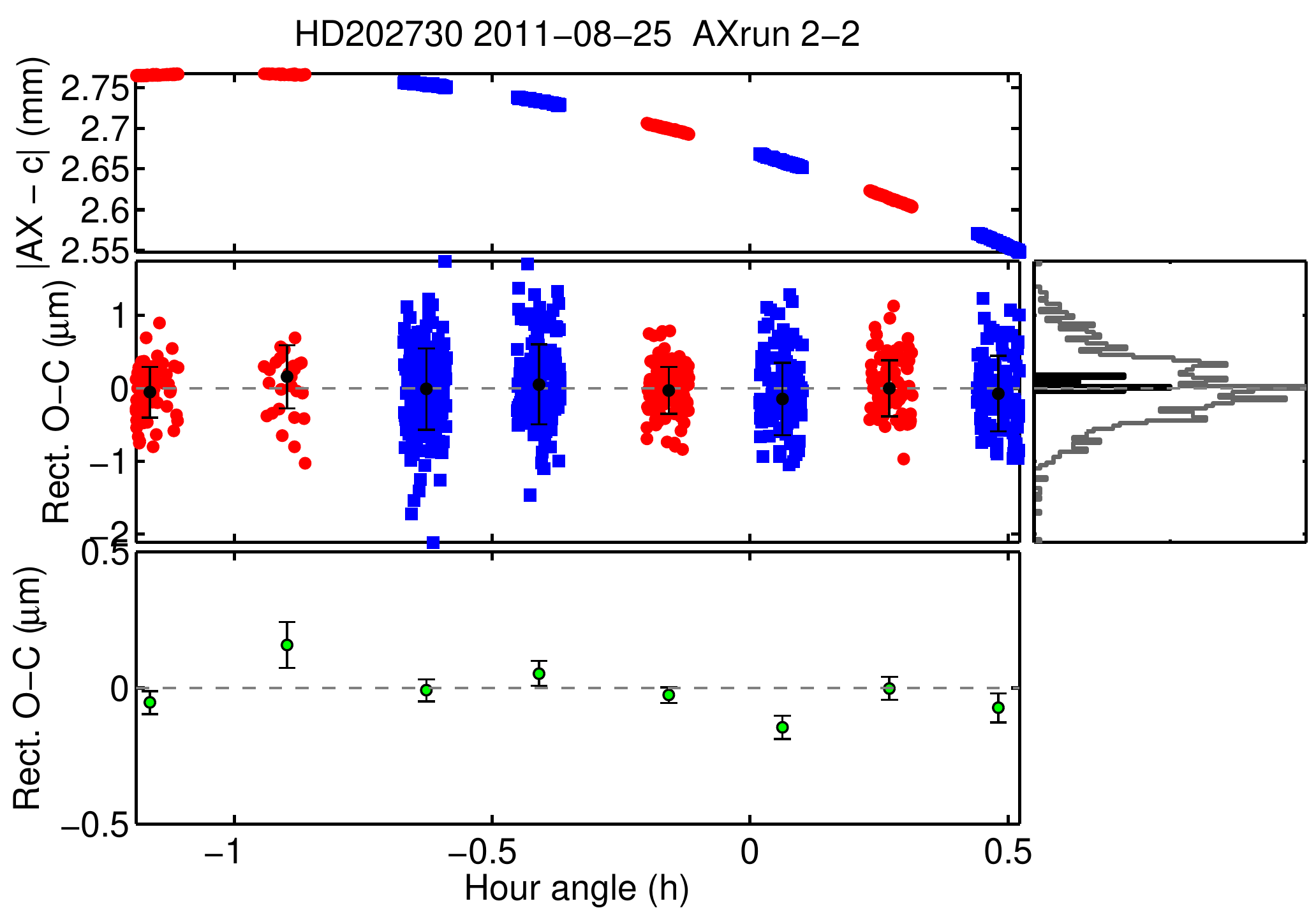}
\caption{Fit residuals for different target separations. \emph{Top left:} HD\,66598 ($\rho\sim36\arcsec$, $\sigma_\mathrm{O-C, binned} = 7.9~\mu$m). \emph{Top right:} HD\,10268 ($\rho\sim20\arcsec$, $\sigma_\mathrm{O-C, binned} = 1.9~\mu$m). \emph{Bottom left:} HD\,10360 ($\rho\sim11\arcsec$, $\sigma_\mathrm{O-C, binned} = 0.33~\mu$m). \emph{Bottom right:} HD\,202730 ($\rho\sim7\arcsec$, $\sigma_\mathrm{O-C, binned} = 0.08~\mu$m).}
\label{fig:rectified2}\end{center}
\end{figure*}

\end{appendix}

\end{document}